\documentclass{JHEP3}

\usepackage{amsfonts,amsmath,amssymb}
\usepackage{epsfig,multicol,bbm}
\usepackage{colonequals}
\usepackage[nice]{nicefrac}
\usepackage{cite}
\usepackage{tabularx}

\newcommand\fverb{\setbox\fverbbox=\hbox\bgroup\verb}
\newcommand\fverbdo{\egroup\medskip\noindent
			\fbox{\unhbox\fverbbox}\ }
\newcommand\fverbit{\egroup\item[\fbox{\unhbox\fverbbox}]}
\newbox\fverbbox

\newcommand{\calN}{\mathcal{N}}
\newcommand{\calS}{\mathcal{S}}
\newcommand{\calM}{\mathcal{M}}

\newcommand{\calR}{\mathcal{R}}
\newcommand{\ft}{\mathfrak{t}}\newcommand{\fq}{\mathfrak{q}}
\newcommand{\fx}{\textsf{x}}\newcommand{\fy}{\textsf{y}}
\newcommand{\SO}{\text{SO}}
\newcommand{\SU}{\text{SU}}
\newcommand{\SP}{\text{SP}}
\newcommand{\E}{\text{E}}
\newcommand{\U}{\text{U}}

\def\be{\begin{equation}}
\def\ee{\end{equation}}
\def\bea{\begin{eqnarray}}
\def\eea{\end{eqnarray}}

\title{Fiber-Base Duality and Global Symmetry Enhancement}

\preprint{DESY 14-212
\\
HU-Mathematik - 2014-32
\\
HU-EP-14/48
\\
KIAS-P14066
}

\author{Vladimir Mitev$^{a}$\footnote{Email: mitev@math.hu-berlin.de}
$\,$, Elli Pomoni$^{b,c}$\footnote{Email: elli.pomoni@desy.de}
$\,$, Masato Taki$^{d}$\footnote{Email: taki@riken.jp}
$\,$  and Futoshi Yagi$^{e}$\footnote{Email: fyagi@kias.re.kr}
\\
\\
\\
\it $^a$ Institut f\"ur Mathematik und Institut f\"ur Physik, Humboldt-Universit\"at zu Berlin, IRIS Haus, Zum Gro{\ss}en Windkanal 6,  12489 Berlin, Germany
\\
\it $^b$ DESY Theory Group, Notkestra{\ss}e 85, 22607 Hamburg, Germany
\\
\it $^c$ Physics Division, National Technical University of Athens,
15780 Zografou Campus, Athens, Greece
\\
\it $^d$ iTHES Research Group and Mathematical Physics Laboratory, RIKEN Nishina Center, Saitama 351-0198, Japan
\\
\it $^e$ Korea Institute for Advanced Study (KIAS)
85 Hoegiro Dongdaemun-gu, 130-722, Seoul, Korea
}

\abstract{

\bigskip

We show that the 5D Nekrasov partition functions
enjoy the enhanced global symmetry of the UV fixed point.
The fiber-base duality is responsible for the global symmetry enhancement.
For $\SU(2)$ with $N_f\leq 7$ flavors the fiber-base symmetry together with the manifest flavor $\SO(2N_f)$ symmetry generate the $\E_{N_f+1}$ global symmetry, while in the higher rank case the manifest global symmetry of the two dual theories related by the fiber-base duality map generate the symmetry enhancement. The symmetry enhancement at the level of the partition function is manifest once we chose an appropriate reparametrization for the Coulomb moduli.

}


\begin{document}


\tableofcontents
\addtolength{\baselineskip}{5pt}

\section{Introduction}

Gauge theories in five dimensional (5D) spacetime are perturbatively non-renormalizable,  infrared (IR) free and may have a Landau pole singularity in the ultraviolet (UV) region. 
Thus, usually, they can only be thought of as the effective low energy limit of some other theory. 
However, in the seminal paper \cite{Seiberg:1996bd}, Seiberg argued using string theory constructions that the 5D $\mathcal{N}=1$ $\SU(2)$ gauge theories 
coupled to $N_f\leq 7$ hypermultiplets\footnote{For $N_f> 8$ there is a Landau pole in the Coulomb branch at $a \sim \frac{1}{g^2_{cl}(N_f - 8)}$, while for $N_f=8$
the effective coupling is a constant and it is impossible to take the strong coupling limit.} have UV fixed points with enhanced global symmetry $\E_{N_f+1}$.
The presence of these non-trivial UV fixed points suggest that these
gauge theories do really ``exist''. The IR free theory with finite coupling should be thought of as a
perturbation away from the UV fixed point triggered by
the relevant operator $\frac{1}{g^2}F_{\mu \nu}^2$ as well as its supersymmetry partner.
Seiberg's arguments were shortly thereafter extended to other 5D theories with various color groups and matter content, see \cite{Morrison:1996xf,Intriligator:1997pq} for the
necessary conditions for the existence of non-trivial UV fixed points.

Until recently, there was difficulty in verifying with a field theory argument or calculation the existence of such UV fixed points with enhanced global symmetry since the corresponding UV fixed points are strongly coupled. However, the recent developments in supersymmetric localization technique
open up the possibility of investigating such SCFTs directly and quantitatively.
In the pioneering paper \cite{Kim:2012gu} the enhancement to $\E_{N_f+1}$ symmetry was directly checked by showing that the superconformal indices of the $\SU(2)$ gauge theories have $\E_{N_f+1}$ symmetry. 
In this paper, we push this even further by showing that the Nekrasov partition functions are invariant under the enhanced  $\E_{N_f+1}$ symmetry as well.

The superconformal index is the partition function of the protected operators  \cite{Romelsberger:2005eg,Kinney:2005ej} of a given theory, up to a sign for the fermionic states. It counts the multiplets obeying shortening conditions, up to equivalence relations setting to zero all combinations of short multiplets that may in principle recombine into long multiplets \cite{Gadde:2009dj}.
As such it is independent of the coupling constants of the theory, is invariant under continuous deformations of the theory,  S-duality \cite{Gadde:2009kb}
and can therefore be evaluated in the free field limit if a Lagrangian description is known.
The superconformal index  has a path integral representation \cite{Kinney:2005ej} as the partition function of the theory on $S^{d-1}\times S^1$, twisted by various chemical potentials, and can be evaluated using localization techniques \cite{Kim:2009wb,Imamura:2011su,Nawata:2011un,Kim:2012gu,Peelaers:2014ima}.

The 5D superconformal index was computed via localization on $S^4\times S^1$ in  \cite{Kim:2012gu} where it was shown that
\be
\mathcal{I}^{\text{5D}}= \int [d a] \, \text{PE}[i(\text{quiver})] \, |Z_{\textrm{inst}}^{\text{5D}}|^2  \, ,
\ee
where $[d a]$ is the invariant Haar measure, PE the Plethystic exponential that gives the multi-particle index from the free single-particle index $i(\text{quiver})$ of a given quiver and $Z_{\textrm{inst}}^{\text{5D}}$ the K-theoretic instanton partition function that localizes on the north and south poles of the $S^4$.
The 5D theories in the IR  are weakly coupled and have a  Lagrangian description. The index computed in the IR, as long as no protected states are ``lost''  while flowing from the UV fixed point to the IR, will
 reorganize itself into characters of the enhanced symmetry of the UV fixed point.
Kim, Kim and Lee in \cite{Kim:2012gu} checked that indeed the index of \SU(2) with $N_f\leq 7$ flavors can be expressed by characters of the groups $\E_{N_f+1}$.\footnote{See also \cite{Hwang:2014uwa}, especially for the cases $N_f=6, 7$.}
Similarly, symmetry enhancement was seen in a few more cases using superconformal indices \cite{Bergman:2013aca,Bao:2013pwa,Hayashi:2013qwa,Taki:2013vka}.\footnote{Computing the superconformal index for theories of higher rank gauge groups  is technically hard because one has to perform a number of integrals $\int [d a]$ (at $N\rightarrow \infty$ things simplify again).
}

In \cite{Iqbal:2012xm} Iqbal and Vafa  pointed out that  the 5D superconformal index is given by a contour integral of the square of  the 5D Nekrasov partition function which can be computed using the topological string partition function
\be
\label{indexNekTop}
\mathcal{I}^{\text{5D}}= \int d a \, |Z_{\textrm{Nek}}^{\text{5D}} (a)|^2\propto \int d a \, |Z_{\textrm{top}} (a)|^2 \, .
\ee
It is natural to expect that the holomorphic part of the integrand $Z_{\textrm{Nek}}^{\text{5D}} (a) \propto Z_{\textrm{top}} (a)$ will enjoy this extended symmetry too, since the 5D Nekrasov partition function counts the BPS spectrum of the low energy theory on $R^4 \times S^1$ in the Coulomb phase and it is natural to expect that the global symmetry structure is encoded at the spectrum of the Coulomb branch.

It is the purpose of this paper to show that this symmetry enhancement is indeed visible already in the holomorphic part of the partition function $Z_{\textrm{Nek}}^{\text{5D}} (a)$ and to explain that this is possible due to the the fiber-base duality \cite{Katz:1997eq,Aharony:1997bh,Bao:2011rc}.
Especially, for the $\SU(2)$ theories, the fiber-base duality should be interpreted as a symmetry of the theory and thus as an extension of the global symmetry, which we denote also as ``fiber-base symmetry'' in this paper. By combining the generators of this fiber-base symmetry with the rest of the global symmetry one obtains the full extended symmetry.

Although from a purely field theoretic point of view the fiber-base duality may seem mysterious,  it can be easily understood if the gauge theory is embedded in string/M-theory\footnote{The two different descriptions below are related with string duality and the Toric diagram is the same as the web diagram \cite{Leung:1997tw}. }:
\begin{itemize}
\item
Geometric engineering on $CY_3$: exchanging the base $\mathbb{P}^1$s with the fiber $\mathbb{P}^1$s is a symmetry of the Calabi-Yau manifold \cite{Katz:1997eq},
\item
$(p,q)$ 5-brane web diagrams in type IIB:  S-duality exchanges D5-branes with NS5-branes  \cite{Aharony:1997bh},
\end{itemize}
with the Coulomb moduli, the coupling constants and the masses being exchanged in a non-trivial manner.

The (refined) topological string partition function can be computed by using the (refined) topological vertex method \cite{Aganagic:2003db,Iqbal:2007ii}.
It is read off from the toric diagram as a function of the K\"ahler parameters for the 2-cycles and can be interpreted as the Nekrasov partition function of the corresponding gauge theory after the appropriate identification of the K\"ahler parameters with the gauge theory parameters \cite{Iqbal:2003ix,Iqbal:2003zz,Taki:2007dh} has been made.

In \cite{Bao:2011rc,Bao:2013wqa} we studied the fiber-base duality  between the low energy effective theory of the 5D $\mathcal{N}=1$ $\SU(N)^{M-1}$ and the $\SU(M)^{N-1}$ linear quiver gauge theories compactified on $S^1$.
The two different gauge theories that are related  by fiber-base duality have the same toric diagram, up to a 90 degree rotation.
The duality map  \cite{Bao:2011rc}  is obtained by comparing these two parametrizations for the dual theories. In addition, the same result is obtained by comparing the SW curves.
For  the unrefined topological string, the fiber-base duality at the level of the 5D Nekrasov partition function is immediately apparent by using the  topological vertex method.
For the refined case, this also holds if we assume that the partition function does not depend on the choice of the preferred direction (slicing invariance).
This case was studied in a subsequent paper by Ito \cite{Ito:2012bb} who, compared the refined Nekrasov partition functions of the two dual theories and obtained the refined map between the gauge theory parameters. As we will explain in section \ref{sec:fiberbaseinvariance},
the refined map that he found in \cite{Ito:2012bb} is the same as our unrefined map \cite{Bao:2011rc}, up to a shift of the
masses   by $\epsilon_+/2$. The  fiber-base duality can be further understood at the level of the integrable models that are dual to the gauge theories under consideration as a spectral duality between the two different integrable systems \cite{Mironov:2013xva}.

Furthermore, a very useful way that has been used recently  to see the symmetry enhancement is through the manipulation of $7$-branes  \cite{DeWolfe:1999hj, Bao:2013pwa,Taki:2013vka,Taki:2014pba}, see also the works  \cite{Gaberdiel:1997ud,Gaberdiel:1998mv,DeWolfe:1998zf,Iqbal:1998xb,DeWolfe:1998bi,DeWolfe:1998eu,DeWolfe:1998pr} as well as \cite{Ganor:1996mu, Seiberg:1996vs, Witten:1996qb, Morrison:1996na, Morrison:1996pp, Ganor:1996gu, Klemm:1996hh, Ganor:1996xd, Ganor:1996pc, Minahan:1998vr}.  Using 7-branes, we show in section \ref{sec:7brane} that the $\SU(2)$  gauge theory with $N_f$ flavors has the enhanced symmetry $\E_{N_f +1}$, that the $\SU(N)$ and its fiber-base dual $\SU(2)^{N-1}$ gauge theories have  $\SU(2N) \times \SU(2)^2$ enhanced symmetry and finally that
the $\SU(N)^{M-1}$ theory and its fiber-base dual $\SU(M)^{N-1}$ enjoys an $\SU(N)^2 \times \SU(M)^2$ global symmetry. The 7-brane technique captures only the non-abelian part of the symmetry and in section \ref{sec:symmetryenhancement}, we show that the symmetry of the last case is actually $\SU(N)^2 \times \SU(M)^2 \times \U(1)$.

 An important outcome of our work is  the  understanding that the fiber-base duality leads to the symmetry enhancement of the 5D theory.  For the simplest case of the pure $\SU(2)$ gauge theory the fiber-base duality leads to an $\E_1$ symmetry enhancement,  as was already pointed out by \cite{Aharony:1997bh}. In section \ref{sec:fiberbaseinvariance},
we  generalize for $\SU(2)$ gauge theories with $N_f \leq 7$ flavors. For these theories, the Weyl symmetry of $\SO(2N_f)$ is manifest in the Nekrasov partition\footnote{Only $\SU(N_f)$ is manifest for  the $U(2)$  Nekrasov partition function. $\SO(2N_f)$ is manifest for the $\SP(1)=\SU(2)$ partition function \cite{Bao:2013pwa,Hayashi:2013qwa, Bergman:2013aca, Bergman:2013ala}} function. The duality map generates the Weyl symmetry of $\E_{N_f+1}$ and thus the enhanced global symmetry.
We will also see that the fiber-base duality plays important role to understand the enhanced global symmetry also for higher rank case\footnote{Analogous observation has been done also in \cite{Zafrir:2014ywa}.}.

We look for their symmetry enhancement by studying the holomorphic half of the integrand of the index \eqref{indexNekTop}, the Nekrasov partition function $Z_{\textrm{Nek}}^{\text{5D}} (a)$.
The Coulomb moduli are parameters that are going to be integrated out anyway, 
so we are allowed to redefine them without changing the know results and we do that in such a way so that the {\it new Coulomb moduli are invariant under the enhanced global symmetry}.
Note that the number of $a$'s is the same in both dual theories. 
In terms of this new Coulomb moduli $\tilde{A}$, we expand the holomorphic piece of the integrand and we discover that the
coefficients of $\tilde{A}^n$ organize themselves in characters of the enhanced symmetry, thus the
enhanced symmetry is manifest. For the $\SU(2)$ theories with $N_f \leq 7$ flavors we find
\bea
Z^{N_f}=1-\left(\sum_{\lambda}\chi_{\lambda}^{\E_{N_f+1}} + c(\ft, \fq) \right)\frac{\ft^{\frac{1}{2}}\fq^{\frac{1}{2}}}{(1-\ft)(1-\fq)}\tilde{A}+\cdots.
\eea
where the coefficients\footnote{The function $c(\ft, \fq)$ is non-vanishing only for $N_f=7$.} of the expansion are characters of the expected enhanced $\E_{N_f+1}$ symmetry. 
The Nekrasov partition functions for IR gauge theories thus carry the hidden $\E_{N_f+1}$ global symmetry at the strongly-coupled UV fixed point.


\section{Enhanced Global Symmetry from 7-branes}
\label{sec:7brane}

In this section, we review the technique to determine the global symmetry 
by using the 7-brane monodromy following 
\cite{Gaberdiel:1997ud,Gaberdiel:1998mv,DeWolfe:1998zf,Iqbal:1998xb,DeWolfe:1998bi,DeWolfe:1998eu,DeWolfe:1998pr}.
Especially, we discuss the $\SU(2)$ theory with $N_f$ flavor \cite{DeWolfe:1999hj} 
as well as the $\SU(N)^{M-1}$ linear quiver theory.


\subsection{7-branes and enhanced symmetry}

Type IIB 7-branes are magnetic sources of the IIB dilaton-axion and can carry $(p,q)$ charges. The usual D7-brane carries an $(1,0)$ charge while a $(p,q)$ 7-brane is obtained by acting on a D7 brane with an $SL(2,\mathbb{Z})$ transformations.
In this paper ${\bf X}_{(p,q)}$ will denote a $(p,q)$ 7-brane,
the $(p,q)$  charge of which  is defined up to an overall sign ambiguity ${\bf X}_{(-p,-q)}\equiv  {\bf X}_{(p,q)}$.
The symplectic inner product between two $(p,q)$ charges is defined as
\begin{align}
z_i\equiv (p_i,q_i),\quad
z_i\wedge z_j
\equiv\det
\left(\begin{array}{cc}p_i\,\,\, &p_j\\ q_i \,\,\,& q_j\end{array}\right).
\end{align}
Since a 7-brane is charged under the IIB dilation and axion,
a 7-brane ${\bf X}_{(p,q)}$ induces a branch cut in the transversal $2$-plane. So a 7-brane changes its charge if it crosses a branch cut coming from another 7-brane. Let us consider adjoining two 7-branes ${\bf X}_{z_1}{\bf X}_{z_2}$. In our convention the branch cuts go downward. Under their reordering, one of these 7-branes will pass a branch cut
and then change its charge as
\begin{align}
\label{BCM}
{\bf X}_{z_1}{\bf X}_{z_2}=
{\bf X}_{z_2+(z_1\wedge z_2)z_1}{\bf X}_{z_1}=
{\bf X}_{z_2}{\bf X}_{z_1+(z_1\wedge z_2)z_2}.
\end{align}

These 7-branes are key to understanding the enhancement of the global symmetry of 5D SUSY theories. Let us consider the gauge symmetry of the world-volume theory on coinciding 7-branes. Not all 7-brane configurations are collapsible and the possible gauge symmetries on 7-branes are already classified \cite{DeWolfe:1998eu,DeWolfe:1998pr}. Since a 7-brane is a source of the dilaton-axion, this scalar experiences a monodromy transformation when moving around a 7-brane.
A 7-brane ${\bf X}_{(p,q)}$ develops the following monodromy
\begin{align}
K_{(p,q)}=
\left(\begin{array}{cc} 1+pq & -p^2 \\ q^2 & 1-pq \end{array}\right).
\end{align}
The possible 7-brane configurations are classified by using this monodromy.
The necessary condition \cite{DeWolfe:1998eu} that the set of 7-branes ${\bf X}_{(p_1,q_1)}\cdots {\bf X}_{(p_n,q_n)}$
can be collapsed to a point\footnote{Beware of the reversing of the order in \eqref{eq:tracesK}.} is
\begin{align}
\label{eq:tracesK}
\textrm{Tr}\left(K_{(p_n,q_n)}\cdots K_{(p_1,q_1)}\right)
\in\{-2,\,
-1,\,0,\,1,\,2\}.
\end{align}
On these coinciding 7-branes, a mysterious enhanced gauge symmetry appears.
Up to an overall $SL(2,\mathbb{Z})$ transformation and reordering,
all the collapsible 7-brane configurations and the corresponding gauge symmetry are
\begin{align}
&A_n:& & {\bf X}_{(1,0)}^{n+1},&\\
&D_{n+4}:& &{\bf X}_{(1,0)}^{n+4}\,{\bf X}_{(1,-1)}\,{\bf X}_{(1,1)},&\\
&\E_{n=6,7,8}:& &{\bf X}_{(1,0)}^{n-1}\,{\bf X}_{(1,-1)}\,{\bf X}_{(1,1)}^2,&\\
&H_{n=0,1,2}:&  &{\bf X}_{(1,0)}^{n+1}\,{\bf X}_{(1,1)},&
\end{align}
where the world-volume of an $H_0$ 7-brane configuration carries no symmetry, $H_1$ carries $\SU(2)$ and $H_2$ carries $\SU(3)$.
Probing these 7-brane gauge symmetries, we can explain the enhancement of the flavor symmetry in 5d supersymmetric theories.

In the following, we demonstrate this method for various examples.

\subsection{Enhancement in $\SU(2)$ gauge theories}

\FIGURE{
\includegraphics[width=12cm]{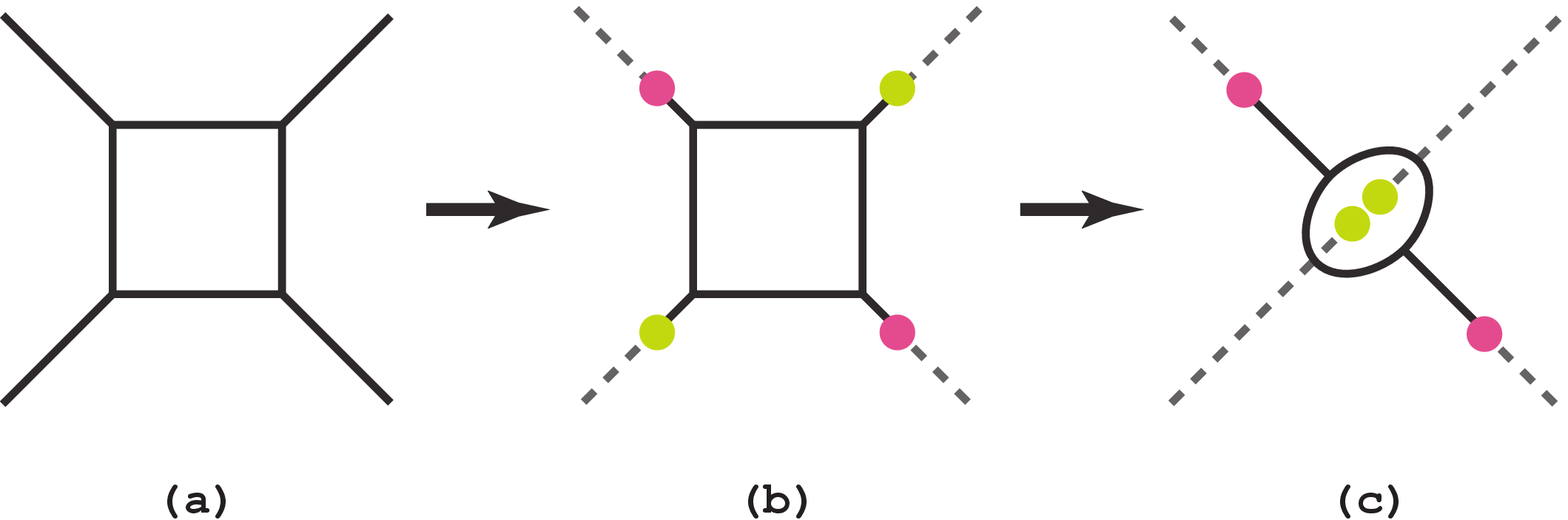}
 \caption{In (a) we see the 5-brane web configuration that gives $\SU(2)$ pure Yang-Mills theory. We can regularize it by introducing 7-branes as (b), where red circles are $(1,-1)$ 7-branes and yellow circles are $(1,1)$ 7-branes. The dashed lines are branch cuts. In (c) we depict a possible collapsed configuration that describes the UV fixed point.}
 \label{fig:SU2YM}
}

The 5D $\SU(2)$ pure Super Yang-Mills theory is given by the 5-brane configuration of figure~\ref{fig:SU2YM}(a).
This web is equivalent to part (b) of this figure since the positions of the 7-branes along the diagonal geodesics do not affect the 5D physics.
To study the UV fixed point theory, we need to consider collapsed configuration because fixed point do not have any scale-full parameter.
However this ${\bf X}_{(1,-1)}{\bf X}_{(1,1)}{\bf X}_{(1,-1)}{\bf X}_{(1,1)}$ configuration is not collapsible,
and the maximally collapsed configuration is shown in part (c) of figure~\ref{fig:SU2YM}.
The coinciding two 7-branes carry an $\SU(2)$ gauge symmetry,
and this symmetry becomes the enhanced flavor symmetry on the probing 5-brane configuration.

\begin{figure}[h]
 \begin{center}
\includegraphics[width=14cm]{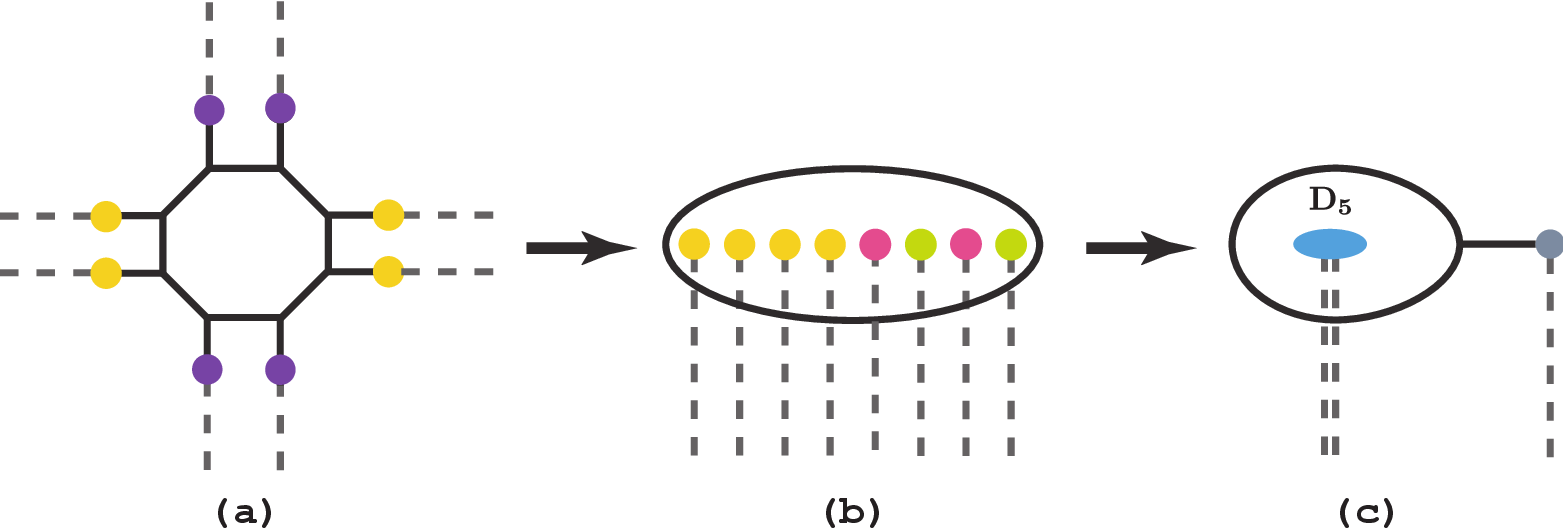}
 \end{center}
 \caption{The 5-brane web configuration that gives $\SU(2)$ gauge theory with 4 flavors (a).
 We can translate it into the 5-brane probe of the 7-brane configuration ${\bf X}_{(1,0)}^4{\bf X}_{(1,-1)}{\bf X}_{(1,1)}{\bf X}_{(1,-1)}{\bf X}_{(1,1)}$ as (b).
 Then (c) is a maximally collapsed configuration that describes the UV fixed point.}
 \label{fig:SU2Nf4}
\end{figure}
The $\SU(2)$ gauge theory with 4 flavors given by the 5-brane configuration of part (a) of figure~\ref{fig:SU2Nf4} is a more interesting case.
We can recast this web into the 5-brane loop on the 7-brane background $\hat{{\bf E}}_{5}={\bf X}_{(1,0)}^4{\bf X}_{(1,-1)}{\bf X}_{(1,1)}{\bf X}_{(1,-1)}{\bf X}_{(1,1)}$,
but these 7-branes are not collapsible coincidentally.
We can find a  ${\bf D}_5$ configuration in this background \cite{DeWolfe:1999hj}
\begin{align}
{\bf X}_{(1,0)}^4{\bf X}_{(1,-1)}{\bf X}_{(1,1)}{\bf X}_{(1,-1)}{\bf X}_{(1,1)}\simeq
{\bf D}_5 {\bf X}_{(-1,2)},
\end{align}
where $\simeq$ implies equality of the corresponding products of $K$'s, up to a conjugation by $K_{(1,0)}$.
The UV fixed point is therefore described by figure~\ref{fig:SU2Nf4}(c),
and the enhanced symmetry in this case is ${\textrm{\bf D}}_5\cong$ \SO(10).
\begin{figure}[ht]
 \begin{center}
\includegraphics[width=6.5cm]{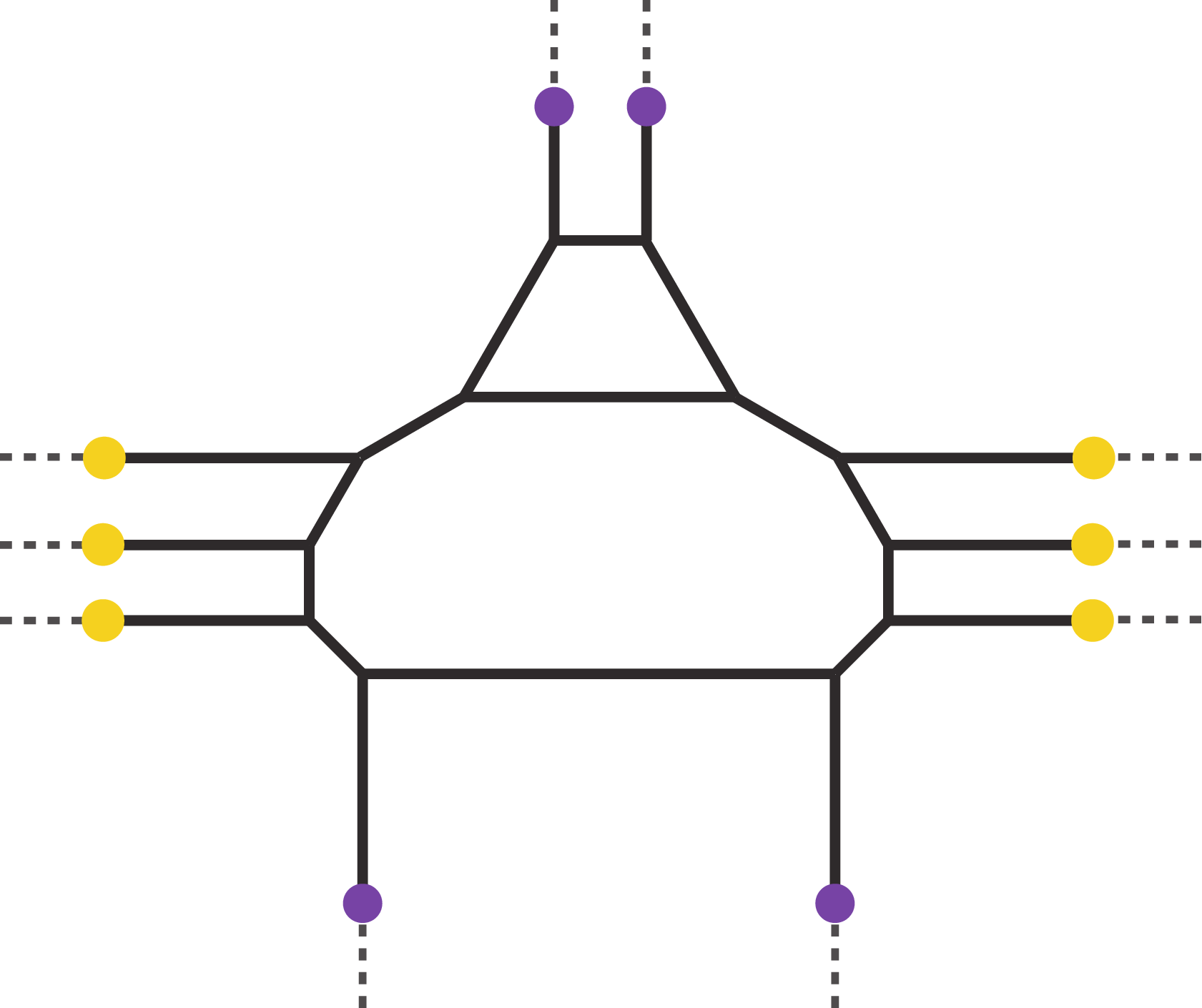}
 \end{center}
 \caption{The web diagram for $\SU(3)$ $N_f=6$ SQCD.
 All external legs are regularized by 7-branes.}
 \label{fig:SU3}
\end{figure}

The $\SU(2)$ theories with other $N_f$ flavor can be also discussed analogously \cite{DeWolfe:1999hj}
and their global turns out to be $\E_{N_f+1}$ as expected.

\subsection{$\SU(3)$ $N_f=6$ SQCD}
\begin{figure}[ht]
 \begin{center}
\includegraphics[width=\linewidth]{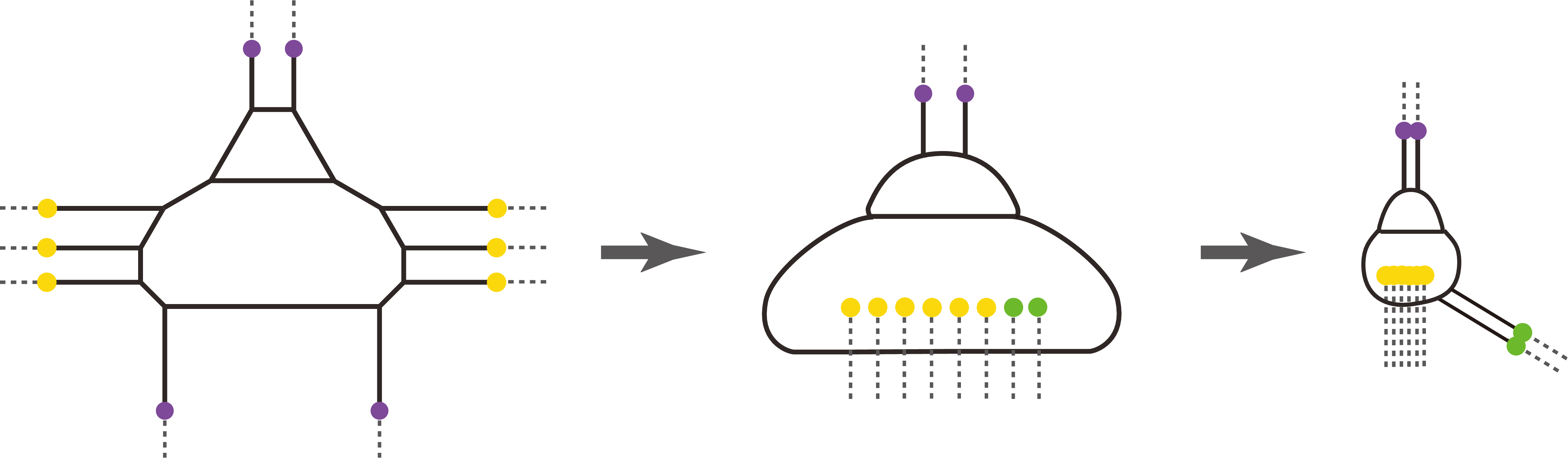}
 \end{center}
 \caption{Moving 7-branes inside and reordering them yields the middle configuration. The maximally enhanced symmetry at the UV fixed point is realized by colliding the $(0,1)$ 5-branes, ${\textrm{\bf{X}}}_{(1,0)}^6$ 7-branes,  and the two 5-branes attached to the green 7-branes $\textrm{\bf{X}}_{(3,-1)}^2$.}
 \label{fig:SU37}
\end{figure}

As in the case of $\SU(2)$ theories, the 7-brane technique works well for the $\SU(3)$ $N_f=6$ SQCD depicted in figure~\ref{fig:SU3}.
This web can be modified by using two types of 7-branes without changing the world-volume 5d gauge theory and, by moving them inside of the 5-brane loop, we obtain the 7-brane configuration illustrated in the middle line of figure~\ref{fig:SU37}.
This configuration shows the following enhanced gauge symmetry
\begin{align}
{\textrm{two $(0,1)$ external 5-branes}}\,{\textrm{\bf{X}}}_{(1,0)}^6{\textrm{\bf{X}}}_{(3,-1)}^2\quad
\rightarrow\quad
\SU(2)\times \SU(6)\times \SU(2).
\end{align}
The configuration $\textrm{\bf{X}}_{(1,0)}^6 \textrm{\bf{X}}_{(3,-1)}^2$ is not collapsible,
and in order to see this enhancement
it is better to move two ${\textrm{\bf{X}}}_{(3,-1)}$ 7-branes outside of the 5-brane loop as shown on the right hand of figure~\ref{fig:SU37}.
Then collapsible $\textrm{\bf{X}}_{(1,0)}^6 $ 7-branes give the $\SU(6)$ symmetry at the UV fixed point
and the coinciding $(3,1)$ flavor 5-branes lead to an $\SU(2)$.

\subsection{$\SU(N)$ gauge theory}

\begin{figure}[t!]
 \begin{center}
\includegraphics[width=\linewidth]{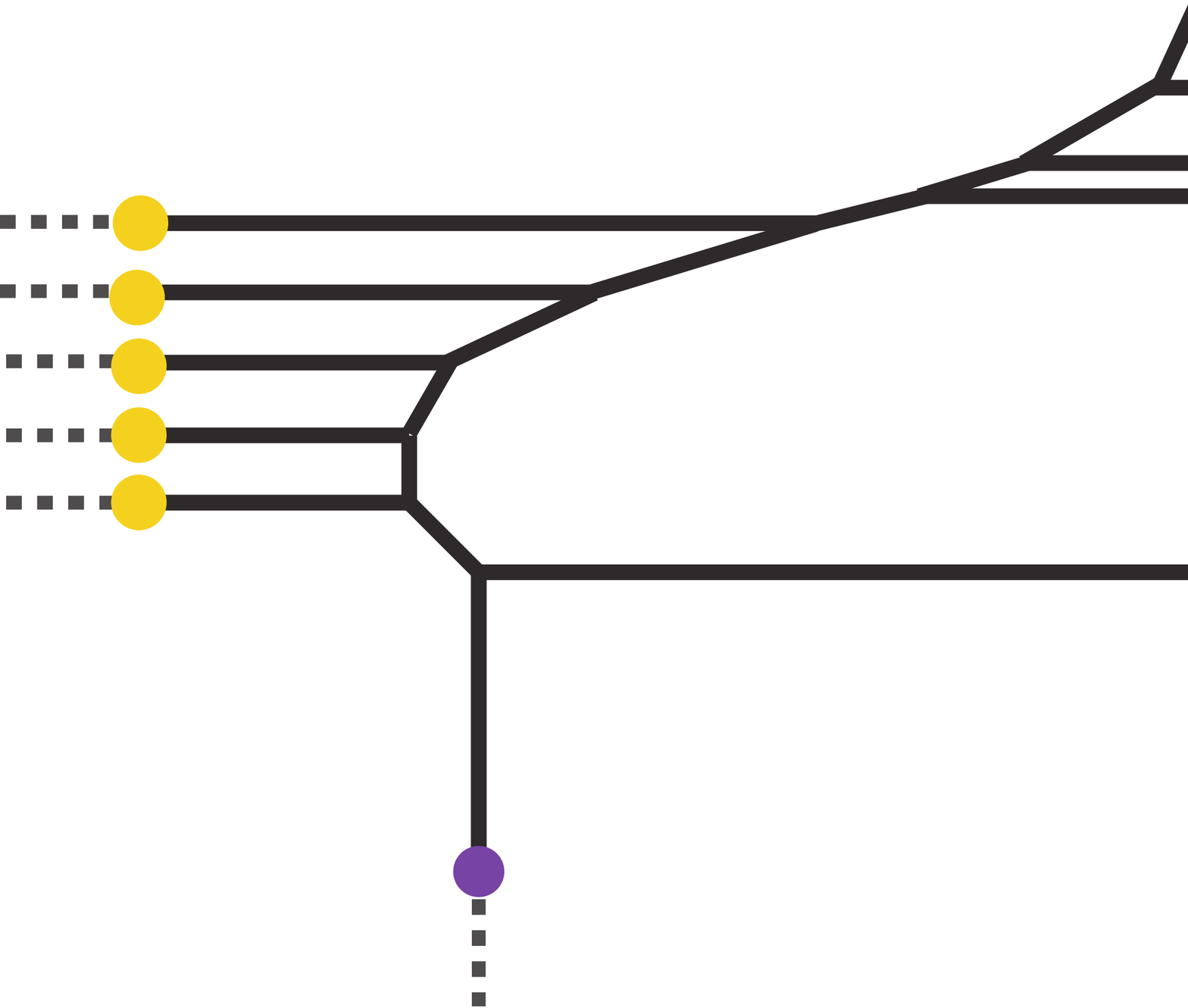}
 \end{center}
 \caption{The web diagram for $\SU(N)$ $N_f=2N$ SQCD (Left).
 By moving $(0,1)$ and lower $(0,1)$ 7-branes inside the 5-brane loop
 and reordering them yields the diagram in the right hand side.
 The blue circles are $(N,-1)$ 7-branes.}
 \label{fig:SUN}
\end{figure}
Similar 7-brane description can be applied to  $\SU(N)$ SQCD with $N_f=2N$ hypermultiplets
to see the global symmetry enhanced at the strongly-coupled UV fixed point.
The web diagram for this $\SU(N)$ theory is illustrated on the left side of figure~\ref{fig:SUN},
and this configuration is equivalent to the 5-brane multi-loop probe of
the 7-brane configuration on the right hand side of this figure.
This collection of 7-branes ${\textrm{\bf{X}}}_{(0,1)}^{2N}$ and flavor 5-branes  carry the enhanced symmetry
$\SU(2)\times \SU(2N)\times \SU(2)$. Thus, this is the enhanced flavor symmetry for $\SU(N)$ SQCD with $N_f=2N$.

\subsection{$\SU(N)$ linear quiver gauge theory}

Let us move on to the $\SU(N)^{M-1}$ linear quiver theory  that is the 5-dimensional uplift of the 4D superconformal quiver.
We focus on the case of $M\geq 3$.

\begin{figure}[ht]
 \begin{center}
\includegraphics[width=\linewidth]{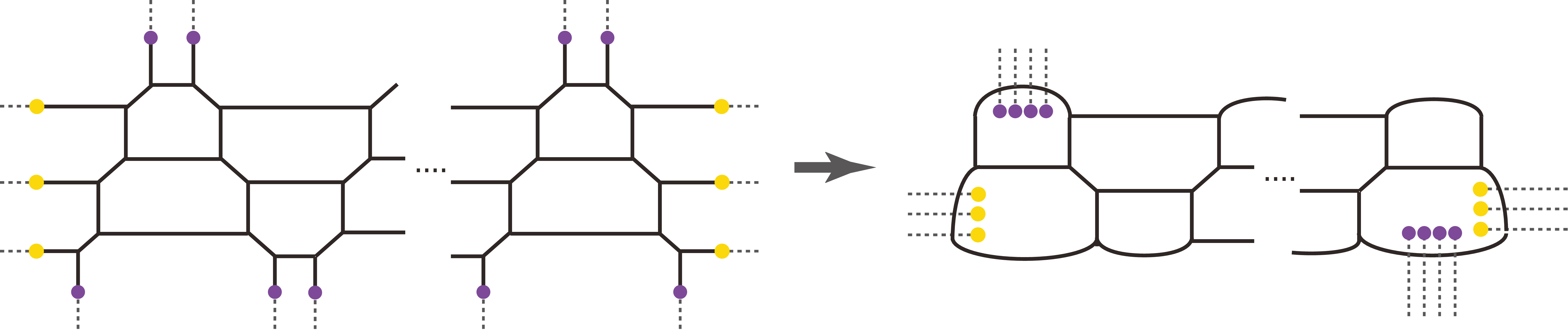}
 \end{center}
 \caption{The web diagram for $\SU(N)^{M-1}$ linear quiver (Left).
 By moving the $(1,0)$ and  the $(0,1)$ 7-branes inside the 5-brane multi-loop,  we obtain the right hand side.  The shape of 5-brane multi-loop is actually warped nontrivially because of the background metric coming from 7-branes.}
 \label{fig:SUNM}
\end{figure}

The web diagram decorated with 7-branes is illustrated in the left side of figure~\ref{fig:SUNM}.
Since a $(p,q)$ 7-brane can pass though $(p,q)$ 5-branes without Hanany-Witten effect,
we can recast this configuration into the right side of figure~\ref{fig:SUNM}.
This system has four stacks of collapsible 7-branes,
which are two ${\bf X}_{(1,0)}^{N}$ and two ${\bf X}_{(0,1)}^{M+1}$.
The non-Abelian part\footnote{We discuss in section \ref{sec:symmetryenhancement} that the actual global symmetry is $\SU(N)^2 \times \SU(M)^2 \times \U(1)$ if we also include the Abelian part.} of the enhanced symmetry for $N\geq 3$ case is therefore $\SU(N)^2\times \SU(M)^2$.
This symmetry is naturally consistent with the fiber-base duality.

The case of $\SU(2)$ quivers is a little exceptional. The configuration for $N=2$ is one of $M$ 5-brane rooms standing side by side, so that the $(0,1)$ 7-branes can be moved to the same 5-brane room without experiencing the Hanany-Witten effect.
This means that two ${\bf X}_{(0,1)}^{M}$ 7-brane stacks are now recombined  in a single stack as ${\bf X}_{(0,1)}^{2M}$,
and that therefore the enhanced symmetry for the $\SU(2)$ quiver is $\SU(2)^2\times \SU(2M)$.
This symmetry, of course, coincides with that of the SQCD with gauge group $\SU(M)$ and $N_f=2M$.


\section{Fiber-base symmetry and symmetry enhancement for $\SU(2)$ theory}
\label{sec:fiberbaseinvariance}

The goal of this section is to demonstrate that for the $\SU(2)$ gauge theories with $N_f<8$, the fiber-base duality can be combined with the manifest $\SO(2N_f)$ flavor symmetry to generate the full enhanced $\E_{N_f+1}$ symmetry seen in the superconformal index. Furthermore, we show that the enhanced symmetry is present in the partition functions as well by expanding them in a power series of the appropriately defined invariant Coulomb modulus $\tilde{A}$.

\subsection{Pure $\SU(2)$}
\label{subsec:puresu2}

We first demonstrate our idea with the simplest example of the
pure $\SU(2)$ super Yang-Mills theory. The pure $\SU(2)$ theory has enhanced
$\E_1 = \SU(2)$ global symmetry at the UV fixed point  \cite{Seiberg:1996bd,Kim:2012gu}. The relevant deformation of the UV fixed point  that will drive\footnote{At the UV fixed point \cite{Seiberg:1996bd} we  have $\frac{1}{g^2}=0$ . To trigger an RG flow we need to add a coupling constant deformation (the supersymmetrization of $\frac{1}{g^2}F_{\mu \nu}^2$). To achieve that we need to begin with the UV fixed point theory and gauge the Cartan part of the global symmetry and then give a vacuum expectation to the scalar of the new vector multiplet. This is also some times referred to the mass deformation of the non-Lagrangian theory. See section 5 in \cite{Intriligator:1997wk}.}  it to the IR respects the Weyl symmetry of $\E_1 = \SU(2)$ and thus it is natural to believe that the holomorphic part of the partition function will enjoy the enhanced symmetry.
In what follows we will try to understand this from the view point
of the fiber-base duality.

\begin{figure}[t]
 \begin{center}
  \includegraphics[width=30mm]{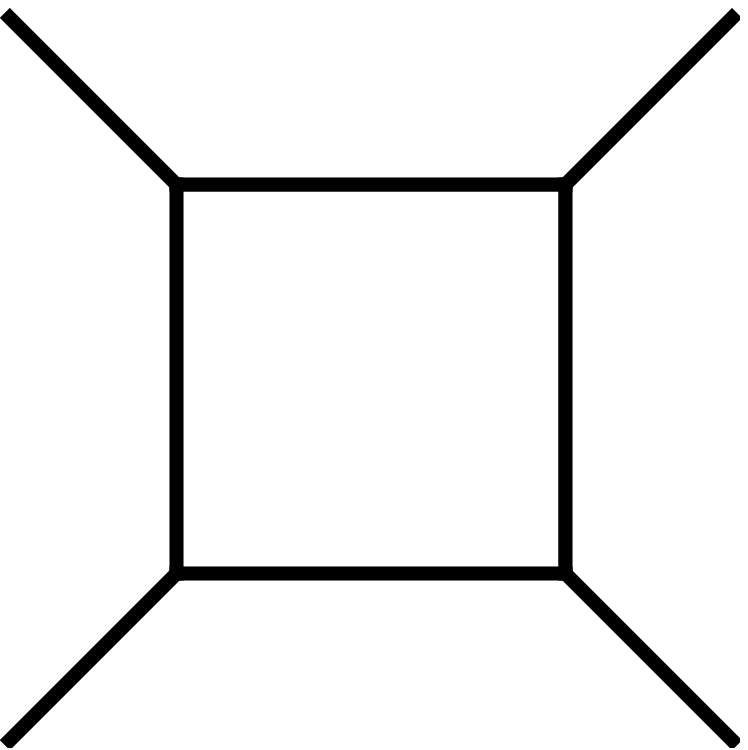}
 \end{center}
 \caption{The toric diagram of the local
$\mathbb{P}^1 \times \mathbb{P}^1$
geometry. }
 \label{toricPureSU(2)}
\end{figure}

The pure $\SU(2)$ theory is realized using the local $\mathbb{P}^1 \times \mathbb{P}^1$ geometry
that is the canonical line bundle over $\mathbb{P}^1 \times \mathbb{P}^1$.
The toric diagram of
 the geometry is depicted in figure \ref{toricPureSU(2)} and can be seen as a $\mathbb{P}^1$ fibered over the base $\mathbb{P}^1$.
We denote the K\"ahler parameter for the base $\mathbb{P}^1$ as $Q_B$
and the K\"ahler parameter for the fiber $\mathbb{P}^1$ as $Q_F$.
At this level it is clear that we can exchange the role of base and the fiber without changing the geometry.
We will have the same Calabi-Yau manifold  with the K\"ahler parameters exchanged
\begin{eqnarray}
\label{eq:fiberbase1}
Q_F \leftrightarrow Q_B.
\label{FB}
\end{eqnarray}
This is known as fiber-base duality
and will refer to the relation (\ref{FB}) and it's generalizations as the ``duality map''.
Roughly speaking, the fiber-base duality is simply understood as just rotating the
toric diagram by 90 degrees.
As discussed in \cite{Leung:1997tw},
the toric diagram can be reinterpreted as a $(p,q)$ 5-brane web.
In this language, the fiber-base duality is translated as
the S-duality which exchanges the D5-branes and the NS5-branes \cite{Aharony:1997bh}.

The ``duality map'' can be checked/derived quantitatively by using the Nekrasov partition function which can be derived using the topological A model on the local toric Calabi-Yau
as the topological string partition function.
The unrefined topological vertex formalism computes the 5D Nekrasov partition function
with self-dual $\Omega$-deformation parameters $\epsilon_1=-\epsilon_2=\hbar$.
The vertex function in the unrefined topological vertex formalism has cyclic invariance $C_{\mu \nu \lambda} = C_{ \nu \lambda \mu} = C_{ \lambda \mu \nu}$.
Using the cyclic invariance together with the duality map \eqref{FB} is enough to show that  fiber-base duality is a symmetry of the unrefined partition function\footnote{See \cite{Bao:2011rc} for conventions.}:
\begin{multline}
\label{pureSYM}
Z^{N_f=0} = \sum_{\mu, \nu, \rho , \sigma}
 \mathfrak{q}^{\frac{\kappa_\mu + \kappa_\nu + \kappa_\rho  + \kappa_\sigma}{2}}
 \left(-Q_F \right)^{|\mu|}  \left(-Q_B \right)^{|\nu|}  \left(-Q_F \right)^{|\rho|}   \left(-Q_B \right)^{|\sigma|}\times
\\ \times C_{\mu^t  \emptyset \nu^t}  C_{\emptyset  \rho \nu }
 C_{  \rho^t  \emptyset \sigma^t} C_{\emptyset  \mu \sigma }
 \, .
\end{multline}
The unrefined vertex $C_{\mu \nu \lambda} \equiv C_{\mu \nu \lambda}\left(\mathfrak{q} \right)$ is a function of the topological string coupling constant $\mathfrak{q}= e^{- \beta \hbar}$.

At the level of the refined topological vertex formalism, the fiber-base duality is less trivial/obvious.
The 5D Nekrasov partition function with generic $\Omega$-deformation is given by the refined topological vertex formalism.
The refined topological vertex function \eqref{eq:topvertex} is a function of two $\Omega$-deformation variables
$C_{\mu \nu \lambda} \equiv C_{\mu \nu \lambda}\left(\mathfrak{q}, \mathfrak{t} \right)\neq  C_{\mu \nu \lambda}\left( \mathfrak{t}, \mathfrak{q}\right)$  and
is not cyclic invariant as one of its legs is special.
The direction of these special legs has to be parallel with each other and is called the preferred direction. In the refined case, the partition function reads
 \begin{eqnarray}
\label{eq:Z0refinedassum}
&& Z^{N_f=0} =  \sum_{\mu, \nu, \rho , \sigma}
 \mathfrak{q}^{\frac{|| \mu ||^2 + || \nu^t ||^2 + || \rho^t ||^2 - || \sigma ||^2 - |\mu| + |\rho| }{2}}
  \mathfrak{t}^{\frac{|| \mu^t ||^2 - || \nu ||^2 - || \rho ||^2 + || \sigma^t ||^2 + |\mu| - |\rho| }{2}}
   \\
  &&
   \times \,
 \left(-Q_F \right)^{|\mu|}  \left(-Q_B \right)^{|\nu|}  \left(-Q_F \right)^{|\rho|}   \left(-Q_B \right)^{|\sigma|}
C_{\mu^t  \emptyset \nu^t}\left(\mathfrak{t} ,\mathfrak{q}\right)  C_{\emptyset  \rho \nu } \left( \mathfrak{t}, \mathfrak{q}\right)
 C_{  \rho^t  \emptyset \sigma^t} \left(\mathfrak{q}, \mathfrak{t} \right) C_{\emptyset  \mu \sigma } \left( \mathfrak{q}, \mathfrak{t} \right).\nonumber
\end{eqnarray}
Nevertheless, it is conjectured\footnote{This assumption is called ``slicing invariance''.
It is worth pointing out that checking the enhanced global symmetry
is equivalent to checking the slicing invariance
for the refined topological string partition function.} that the topological amplitude does not depend on the choice of the preferred direction. Combining this conjecture together with the duality map \eqref{FB} one can show that  fiber-base duality is a symmetry of the refined partition function \cite{Ito:2012bb}.

Comparing the topological string amplitude on $\mathbb{P}^1 \times \mathbb{P}^1$  with the Nekrasov partition function for the pure \SU(2) theory we obtain the relation between the K\"ahler parameters of the CY
on one hand and the Coulomb moduli parameter $a$  and the gauge coupling constant $q= e^{2\pi i \tau}$ on the other:
\begin{eqnarray}
Q_F = e^{- 2 \beta a}, \qquad
Q_B = q e^{- 2 \beta a},
\end{eqnarray}
where $\beta$ is the circumference of the compactified $S^1$.
The duality map (\ref{FB}) can then be rewritten in the
 language of gauge theory parameters as
\begin{eqnarray}
\label{eq:fiberbase2}
q \rightarrow q^{-1},
\qquad
e^{- 2 \beta a} \rightarrow q e^{- 2 \beta a}.
\end{eqnarray}
In what follows we interpret the first equation in
\eqref{eq:fiberbase2} as the Weyl reflection of  the enhanced $\E_1$ global symmetry.

First however, we have to be careful with
the second equation in \eqref{eq:fiberbase2} which
indicates that the Coulomb moduli parameter
 transforms under this Weyl reflection,
a fact very surprising form the
field theory viewpoint.
In order to make the $\E_1$  symmetry  manifest,
we need to introduce a new parameter
that, unlike the original Coulomb moduli parameter,  is invariant under the Weyl reflection. Obviously, the combination
\be
\tilde{A} \equiv (Q_F Q_B)^{\frac{1}{4}} = q^{\frac{1}{4}} e^{ - \beta a }=e^{ -  \beta a_{\text{new}} }
\label{eq:shifted0}
\ee
is invariant under the duality map (\ref{FB}). Since $\tilde{A}$ is given by the exponential of the new Coulomb moduli parameter $a_{\text{new}} \colonequals  a - \frac{\log (q)}{4 \beta} $ that is defined via a shift,
we also refer to it as ``the shifted Coulomb moduli parameter''.

In summary, we learn that in order to see the global symmetry enhancement we should parametrize the
K\"ahler parameters in terms of
the shifted Coulomb moduli parameter as
\be
\label{eq:defQFQB}
Q_F = u^{-1} \tilde{A}^2,
\qquad
Q_B = u \tilde{A}^2 \, .
\ee
We introduce the new variable $u^2=q$ so as to simplify our formulas.

The topological string amplitude \eqref{eq:Z0refinedassum} can be taken from \cite{Bao:2013pwa}, rewritten using the function of \eqref{eq:deffunctionS} and the parametrization of \eqref{eq:defQFQB}
\begin{align}
\label{eq:partfunction0}
Z^{N_f=0}(Q_F,Q_B;\ft,\fq)
=\sum_{\mu_1,\mu_2}\frac{\prod_{i=1}^2u^{2|\mu_i|}\ft^{||\mu_i^t||^2}\fq^{||\mu_i||^2}\tilde{Z}_{\mu_i}(\ft,\fq)\tilde{Z}_{\mu_i^t}(\fq,\ft)}{\calS_{\mu_2\mu_1}\big(\sqrt{\frac{\fq}{\ft}}u^{-1} \tilde{A}^2\big)\calS_{\mu_2\mu_1}\big(\sqrt{\frac{\ft}{\fq}}u^{-1} \tilde{A}^2\big)},
\end{align}
where we remind that $u^2=q$. The functions $\tilde{Z}$ and $\calS_{\mu \lambda}$ are defined in equation \eqref{eq:mainfunctionsdefinitions} in the appendix \ref{app:special}.
Inserting \eqref{eq:defQFQB} and expanding $Z^{N_f=0}$ in the modulus $\tilde{A}$ leads to the expression\footnote{In what follows characters are labeled by their dimension. }
\bea
\label{expansionSU(2)}
Z^{N_f=0}&=&1+\frac{\ft+\fq}{(1-\ft)(1-\fq)}\chi_{2}^{\E_1}(u)\tilde{A}^2+\Bigg[\frac{ (\fq^2+\ft^2)(\fq+\ft+\fq^2+\ft^2+\fq\ft(1+\fq+\ft))}{\fq\ft(1-\fq)  (1+\fq) (1-\ft)  (1+\ft)}\nonumber\\&&+\frac{(\fq+\ft+\fq^2+\ft^2+\fq\ft(1+\fq+\ft))\chi_{3}^{\E_1}(u)}{(1-\fq)^2  (1+\fq) (1-\ft)^2  (1+\ft)}\Bigg]\tilde{A}^4+\mathcal{O}(\tilde{A}^6) .
\eea
We  thus discover that it organizes in terms of characters of the enhanced $\E_1$ global symmetry.

To obtain the expansion \eqref{expansionSU(2)} we have assumed that the coefficient of $\tilde{A}^{2n}$ is completely determined by $k$ instanton contributions with $k \le n$ in \eqref{eq:partfunction0}. We can check this assumption experimentally by expanding to a few more orders but we also justify it in the appendix \ref{app:notecoulombmoduliexpansion}.

Let us finish this subsection by making some remarks on the superconformal index. The index is defined as the contour integral over the Coulomb modulus \cite{Iqbal:2012xm}
\begin{equation}
\label{eq:defindex}
\mathcal{I}_0=\frac{1}{2}\oint_{|\tilde{A}|=\epsilon}\frac{d\tilde{A}}{2\pi i \tilde{A}}M(x,y)\overline{M}(x,y) Z^{N_f=0}(\tilde{A},u;x,y)\overline{Z}^{N_f=0}(\tilde{A},u;x,y),
\end{equation}
where $\epsilon>0$ is small enough so that the contour integral only picks up the residue at zero, we have set $\fq= xy$, $\ft=\frac{y}{x}$ and $M(x,y)$ is the refined McMahon function\footnote{This is the contribution from the constant map. In our convention, this contribution is not included in the partition function $Z^{N_f=0}$.}. The ``complex conjugation'' acts by inverting\footnote{Of course, we need to use the proper analytic continuation to reframe $\bar{Z}^{N_f=0}$ in such a way as to be able to expand in small $x$.} $\tilde{A}$, $u$, $x$ and $y$. One notes that the expansion of $Z^{N_f=0}$ in powers of $\tilde{A}$ \textit{does not} commute with the expansion in powers of $x$ due to the presence in the sum over partitions $\mu_i$ of \eqref{eq:partfunction0} of terms like $\big(1-\frac{\tilde{A}}{ux^2}\big)^{-1}$. Thus, we cannot extract the index
\be
\label{eq:indexNf0}
\mathcal{I}_0=1 + \chi^{\E_1}_{3} x^2 + \chi^{\SU(2)}_{2}(y)\big[1 +  \chi^{\E_1}_{3}\big] x^3 + \big[1 +  \chi_{5}^{\E_1} +  \chi_{3}^{\SU(2)}(y)(1+ \chi_{3}^{\E_1})\big] x^4+\mathcal{O}(x^5),
\ee
directly by plugging the expansion \eqref{expansionSU(2)} into \eqref{eq:defindex}. In the above, all characters of $\E_1$ are functions of the variable $u$. In order to get the result \eqref{eq:indexNf0}, we need to expand $Z^{N_f=0}$ in a power series in $x$ first.

\subsection{The case of $\SU(2)$ with $N_f\leq 4$ fundamental flavors}

 We now proceed to the cases with fundamental flavors. As we will see, our idea will generalize straightforwardly, up to two additional features/points that will play an important role.

The first point is concerned with  the mass parametrization in the refined case.
As was already discussed  in  \cite{Okuda:2010ke}, starting with the $\mathcal{N}=2^*$ Nekrasov partition function, one has to
 shift the mass parameter for the adjoint hypermultiplet as
\begin{eqnarray}
\label{massShift}
m_{\text{new}} = m_{\text{old}} + \frac{\epsilon_+}{2} \, .
\end{eqnarray}
Only then  does one get the correct  Nekrasov partition function for $\mathcal{N}=4$ SYM (which is 1) by sending to zero the mass deformation
 $m_{\text{new}}=0$.
This shift is convenient also for the fundamental hypermultiplets due to the following reason.
The explicit duality map for the theory with matter in the refined case
is discussed in \cite{Ito:2012bb}, where
it is pointed out that the dependence
on the Omega deformation parameter (in the combination of
$\epsilon_+ \colonequals \epsilon_1 + \epsilon_2$ or $\fq/\ft$)
appears when we parametrize the K\"ahler parameters
in terms of gauge theory parameters.
However,  such dependence disappears
when we shift the masses as in \eqref{massShift}
and  the parametrization (the duality map) becomes exactly the same as in the unrefined case \cite{Bao:2011rc}.
Moreover, this shift of the mass parameters \eqref{massShift} is  motivated from the Weyl symmetry  that acts as
\begin{eqnarray}
m_{\text{new}} \to -m_{\text{new}} \, ,
\end{eqnarray}
which is more natural than the transformation to the original mass parameter
$m_{\text{old}} \to \epsilon_+ - m_{\text{old}}$.
Therefore, the new shifted mass parameters should be used
and the duality map for the refined case is exactly the same as in the unrefined case.

\begin{figure}[h]
\begin{center}
\includegraphics[height=2.5cm]{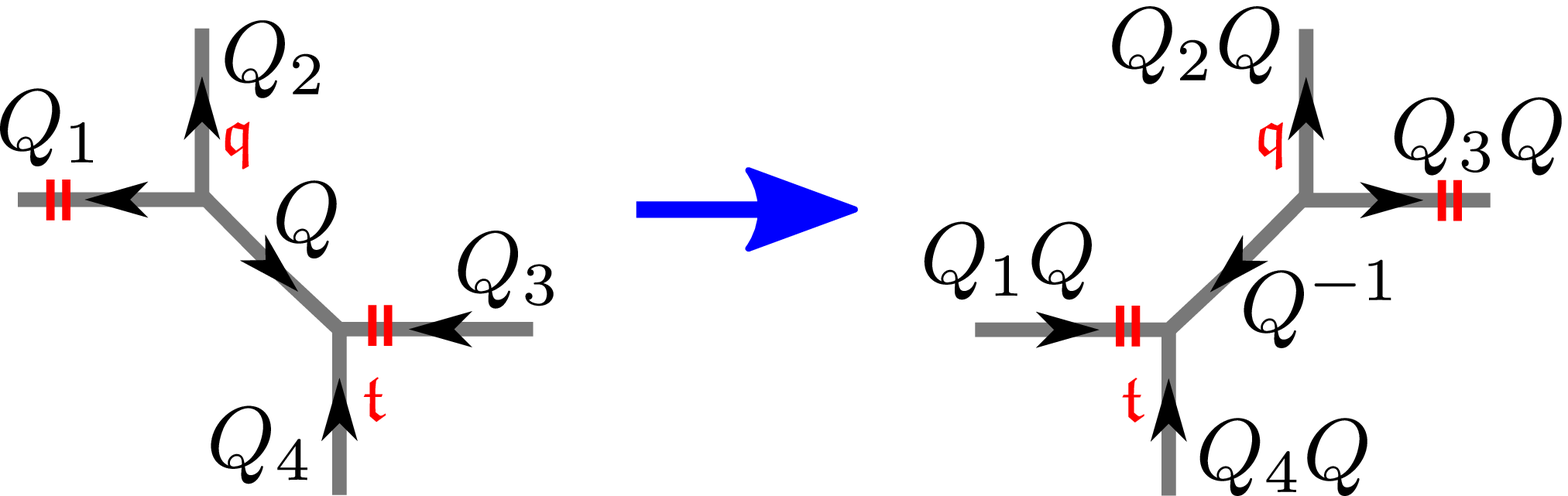}
\end{center}
\caption{This figure illustrates the way the K\"ahler parameters are modified by flopping.}
\label{fig:floppingillustrated}
\end{figure}

The second point is about flopping. The procedure of flopping involves sending the K\"ahler parameter $Q$ of one of the branes of the web diagram to $Q^{-1}$ while changing  the geometry of the web diagram and  the  K\"ahler parameters $Q_i$ of all the branes adjacent to the one being flopped are sent to $Q_iQ$, see figure~\ref{fig:floppingillustrated}.
In our previous article \cite{Bao:2013pwa}, we use the functions $\calR_{\lambda\mu}$ defined in \eqref{eq:mainfunctionsdefinitions} to write the topological string partition functions. In order to make the invariance under flopping $Q\rightarrow Q^{-1}$ as nice as possible, we now introduce a new one, namely
\be
\label{eq:deffunctionSmain}
\calS_{\lambda\mu}(Q;\ft,\fq)=(-1)^{|\lambda|}Q^{-\frac{|\lambda|+|\mu|}{2}}\ft^{\frac{||\lambda^t||^2}{2}}\fq^{\frac{||\mu||^2}{2}}\calR_{\lambda^t\mu}(Q;\ft,\fq).
\ee
As we  show in appendix \ref{app:special}, under flopping\footnote{In all our considerations,  the phase $(-Q)^{\frac{1}{12}}$ can be ignored.} the new function behaves as
\be
\label{eq:Sfloppingmain}
\calS_{\lambda\mu}(Q^{-1};\ft,\fq)\rightarrow (-Q)^{\frac{1}{12}}\calS_{\mu\lambda}(Q;\ft,\fq).
\ee

When we couple the vector multiplet to hypermultiplets, there are several
 toric diagrams that correspond to the gauge theory under consideration, but they are all related with each other by a flop \cite{Aharony:1997bh,Iqbal:2004ne,Konishi:2006ev,Gukov:2007tf,Taki:2008hb}.
In order to expand the topological string partition function
in terms of the \textit{positive} powers of the shifted Coulomb moduli parameters,
we need to choose the proper one, for which the K\"ahler parameters
for each 2-cycle does not include
the negative power of the shifted Coulomb moduli parameters. The topological amplitudes of the original and the flopped diagram are related to each other by the replacement \eqref{eq:Sfloppingmain} with respect to the K\"ahler parameter being flopped, as we illustrate in the next subsections.

\subsubsection{$\SU(2)$ with $N_f=1$}

\begin{figure}[h]
\begin{center}
\includegraphics[height=4.5cm]{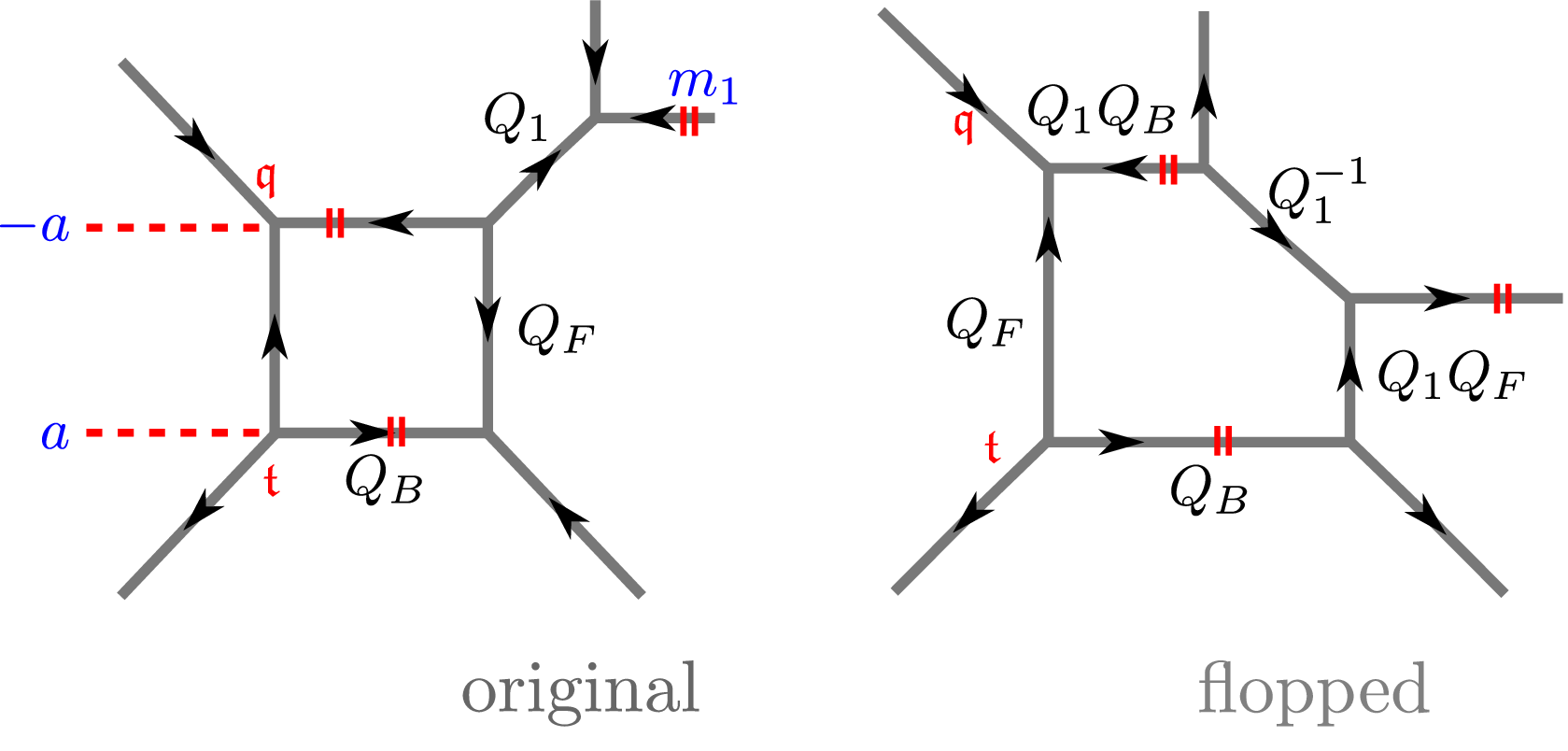}
\end{center}
\caption{The left hand side is the original $\SU(2)$ one flavor case, while for the right diagram we have flopped $Q_1$. }
\label{fig:1flopped}
\end{figure}
In figure \ref{fig:1flopped} we see on the left hand side the original flavor $\SU(2)$ dual toric diagram while the right one is the flopped one. 
The expected global symmetry is $\E_2 = \SU(2) \times \U(1)$,
whose Weyl transformation is given by
\begin{eqnarray}
u_1 \to u_1{}^{-1}, \qquad u_2 \to u_2,
\label{E2}
\end{eqnarray}
which are related to the instanton factor $q$
and the fundamental mass $m_1$ as (4.10) in \cite{Kim:2012gu}
\be
u_1 = q^{\frac{1}{2}} e^{ - \frac{1}{4} \beta m_1 },
\qquad
u_2 = q^{-\frac{1}{2}} e^{ - \frac{7}{4} \beta m_1 }.
\label{u1u2}
\ee
The fiber-base duality map is, as before,  given by the exchange\footnote{This map is obtained by the reflection along the diagonal axis rather than the 90 degree rotation.
As discussed in \cite{Bao:2011rc}, they are essentially equivalent up to trivial reflection along the vertical axis, which does not change the partition function.}
\be
\label{eq:fiberbaseNf1}
Q_B \leftrightarrow Q_F, \qquad  Q_1 \leftrightarrow Q_1,
\ee
which can be read off from either the original or the flopped diagram, see figure~\ref{fig:1flopped}. The parametrization can also be read off from the diagram and is given by
\begin{equation}
Q_F = e^{- 2 \beta a}, \quad Q_B = q e^{- \beta (2a + \frac{1}{2} m_1) }
= u_1^2 e^{- 2 \beta a }, \quad Q_1 = e^{- \beta (-a-m_1) }
= u_1^{-\frac{1}{2}} u_2^{-\frac{1}{2}} e^{\beta a },
\end{equation}
where we used \eqref{u1u2}. The duality map \eqref{eq:fiberbaseNf1} leads to the transformation rules
\be
\label{eq:Nf1transformation}
u_1 \to u_1{}^{-1},
\qquad
u_2 \to u_2,
\qquad
e^{-\beta a} \to u_1 e^{-\beta a}
\ee
Again, we can identify the first two as the Weyl transformation
of the enhanced symmetry \eqref{E2}.
From the transformations \eqref{eq:Nf1transformation}, we identify the invariant Coulomb modulus as
\begin{eqnarray}
\tilde{A}= u_1^{\frac{1}{2}} u_2^{\frac{c}{2}} e^{- \beta a},
\label{eq:shifted1}
\end{eqnarray}
where $c$ is an arbitrary constant.
For reasons that we shall explain in subsection \ref{subsec:effectivecoupling}, here we choose it to be $c=-\nicefrac{1}{7}$, indicating that $\tilde{A}=q^{\frac{2}{7}}e^{-\beta a}$
in which case the K\"ahler parameters are rewritten as
\be
\label{eq:param1flavor}
Q_F = u_1^{-1} u_2^{\frac{1}{7}} \tilde{A}^2,
\qquad
Q_B = u_1 u_2^{\frac{1}{7}} \tilde{A}^2,
\qquad
Q_1 = u_2^{-\frac{4}{7}} \tilde{A}^{-1}.
\ee
The frame in which the flavor symmetry is apparent corresponds to the right diagram of figure~\ref{fig:1flopped} and the corresponding topological string partition function can be computed in two ways. The first is to directly apply the refined topological string formalism on the right diagram of figure~\ref{fig:1flopped}. The second way is to take the result of \cite{Bao:2013pwa}, equation (4.27), that was computed for the the left diagram of figure~\ref{fig:1flopped}, rewrite the results using the functions $\calS$ instead of $\calR$ and then use the flopping rule \eqref{eq:Sfloppingmain}, which in this case implies $\calS_{\mu_1\emptyset}(Q_1)\rightarrow \calS_{\emptyset\mu_1}(Q_1^{-1})$.
The final result obtained using either method reads
\begin{align}
\label{eq:partfunction1flopped}
Z^{N_f=1}=&
\sum_{ \mu_1,\mu_2}\big(-Q_1^{\frac{1}{2}}Q_BQ_F^{-1}\big)^{|\mu_1|}\big(-Q_1^{\frac{1}{2}}Q_BQ_F^{-\frac{1}{2}}\big)^{|\mu_2|} \prod_{i=1}^2\ft^{\frac{||\mu_i^t||^2}{2}}\fq^{||\mu_i^t||}\tilde{Z}_{\mu_i}(\ft,\fq)\tilde{Z}_{\mu_i^t}(\fq,\ft)\nonumber\\&\times \frac{\calS_{\emptyset\mu_1}(Q_1^{-1}) \calS_{\mu_2\emptyset}(Q_1 Q_F)}{\calS_{\mu_2\mu_1}(Q_F\sqrt{\frac{\ft}{\fq}})\calS_{\mu_2\mu_1}(\sqrt{\frac{\fq}{\ft}}Q_F)}\ .
\end{align}
Using the expression \eqref{eq:partfunction1flopped}, setting the correct Coulomb modulus \eqref{eq:param1flavor} and
expanding in a power series in $\tilde{A}$, we obtain the expression
\be
\begin{split}
Z^{N_f=1}&=1-\frac{\ft^{\frac{1}{2}}\fq^{\frac{1}{2}}(\chi_{2}^{\SU(2)}u_2^{-\frac{3}{7}}+u_2^{\frac{4}{7}})}{(1-\ft)(1-\fq)}\tilde{A}+\Big [u_2^{-\frac{6}{7}}\fq\ft(1+\fq\ft)\\&+\fq\ft(\fq+\ft)(\chi_{3}^{\SU(2)}u_2^{-\frac{6}{7}}+u_2^{\frac{8}{7}})+\big(\fq(1-\fq^2)+\ft(1-\ft^2)\\&+\fq\ft(1+\fq\ft+\fq^2\ft+\fq\ft^2)\big)\chi_{2}^{\SU(2)}u_2^{\frac{1}{7}}\Big]\frac{\tilde{A}^2}{(1-\ft)^2(1+\ft)(1-\fq)^2(1+\fq)}+\mathcal{O}(\tilde{A}^3).
\end{split}
\ee
In the above the dependence of the fugacity $u_1$ is contained in the $\SU(2)$ characters $\chi_{\text{dim}}^{\SU(2)}$. Thus, in this frame the enhanced global symmetry $E_2\cong \SU(2)\times \U(1)$ is apparent.

\subsubsection{$\SU(2)$ with $N_f=2$}

The enhanced symmetry in the case of two flavors is $\E_3 = \SU(2) \times \SU(3)$,
whose Weyl group is $S_2\times S_3$ with transformations that act on the fugacities $u$ and $y_j$ by either $u \to u^{-1}$ or by permuting the $y_j$.
We define these fugacities to be the following functions
\begin{align}
\label{eq:2flavorfugacities}
u =e^{ \beta (m_1 + m_2)},\quad  y_1 = q^{\frac{2}{3}}, \quad
y_2 =q^{-\frac{1}{3}} e^{+ \frac{1}{2} \beta (m_1 - m_2)},\quad
y_3 = q^{-\frac{1}{3}} e^{- \frac{1}{2} \beta (m_1 - m_2)},
\end{align}
where $q$ is the instanton factor and $m_i$ are the masses of the fundamental flavors\footnote{This is slightly different from equations (4.13) and (4.14) of \cite{Kim:2012gu} because of a change in the sign of $m_2$.}. By construction, the fugacities obey the constraint $y_1y_2y_3 = 1$.
\begin{figure}[h]
\begin{center}
\includegraphics[height=5cm]{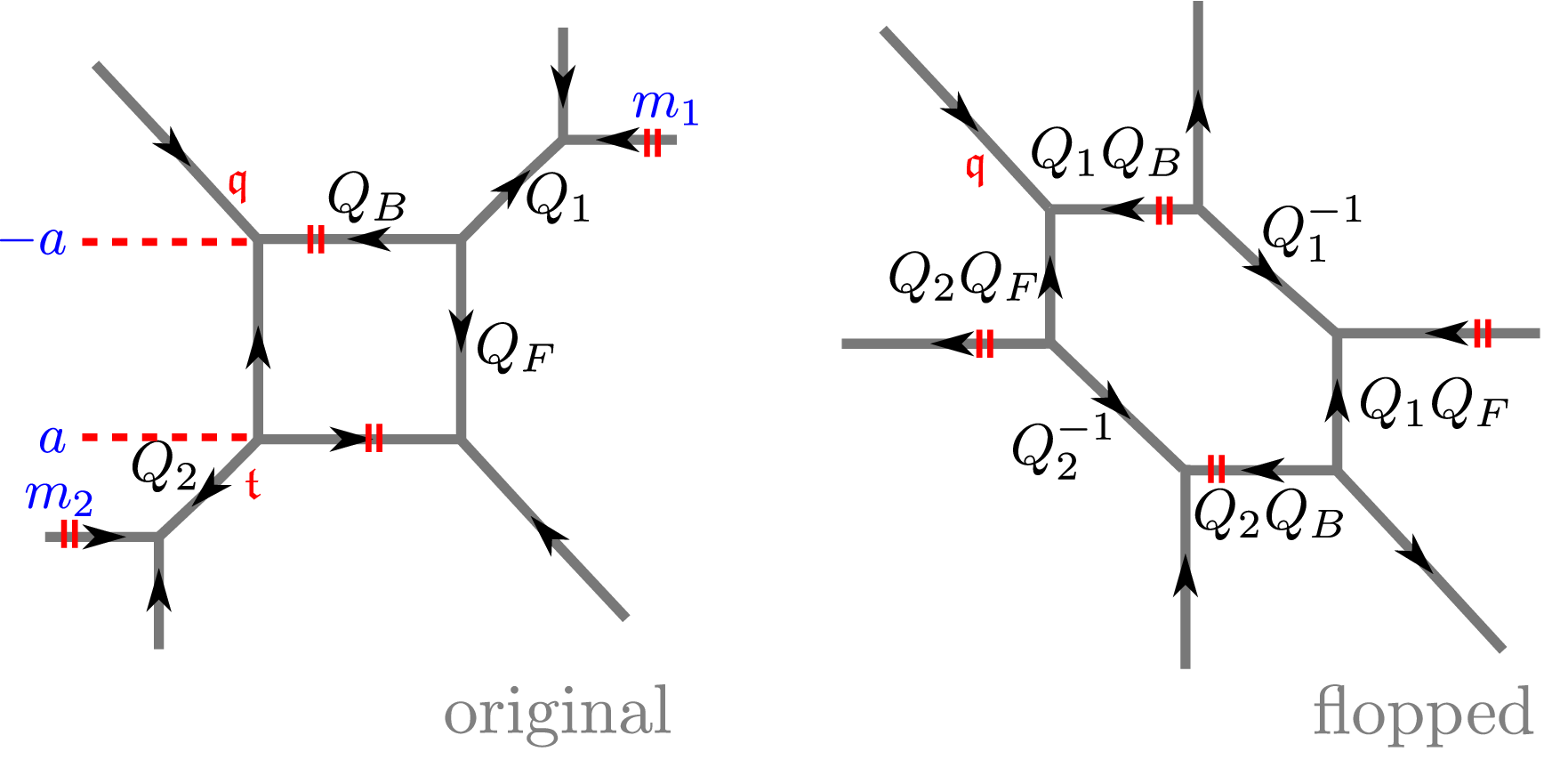}
\end{center}
\caption{The left hand side is the $\SU(2)$ with two flavor, while the right hand side is the flopped version. We choose to put the two exterior branes diagonally opposite to each other because that way there is no reason to remove contributions coming from parallel exterior branes. }
\label{fig:2flopped}
\end{figure}
The parametrization of the moduli can be taken from figure~\ref{fig:2flopped} and reads
\begin{align}
&Q_F = e^{- 2 \beta a}, &
&Q_B = q e^{- \beta (2a + \frac{1}{2} m_1 - \frac{1}{2} m_2)}
= \frac{y_1}{y_2} e^{- 2 \beta a},&
\nonumber \\
&Q_1 = e^{- \beta (-a-m_1) }
= \sqrt{\frac{y_2}{y_3}} u e^{ \beta a },& &
Q_2 =e^{- \beta (m_2-a) }
= \sqrt{\frac{y_2}{y_3}} u^{-1} e^{ \beta a }.&
\end{align}
The fiber-base duality map again acts by exchanging $Q_F$ and $Q_B$ while leaving $Q_1$ and $Q_2$ invariant. This translates to the following map
\be
y_1 \leftrightarrow y_2,
\qquad y_3 \to y_3, \qquad u \to u,
\qquad e^{- 2 \beta a }
\to y_1 y_2{}^{-1} e^{- 2 \beta a}.
\label{2f-map}
\ee
In order to find the invariant Coulomb modulus, we consider the Weyl transformations of the flavor symmetry $\SO(4) = \SU(2) \times \SU(2)$, given by  $m_1 \leftrightarrow m_2$ or $m_1 \leftrightarrow -m_2$.
By \eqref{eq:2flavorfugacities}, they are equivalent to
\begin{eqnarray}
\label{eq:Nf2SO4flavorWeyl}
 y_2 \leftrightarrow y_3 ,
\qquad
 u \to u^{-1}
\end{eqnarray}
and together with the first three in \eqref{2f-map} generate the full
$E_3$ Weyl symmetry. We thus define the Coulomb modulus
\begin{eqnarray}
\tilde{A}^2 = y_1 e^{- 2 \beta a}\Longrightarrow \tilde{A}=q^{\frac{1}{3}}e^{-\beta a}
\label{eq:shifted2}
\end{eqnarray}
which is invariant under the fiber-base duality and the flavor Weyl symmetry and  is thus also invariant under the complete set of $E_3$ Weyl transformations. We can then express the K\"ahler parameters as
\begin{eqnarray}
\label{eq:param2flavor}
Q_F = \tilde{A}^2 y_1{}^{-1} ,
\qquad
Q_B = \tilde{A}^2 y_2{}^{-1},
\qquad
Q_1 = \tilde{A}^{-1} y_3{}^{-1} u,
\qquad
Q_2 = \tilde{A}^{-1} y_3{}^{-1} u^{-1}.
\end{eqnarray}
In figure \ref{fig:2flopped} we see on the left hand side the original two flavor $\SU(2)$ dual toric diagram and on the right hand side the flopped version. 
For the flopped diagram, we obtain using the parametrization \eqref{eq:param2flavor}
\bea
\label{eq:partfunction2flopped}
Z^{N_f=2}&=&\sum_{ \boldsymbol{\mu}}\prod_{i=1}^2 \left(-Q_1Q_2Q_F^{-1}Q_B^2\right)^{\frac{|\mu_i|}{2}}\fq^{\frac{||\mu_i||^2}{2}}\ft^{\frac{||\mu_i^t||^2}{2}}\tilde{Z}_{\mu_i}(\ft,\fq)\tilde{Z}_{\mu_i^t}(\fq,\ft)\nonumber\\&&\times \frac{\calS_{\mu_2\emptyset}(Q_2^{-1})\calS_{\mu_2\emptyset}(Q_1Q_F) \calS_{\emptyset\mu_1}(Q_1^{-1})\calS_{\emptyset\mu_1}(Q_2 Q_F)}{\calS_{\mu_2\mu_1}(\sqrt{\frac{\ft}{\fq}}Q_F)\calS_{\mu_2\mu_1}(\sqrt{\frac{\fq}{\ft}}Q_F)}\nonumber\\
&=&\sum_{ \boldsymbol{\mu}}\prod_{i=1}^2 \left(-q\right)^{|\mu_i|}\fq^{\frac{||\mu_i||^2}{2}}\ft^{\frac{||\mu_i^t||^2}{2}}\tilde{Z}_{\mu_i}(\ft,\fq)\tilde{Z}_{\mu_i^t}(\fq,\ft)\nonumber\\&&\times \frac{\calS_{\mu_2\emptyset}(\tilde{A}y_2u)\calS_{\mu_2\emptyset}(\tilde{A}y_3u) \calS_{\emptyset\mu_1}(\tilde{A}y_2u^{-1})\calS_{\emptyset\mu_1}(\tilde{A}y_3u^{-1})}{\calS_{\mu_2\mu_1}(\sqrt{\frac{\ft}{\fq}}\tilde{A}^2y_1^{-1})\calS_{\mu_2\mu_1}(\sqrt{\frac{\fq}{\ft}}\tilde{A}^2y_1^{-1})},
\eea
where we remind \eqref{eq:2flavorfugacities} that $q=y_1^{\frac{3}{2}}$ is the instanton factor.
Since there are no exterior parallel branes, there is no full spin content to remove. Expanding \eqref{eq:partfunction2flopped} in the invariant Coulomb modulus, we get
\bea
Z^{N_f=2}&=&1-\frac{\ft^{\frac{1}{2}}\fq^{\frac{1}{2}}}{(1-\ft)(1-\fq)}\chi_{3}^{\SU(3)}\chi_{2}^{\SU(2)}\tilde{A}+\Big[\fq\ft(1+\fq\ft)\chi_{6}^{\SU(3)}\nonumber\\&&+(\fq(1-\fq^2)+\ft(1-\ft^2)+\fq^2\ft^2(\fq+\ft))\chi_{\bar{3}}^{\SU(3)}
+\fq\ft(\fq+\ft)\chi_{6}^{\SU(3)}\chi_{3}^{\SU(2)}\nonumber\\&&+\fq\ft(1+\fq\ft)\chi_{\bar{3}}^{\SU(3)}\chi_{3}^{\SU(2)}\Big]\frac{\tilde{A}^2}{(1-\fq)^2 (1+\fq) (1-\ft)^2 (1+\ft)}+\cdots.
\eea
In the above, the characters are again labeled by its dimension. Note that the invariance under the Weyl transformations \eqref{eq:Nf2SO4flavorWeyl} of the flavor symmetry $\SO(4) = \SU(2) \times \SU(2)$ is directly visible in the Nekrasov partition function \eqref{eq:partfunction2flopped}. Checking the invariance under the exchange $u\leftrightarrow u^{-1}$ requires using \eqref{eq:exchangerelationMN}, relabeling the partitions and using the fact that the topological amplitude is invariant under $\ft\leftrightarrow \fq$.

\subsubsection{$\SU(2)$ with $N_f=3$}

In the case of three hypermultiplets, the enhanced symmetry is $\E_4 = \SU(5)$, whose Weyl transformations are given by permutations $y_i \leftrightarrow y_j$ of the fugacities $y_i$ subject to $\prod_iy_i=1$. They are parametrized by
\begin{align}
&y_1 = q^{-\frac{4}{5}},& &y_2 = q^{\frac{1}{5}} e^{ \frac{\beta}{2}  (-m_1 + m_2 - m_3)},& &&
 \\
&y_3 = q^{\frac{1}{5}} e^{ \frac{\beta}{2}  (m_1 - m_2 - m_3)},&
&y_4 = q^{\frac{1}{5}} e^{ \frac{\beta}{2}  (m_1 + m_2 + m_3)},&
& y_5 = q^{\frac{1}{5}} e^{ \frac{\beta}{2}  (-m_1 - m_2 + m_3)},&\nonumber
\end{align}
where as before $q$ is the instanton factor and $m_i$ are the masses of the fundamental flavors. These fugacities were chosen in such a way that the obvious flavor symmetry exchanging the masses only affects the $y_i$ for $i=2,\ldots, 5$. The Weyl transformation involving $y_1$ will, as we shall see in the following, arise from the fiber-base duality.
\begin{figure}[h]
\begin{center}
\includegraphics[height=5cm]{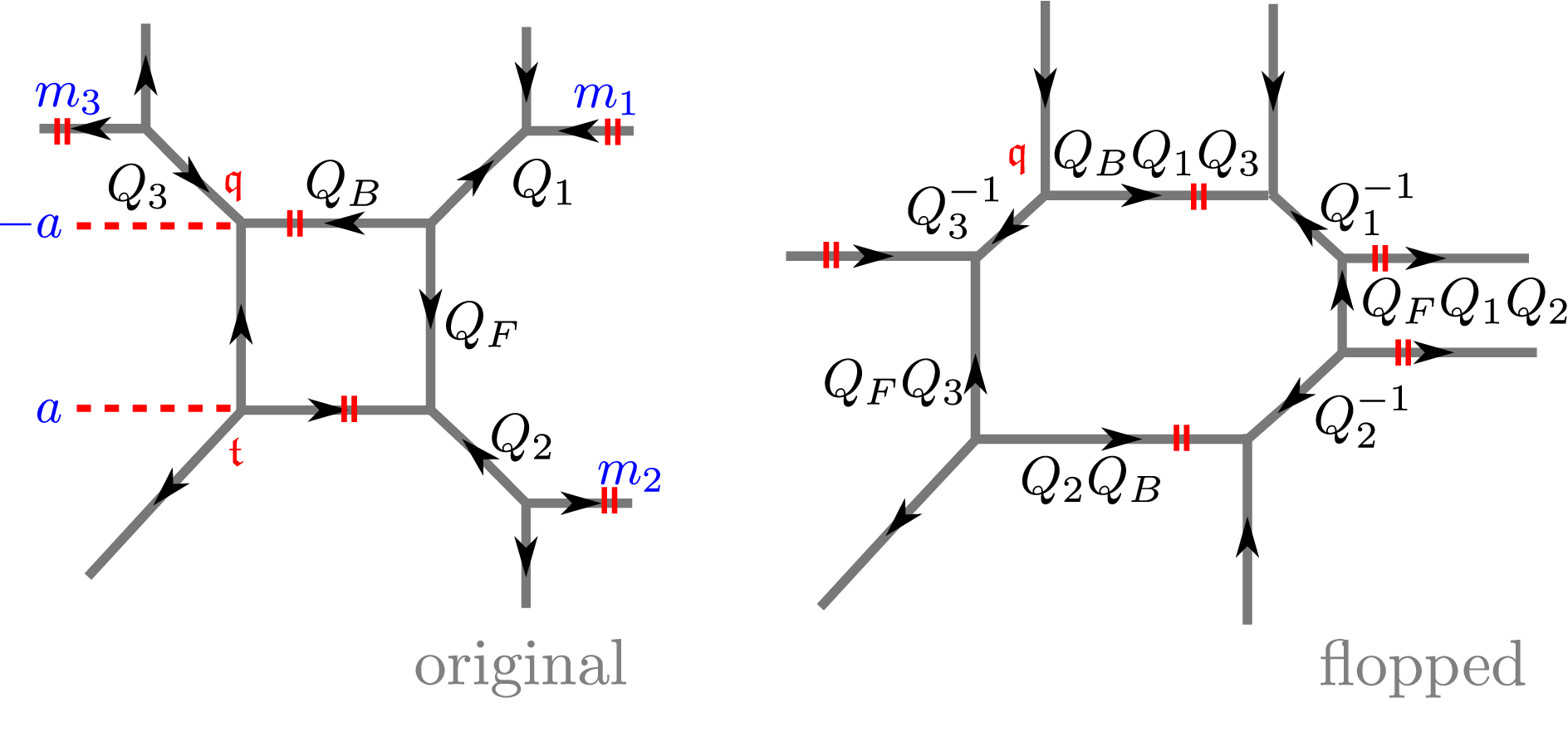}
\end{center}
\caption{The left hand side is the $\SU(2)$ with three flavor, while the right hand side is the flopped version. }
\label{fig:3flopped}
\end{figure}
The parametrization of the moduli is read off from figure~\ref{fig:3flopped} and can be expressed using the fugacities as
\begin{align}
&Q_F=e^{- 2 \beta a } ,& &Q_B
= q e^{- 2 \beta (a + \frac{1}{2} m_1-\frac{1}{2}m_2+\frac{1}{2}m_3) }
= e^{- 2 \beta a } \frac{y_2}{y_1},&
\nonumber \\
&Q_1
=e^{- \beta (-a-m_1) }
=e^{ \beta a } \sqrt{\frac{y_3 y_4}{y_2 y_5}},& &Q_2
=e^{- \beta (m_2-a) }
=e^{ \beta a } \sqrt{\frac{y_3 y_5}{y_2 y_4}},&
 \\
&Q_3
=e^{- \beta (-a-m_3) }
=e^{ \beta a } \sqrt{\frac{y_4 y_5}{y_2 y_3}}.& &&\nonumber
\end{align}
The fiber-base duality map now acts by exchanging not just $Q_B$ and $Q_F$ but also $Q_2$ and $Q_3$, which translates to
\be
\label{3f-map}
y_1 \leftrightarrow y_2,
\qquad y_3 \leftrightarrow y_4,
\qquad e^{- 2 \beta a }
\to y_1^{-1}  y_2e^{- 2 \beta a }.
\ee
The Weyl transformations of the non-enhanced flavor symmetry $\SO(6) = \SU(4)$ is given by
 $ m_i \leftrightarrow m_j,\, m_i \leftrightarrow - m_j$ .
They are equivalent to the set of transformations
\begin{eqnarray}
y_k \leftrightarrow y_{l} ,
\qquad
k,l=2,3,4,5.
\end{eqnarray}
These together with the first two transformations in \eqref{3f-map} generate the full
 Weyl group of the enhanced $\E_4$ symmetry. Note that $a$ also transforms non-trivially under Weyl reflections that involve $y_1$. One easily see that the invariant Coulomb modulus should be defined as
\begin{eqnarray}
\tilde{A}^2 = y_1^{-1} e^{- 2 \beta a}\Longrightarrow \tilde{A}=q^{\frac{2}{5}}e^{-\beta a},
\label{eq:shifted3}
\end{eqnarray}
leading to the following parametrization of the moduli
\begin{align}
\label{eq:param3flavor}
&Q_F=\tilde{A}^2 y_1,&
&Q_B=\tilde{A}^2 y_2,& &&\nonumber\\
&Q_1=\tilde{A}^{-1}y_3y_4,&
&Q_2=\tilde{A}^{-1}y_3y_5,&
&Q_3=\tilde{A}^{-1}y_4y_5.&
\end{align}
Taking the flopped three flavor diagram, computing the topological string partition function and dividing out \cite{Bao:2013pwa, Hayashi:2013qwa, Bergman:2013ala} the decoupled non-full spin content $\calM(Q_1Q_2Q_F)\calM(Q_1Q_3Q_B\frac{\ft}{\fq})$ leads to the partition function
\bea
\label{eq:partfunction3}
Z^{N_f=3}&=&\frac{\calM\big(Q_F\big)\calM\big(Q_F\frac{\ft}{\fq}\big)}{\calM(Q_1Q_3Q_B\frac{\ft}{\fq})\prod_{i=1}^3\calM\big(Q_i^{-1}\sqrt{\frac{\ft}{\fq}}\big)\calM\big(Q_iQ_F\sqrt{\frac{\ft}{\fq}}\big)}\nonumber\\&&\times \sum_{\boldsymbol{\mu}}(Q_1 Q_3 Q_B)^{|\mu_1|} (-Q_2 Q_B)^{|\mu_2|}\ft^{||\mu_1^t||^2+\frac{||\mu_2^t||^2}{2}}\fq^{\frac{||\mu_2||^2}{2}}\prod_{i=1}^2\tilde{Z}_{\mu_i}(\ft,\fq)\tilde{Z}_{\mu_i^t}(\fq,\ft)\\&&\times \frac{\left(\prod_{i=1,3}\calN_{\mu_1\emptyset}\big(Q_i^{-1}\sqrt{\frac{\ft}{\fq}}\big)\calN_{\emptyset\mu_2}\big(Q_i Q_F\sqrt{\frac{\ft}{\fq}}\big)\right)\calN_{\emptyset\mu_2}\big(Q_2^{-1}\sqrt{\frac{\ft}{\fq}}\big)\calN_{\mu_1\emptyset}\big(Q_2 Q_F\sqrt{\frac{\ft}{\fq}}\big)}{\calN_{\mu_1\mu_2}\big(Q_F\big)\calN_{\mu_1\mu_2}\big(Q_F\frac{\ft}{\fq}\big)}.\nonumber
\eea
Using \eqref{eq:param3flavor},
we obtain the following expansion of the topological string partition function
\begin{align}
Z^{N_f=3}=&1-\frac{\ft^{\frac{1}{2}}\fq^{\frac{1}{2}}\chi_{\overline{10}}^{\E_4}}{(1-\ft)(1-\fq)}\tilde{A}+\Big(\fq\ft\left(\fq+ \ft\right)\chi_{50}^{\E_4}+\fq \ft (1+\fq \ft)\chi_{45}^{\E_4}\\&+(\fq(1-\fq^2)+\ft(1-\ft^2)+\fq^2\ft^2(\fq+\ft))\chi_{5}^{\E_4}\Big)\frac{\tilde{A}^2}{(1-\fq)^2 (1+\fq) (1-\ft)^2 (1+\ft)}+\cdots\nonumber
\end{align}
where the characters are labeled by their dimensions.

\subsubsection{$\SU(2)$ with $N_f=4$}

\begin{figure}[h]
\begin{center}
\includegraphics[height=5cm]{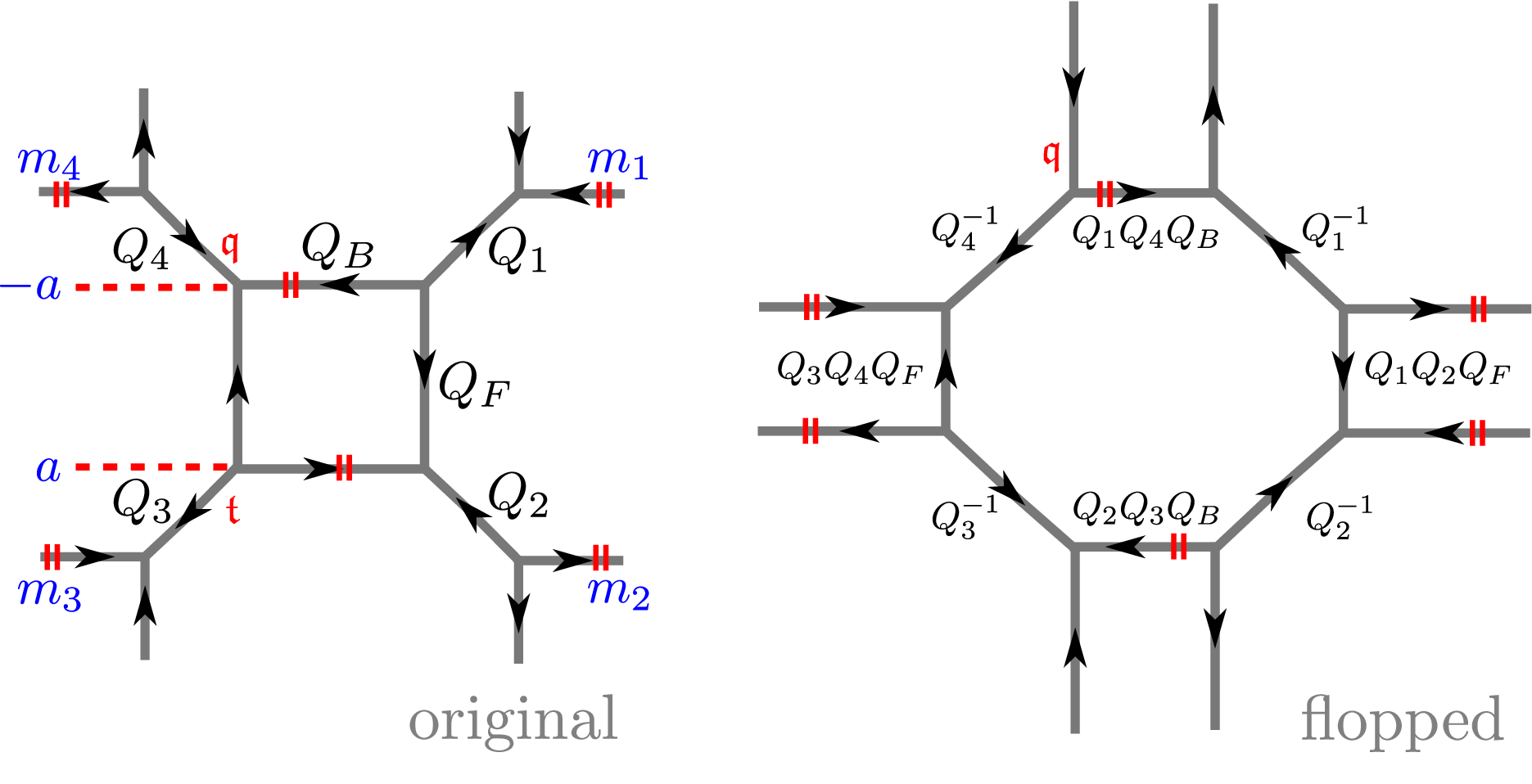}
\end{center}
\caption{This figure shows on the right hand side the web diagram of $\SU(2)$ with four flavors that is related to the original one by four floppings.}
\label{fig:4fullyflopped}
\end{figure}

In the case of four flavors, the enhanced symmetry is $\E_5 = \SO(10)$,
whose Weyl transformations are $ y_i \leftrightarrow y_j $ and
$y_i \leftrightarrow y_j{}^{-1}$. The five independent fugacities $y_i$ are given by
\begin{align}
&y_1 = q, &
&y_2 = e^{- \frac{1}{2} \beta (-m_1 + m_2 + m_3 - m_4)},&
&y_3 = e^{- \frac{1}{2} \beta (-m_1 - m_2 + m_3 + m_4)},&\nonumber \\
&y_4 = e^{- \frac{1}{2} \beta (m_1 - m_2 + m_3 - m_4)},&
&y_5 = e^{- \frac{1}{2} \beta ( m_1 + m_2 + m_3 + m_4)}.&&&
\end{align}
The parametrization of the moduli is given by
\begin{align}
\label{eq:paramfourflavor}
&Q_F=e^{- 2 \beta a },  &
&Q_B
= q e^{-  \beta (2a + \frac{1}{2} m_1-\frac{1}{2}m_2
-\frac{1}{2}m_3+\frac{1}{2}m_4) }
= e^{- 2 \beta a }\frac{y_1}{y_2},&
\nonumber \\
&Q_1
=e^{- \beta (-a-m_1) }
=e^{ \beta a } \sqrt{\frac{ y_2 y_3}{y_4 y_5}},&
&Q_2=e^{- \beta (m_2-a) }
=e^{ \beta a } \sqrt{\frac{ y_2 y_5}{y_3 y_4}},&
\\
&Q_3
=e^{- \beta (m_3-a) }
=e^{ \beta a } \sqrt{y_2 y_3y_4y_5},&
&Q_4
=e^{- \beta (-a-m_4) }
=e^{ \beta a } \sqrt{\frac{ y_2 y_4}{y_3 y_5} },&\nonumber
\end{align}
as  illustrated in figure~\ref{fig:4fullyflopped}.
The fiber-base duality map for our parametrization is
\begin{eqnarray}
Q_F \leftrightarrow Q_B,
\qquad Q_1 \to Q_1, \qquad Q_2 \leftrightarrow Q_4, \qquad Q_3 \to Q_3
\end{eqnarray}
which is translated into
\begin{eqnarray}
 y_1 \leftrightarrow y_2,
\quad y_4 \leftrightarrow y_5,
\quad e^{- 2 \beta a }
\to \frac{y_1}{y_2}  e^{- 2 \beta a }
\label{4f-map}
\end{eqnarray}
These together with the first two in (\ref{4f-map}) generates the
$\E_5$ Weyl symmetry. If we define
\begin{eqnarray}
\tilde{A}^2 = y_1 e^{- 2 \beta a}\Longrightarrow \tilde{A}=q^{\frac{1}{2}}e^{-\beta a}
\label{eq:shifted4}
\end{eqnarray}
which is invariant under the $\E_4$ Weyl transformation, we find the following parametrization
\begin{align}
\label{eq:param4flavor}
&Q_F=\tilde{A}^2 y_1^{-1},&
&Q_B=\tilde{A}^2 y_2^{-1},&
&Q_1=\tilde{A}^{-1}\sqrt{\frac{y_1 y_2 y_3}{y_4 y_5}},&\\
&Q_2=\tilde{A}^{-1}\sqrt{\frac{y_1 y_2 y_5}{y_3 y_4}},&
&Q_3=\tilde{A}^{-1}\sqrt{y_1 y_2 y_3 y_4 y_5},&
&Q_4=\tilde{A}^{-1}\sqrt{\frac{y_1 y_2 y_4}{ y_3 y_5}},&
\nonumber
\end{align}

In the four flavor case, as in the previous cases, we go to the frame that allows for an expansion in positive powers of the Coulomb modulus.
This frame is depicted in figure~\ref{fig:4fullyflopped} and its partition function is given, after dividing out the decoupled non-full spin content $\calM(Q_1Q_4Q_B)\calM(Q_2Q_3Q_B\frac{\ft}{\fq})\calM(Q_1Q_2Q_F)\calM(Q_3Q_4Q_F)$, by
\begin{align}
\label{eq:partfunction4fullyfloppedrenormalized}
Z^{N_f=4}=&\frac{\calM\big(Q_F\big)\calM\big(Q_F\frac{\ft}{\fq}\big)}{\calM(Q_1Q_4Q_B)\calM(Q_2Q_3Q_B\frac{\ft}{\fq})\prod_{i=1}^4\calM\big(Q_i^{-1}\sqrt{\frac{\ft}{\fq}}\big)\calM\big(Q_iQ_F\sqrt{\frac{\ft}{\fq}}\big)}\nonumber\\&\times \sum_{\boldsymbol{\mu}}(Q_2 Q_3 Q_B\frac{\ft}{\fq})^{|\mu_1|} (Q_1 Q_4 Q_B)^{|\mu_2|}\ft^{||\mu_1^t||^2}\fq^{||\mu_2||^2}\prod_{i=1}^2\tilde{Z}_{\mu_i}(\ft,\fq)\tilde{Z}_{\mu_i^t}(\fq,\ft)\\&\times \frac{\prod_{i=2,3}\calN_{\mu_1\emptyset}\big(Q_i^{-1}\sqrt{\frac{\ft}{\fq}}\big)\calN_{\emptyset\mu_2}\big(Q_i Q_F\sqrt{\frac{\ft}{\fq}}\big)\prod_{i=1,4}\calN_{\emptyset\mu_2}\big(Q_i^{-1}\sqrt{\frac{\ft}{\fq}}\big)\calN_{\mu_1\emptyset}\big(Q_i Q_F\sqrt{\frac{\ft}{\fq}}\big)}{\calN_{\mu_1\mu_2}\big(Q_F\big)\calN_{\mu_1\mu_2}\big(Q_F\frac{\ft}{\fq}\big)}.\nonumber
\end{align}
Setting the parameters \eqref{eq:param4flavor}
in \eqref{eq:partfunction4fullyfloppedrenormalized}, then expanding $Z^{N_f=4}$ in $\tilde{A}$ leads to
\bea
Z^{N_f=4}&=&1-\frac{\fq^{\frac{1}{2}}\ft^{\frac{1}{2}}\chi^{\E_5}_{\overline{15}}}{(1-\fq)(1-\ft)}\tilde{A}+\Big[\big(\fq(1-\fq^2)+\ft(1-\ft^2)+\fq^2\ft^2(\fq+\ft)\big)\chi^{\E_5}_{10}\nonumber\\&&+\fq\ft(1+\fq\ft)\chi^{\E_5}_{120}+\fq\ft(\fq+\ft)\chi^{\E_5}_{126}\Big]\frac{\tilde{A}^2}{(1-\ft)^2(1+\ft)(1-\fq)^2(1+\fq)}+\cdots
\eea
where $\E_5=\SO(10)$ and the characters are normalized such that the character of the fundamental representation is $\chi^{\E_5}_{10}=\sum_{i=1}^5(y_i+y_i^{-1})$. 

\subsection{The case of $\SU(2)$ with $N_f\geq  5$ fundamental flavors}
\label{Nfg5}

\begin{figure}[h]
\begin{center}
\includegraphics[height=3.5cm]{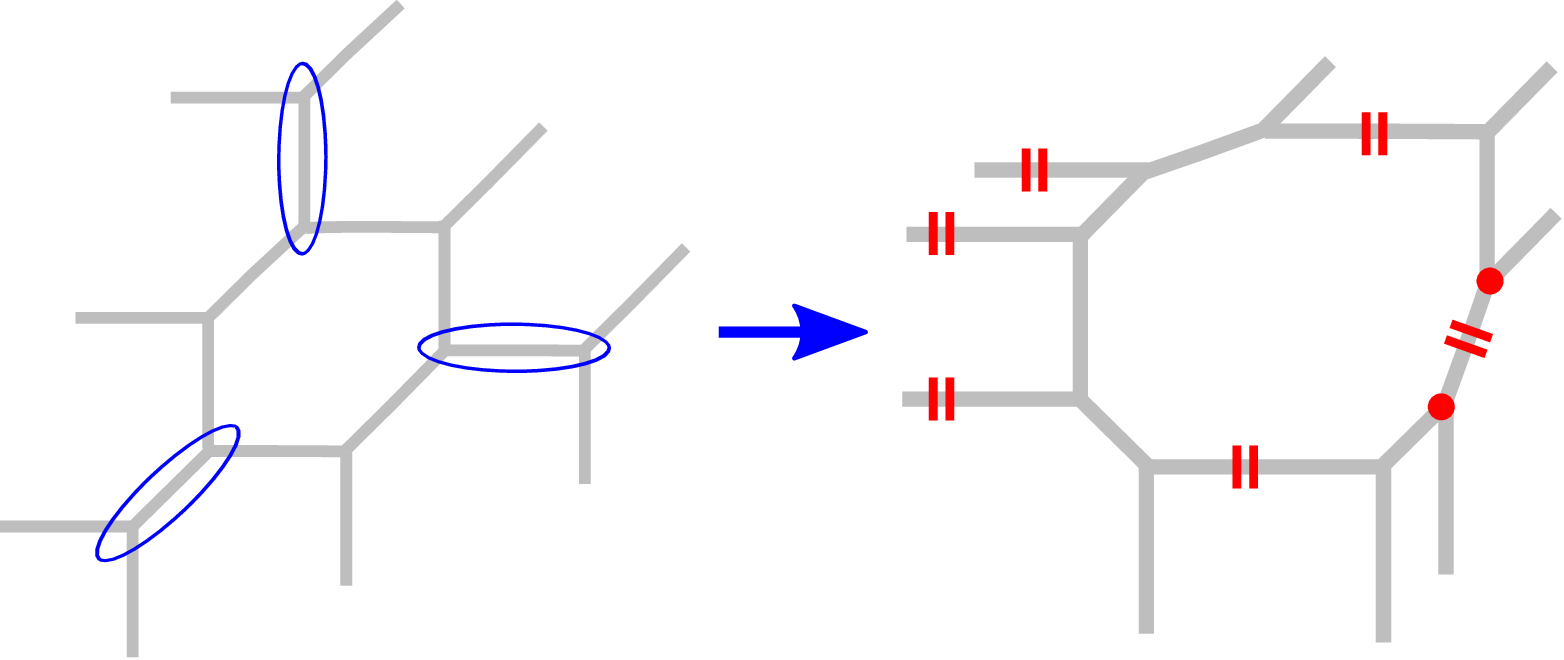}
\end{center}
\caption{The left hand side shows the standard $T_3$ web diagram while the right hand side depicts a flopped version, in which only positive powers of the invariant Coulomb modulus appear. To compute the  topological string amplitude, we need the new topological string vertex introduced by Iqbal and Koz{\c c}az \cite{Iqbal:2012mt}, depicted here by the red circles.}
\label{fig:T3flopped}
\end{figure}

Starting at five flavor, we have the problem that we cannot choose a frame that would simultaneously allow for an expansion in only positive powers of the invariant Coulomb modulus and that would also be treatable using the standard refined topological formalism. The issue being that in the frame in which only positive powers of the invariant Coulomb modulus appear, we cannot choose a preferred direction, see figure~\ref{fig:T3flopped} for an illustration in the $N_f=5$ case. In principle, one could use the new vertex \cite{Iqbal:2012mt} and we leave that for future research.

In \cite{Hayashi:2013qwa, Hayashi:2014wfa}, it is checked that the topological string
partition function for the $\E_{N_f+1}$ theory,
whose corresponding toric diagram is given in \cite{Benini:2009gi}
in the language of $(p,q)$ 5-brane web,
agree with the Nekrasov partition function for $\SP(1)$ theory with $N_f$ flavor.
Taking this result into account, we will now use the results of \cite{Hwang:2014uwa}, equation (3.38ff), to show the symmetry enhancement of the Nekrasov partition function for $N_f= 5, 6$ and $7$. Furthermore, for each case, we will show by using an appropriate $(p,q)$ web diagram that the  fiber-base duality is part of the enhanced symmetry.

In order to compute the instanton contribution,
we first need to include the contribution from the singlet field
(the $\SP(1)$ antisymmetric tensor) to compute the ADHM quantum mechanics and then,
divide by the extra factor which is discussed in section 3.4 in their paper
to obtain the instanton Nekrasov partition function.\footnote{
The computation in this subsection is based on the explicit result of the instanton computation,
which Chiun Hwuang, one of the author of \cite{Hwang:2014uwa}, gave us by Mathematica file.
We appreciate his kindness.
}

As for the perturbative part, we can easily read off from the computation
of the superconformal index.
The perturbative contribution from the vector multiplet to the
superconformal index is given in \cite{Kim:2012gu,Hwang:2014uwa} as
\begin{multline}
 {\rm PE} \left[ \frac{1+\fq\ft}{(1-\fq)(1-\ft)}
(\tilde{a}^{-2} + \tilde{a}^2 + 1) \right]
=\\= {\rm PE} \left[ \left( \frac{\fq+\ft}{(1-\fq)(1-\ft)} \right)
(\tilde{a}^{-2} + \tilde{a}^2 ) \right]
\frac{  M(\ft,\fq)M(\fq,\ft)}{ (2 \sinh (\beta a))^{2}}  \times {\rm PE} [ 1 ]\, ,
\end{multline}
where we have defined $\tilde{a} = e^{- \beta a}$.
The numerator of the second factor corresponds to the constant map, given by the MacMahon function,
which we have omitted in this paper.
The denominator of the second factor cancels with the Haar measure of the Coulomb moduli integral.
The last factor is the unimportant diverging constant, which we discard by hand.
Then, the perturbative contribution to the corresponding topological string
partition function is ``chiral half'' of the first part, which we give
\begin{eqnarray}
{\rm PE} \left[ \left( \frac{\fq+\ft}{(1-\fq)(1-\ft)} \right)
\tilde{a}^2  \right]\, .
\label{vec-pert-top}
\end{eqnarray}
Here we took the positive power of $\tilde{a}$ so that
it agrees with the topological string result which we have used
in the previous subsection.\footnote{
Our choice is slightly different from the
standard expression of perturbative of the Nekrasov partition function.
This difference is addressed in detail in section 3.6.
In the following, we simply use the expression in (\ref{vec-pert-top}).
}
Also for the hypermultiplet contribution, we choose
the chiral half of the perturbative part of the superconformal index
in such a way that the only the positive power of $\tilde{a}$ appears.

Defining $\tilde{m}_i=e^{-\beta m_i}$,
we can give the perturbative part of the partition function for any $N_f$
\begin{eqnarray}
\label{eq:pertpartgeneralNf}
Z_{\rm pert}^{N_f}
&=& \mathrm{PE} \left[
- \frac{\fq^{\frac{1}{2}} \ft^{\frac{1}{2}}}{(1-\fq)(1-\ft)}
\sum_{\ell=1}^{N_f}
(\tilde{m}_{\ell} + \tilde{m}_{\ell}^{-1}) \tilde{a}
+ \frac{\fq+\ft}{(1-\fq)(1-\ft)} \tilde{a}^2
 \right]
\nonumber \\
&=&
1 - \frac{\fq^{\frac{1}{2}} \ft^{\frac{1}{2}}}{(1-\fq)(1-\ft)}
\chi^{\SO(2N_f)}_{\rm fund}\ \tilde{a} + \Bigg[
\frac{\fq+\ft}{(1-\fq)(1-\ft)}
+ \frac{\fq \ft}{(1-\fq^2)(1-\ft^2)} \chi_{\rm antisym}^{\SO(2N_f)}
\nonumber \\
&&
+ \frac{\fq \ft (\fq+\ft)}{(1-\fq)^2(1+\fq)(1-\ft)^2(1+\ft)} \left(\chi_{\rm fund}^{\SO(2N_f)}\right)^2
\Bigg] \tilde{a}^2 + \cdots,
\end{eqnarray}
which is consistent with what we have used in the previous subsection.

\subsubsection{$\SU(2)$ with $N_f=5$}

\begin{figure}
\centering
\includegraphics[width=12cm]{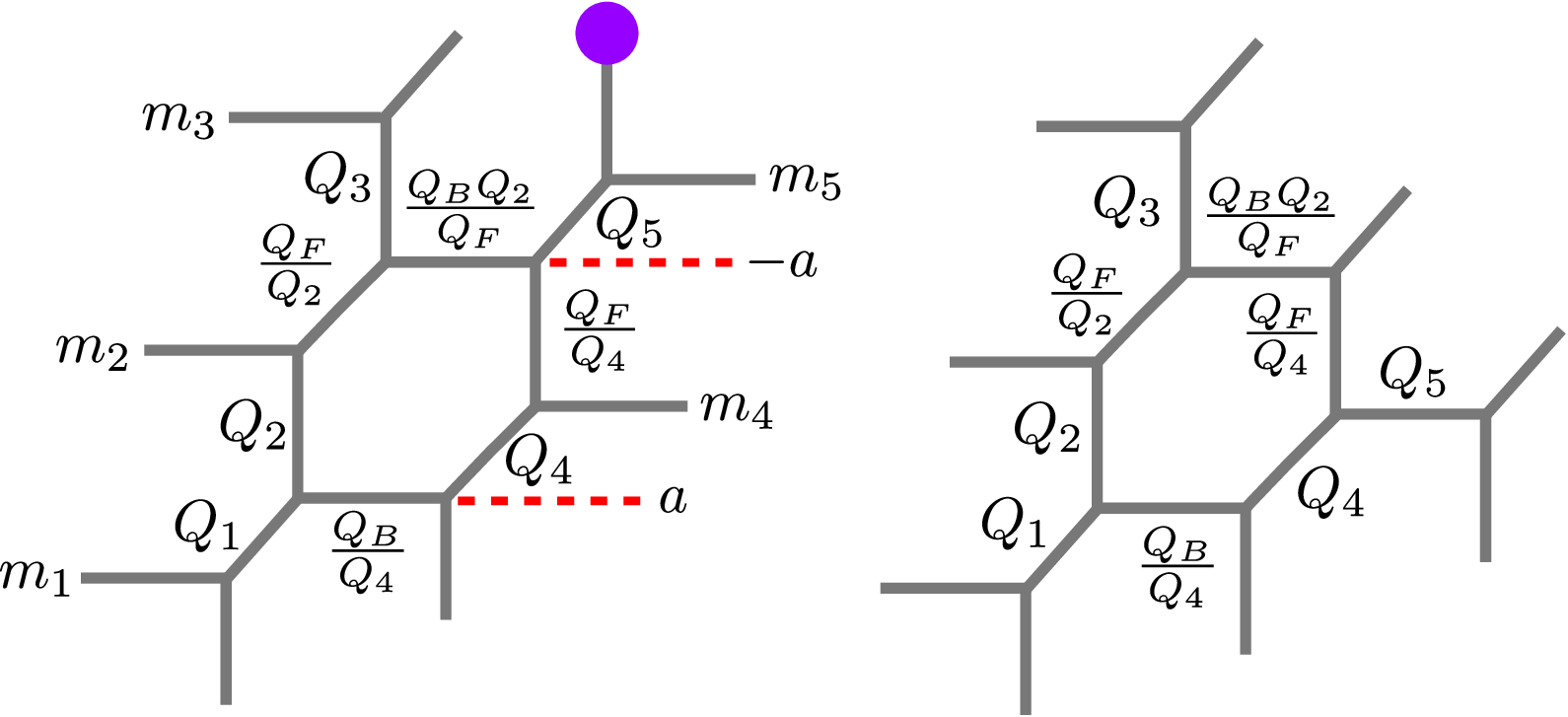}
\caption{The left hand side of this figure illustrates $\SU(2)$ with 5 flavor, where one 5-brane ends on a 7-brane. We then move the 7-brane and by the Hanany-Witten effect obtain the right hand side of the figure which shows the $T_3$ \cite{Benini:2009gi}.  }
 \label{fig:E6}
\end{figure}

\begin{figure}[h]
\begin{center}
\includegraphics[height=3cm]{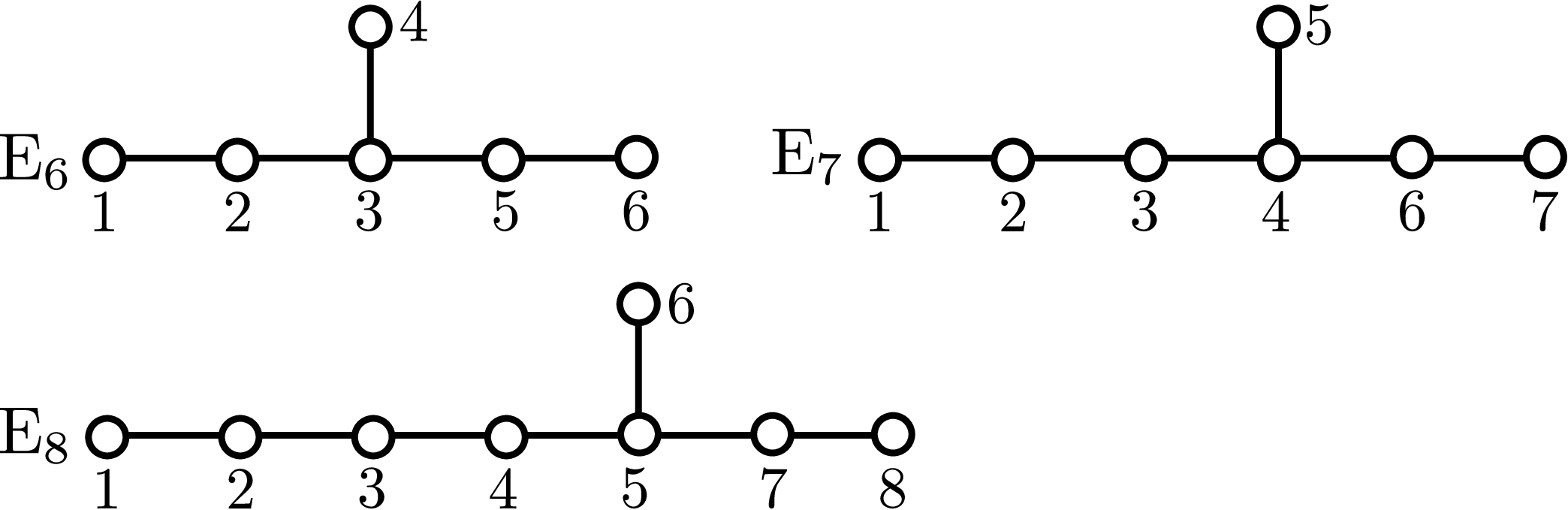}
\end{center}
\caption{This figure shows our labeling of the roots of E$_6$, E$_7$ and E$_8$.}
\label{fig:EDynkin}
\end{figure}

For the case of $\SU(2)$ with five flavors, we start with the left side of figure~\ref{fig:E6} and by the Hanany-Witten effect obtain the right hand side which is the $T_3$ web diagram, after we set the $7$-brane to infinity. We  use the left hand side of figure~\ref{fig:E6} to read off the parametrization and the right hand side to derive the action of the fiber-base duality.
We introduce the fugacities
\be
\label{eq:fugacitiesE6}
y_i=e^{-\beta m_i}, \qquad y_j=e^{\beta m_j}, \qquad y_6=q^{-2}.
\ee
for $i=1, 4$ and $j=2, 3, 5$. The $\SO(10)$ Weyl transformations act as $y_i\leftrightarrow y_j$ and $y_i\leftrightarrow y_j^{-1}$ for $i,j=1,\ldots, 5$. We use these fugacities to parametrize the  K\"ahler
moduli of figure~\ref{fig:E6} as
\begin{align}
&Q_1=e^{\beta a}y_1,& &Q_2=e^{-\beta a}y_2,&  &Q_3=e^{\beta a}y_3,& &&\nonumber\\ &Q_4=e^{-\beta a}y_4^{-1},& &Q_5=e^{\beta a}y_5,& &Q_F=e^{-2\beta a},& &Q_B=e^{-2\beta a}\prod_{i=1}^6y_i^{-\frac{1}{2}}.&
\end{align}
The fiber-base duality map is easily read off from figure~\ref{fig:E6} and leaves $Q_1$ invariant while transforming the rest as
\begin{eqnarray}
Q_3 \leftrightarrow Q_5,
\qquad
Q_2 \leftrightarrow Q_BQ_4^{-1},
\qquad
Q_F \leftrightarrow Q_B.
\end{eqnarray}
In terms of the fugacities \eqref{eq:fugacitiesE6}, we can write the fiber-base duality as
\be
\label{eq:fiberbaseNf5part2}
e^{-\beta a}\rightarrow e^{-\beta a}\prod_{k=1}^6y_k^{-\frac{1}{4}}, \qquad
y_i\rightarrow y_i\prod_{k=1}^6y_k^{-\frac{1}{4}},\qquad y_6\rightarrow y_6\prod_{k=1}^7y_k^{-\frac{3}{4}}.
\ee
combined with the exchange $y_3\leftrightarrow y_5$ and $y_2\leftrightarrow y_4$. 
We would like to see this fiber-base duality as a Weyl reflection of E$_6$. For this purpose, let us number  the simple roots of E$_6$ as in figure~\ref{fig:EDynkin} and, using a system of weights $h_i$ subject to $(h_i,h_j)=\delta_{ij}$, given by
\be
e_i = h_i-h_{i+1}, \qquad e_5=h_4+h_5,\qquad e_6 = -\frac{1}{2}\sum_{k=1}^5 h_k - \frac{\sqrt{3}}{2} h_6.
\ee
One easily sees that the scalar products of the above expressions are $(e_i,e_i)=2$ for all $i$ and $(e_i,e_j)=-1$ iff the nodes $i$ and $j$ are connected in  figure~\ref{fig:EDynkin}. We can identify the fugacities $y_i$ with $e^{h_i}$ for $i=1,\ldots, 5$ and $y_6=e^{\sqrt{3}h_6}$ so that the Weyl reflection corresponding to a root $\alpha$ of E$_6$ is given by $e^{\tilde{\alpha}}\rightarrow e^{\tilde{\alpha}-(\tilde{\alpha},\alpha)\alpha}$. With this identification, we see that the fiber-base duality \eqref{eq:fiberbaseNf5part2} can be understood as a Weyl reflection with respect to $e_6$ followed by two $\SO(10)$ reflections exchanging $y_3\leftrightarrow y_5$ and $y_2\leftrightarrow y_4$. Thus we have uncovered the E$_6$ nature of the fiber-base transformation. Furthermore,  we also read from \eqref{eq:fiberbaseNf5part2} that the combination
\begin{eqnarray}
\label{eq:invariantCoulombmodulusNf5}
\tilde{A} = q^{\frac{2}{3}} e^{-\beta a}
\end{eqnarray}
is invariant under the duality map and in fact under the whole E$_6$ Weyl symmetry group.

We can now turn our attention to the partition function. The instanton contribution is
\begin{eqnarray}
\label{eq:fiveflavorinstantons}
Z_{\text{inst}}^{N_f=5}&=&1+\Bigg( - \frac{\fq^{\frac{1}{2}} \ft^{\frac{1}{2}}}{(1-\ft)(1-\fq)} \chi_{\overline{16}} \tilde{a}
+ \frac{\fq + \ft}{(1-\ft)(1-\fq)} \chi_{16} \tilde{a}^2\Bigg)q+\Bigg( - \frac{\fq^{\frac{1}{2}} \ft^{\frac{1}{2}}}{(1-\ft)(1-\fq)} \tilde{a}\nonumber \\
&&
+ \Bigg[
\frac{\fq +\ft}{(1-\fq)(1-\ft)} \chi_{10}
+ \frac{\fq \ft}{(1 - \fq^2) (1 - \ft^2) } \chi_{120}
\nonumber \\
&&
+ \frac{\fq \ft (\fq + \ft) }{(1 - \fq)^2 (1 + \fq) (1 - \ft)^2 (1 + \ft) } \chi_{\overline{16}}{}^2
\Bigg] \tilde{a}^2 \Bigg)q^2+\Bigg(\frac{\fq \ft \chi_{\overline{16}}}{ (1 - \fq)^2 (1 - \ft)^2 } \tilde{a}^2\Bigg)q^3  \nonumber\\
&&+\Bigg(\frac{\fq \ft (\fq + \ft)}{(1 - \fq)^2 (1 + \fq) (1 - \ft)^2 (1 + \ft)}
\tilde{a}^2\Bigg)q^4+\mathcal{O}(\tilde{a}^3)+\mathcal{O}(q^5).
\end{eqnarray}
Multiplying the above with the perturbative part \eqref{eq:pertpartgeneralNf} and using the invariant Coulomb modulus \eqref{eq:invariantCoulombmodulusNf5}, we arrive at\footnote{We use the LieART \cite{Feger:2012bs} convention of what constitutes $V$ and what is $\overline{V}$.}
\begin{eqnarray}
Z^{N_f=5}
\label{eq:finalNf5expansion}
&=& Z^{N_f=5}_{\text{pert}}Z^{N_f=5}_{\text{inst}} =1 - \frac{\fq^{\frac{1}{2}}\ft^{\frac{1}{2}}}{(1-\fq)(1-\ft)} \chi_{\overline{27}}^{\E_6}\tilde{A}
+ \Bigg[
\frac{\fq + \ft}{(1-\fq)(1-\ft)} \chi_{27}^{\E_6}
 \\
&&
+ \frac{\fq \ft}{(1-\fq^2)(1-\ft^2)} \chi_{351}^{\E_6}
+ \frac{\fq \ft(\fq+\ft)}{(1-\fq)^2(1+\fq)(1-\ft)^2(1+\ft)}
\left( \chi_{\overline{27}}^{\E_6} \right) ^2
\Bigg] \tilde{A}^2 + \cdots.\nonumber
\end{eqnarray}
The precise character identities needed to derive \eqref{eq:finalNf5expansion} from the instanton and perturbative parts are contained in appendix \ref{app:characters}.

\subsubsection{$\SU(2)$ with $N_f= 6$}

\begin{figure}
\centering
\includegraphics[width=12cm]{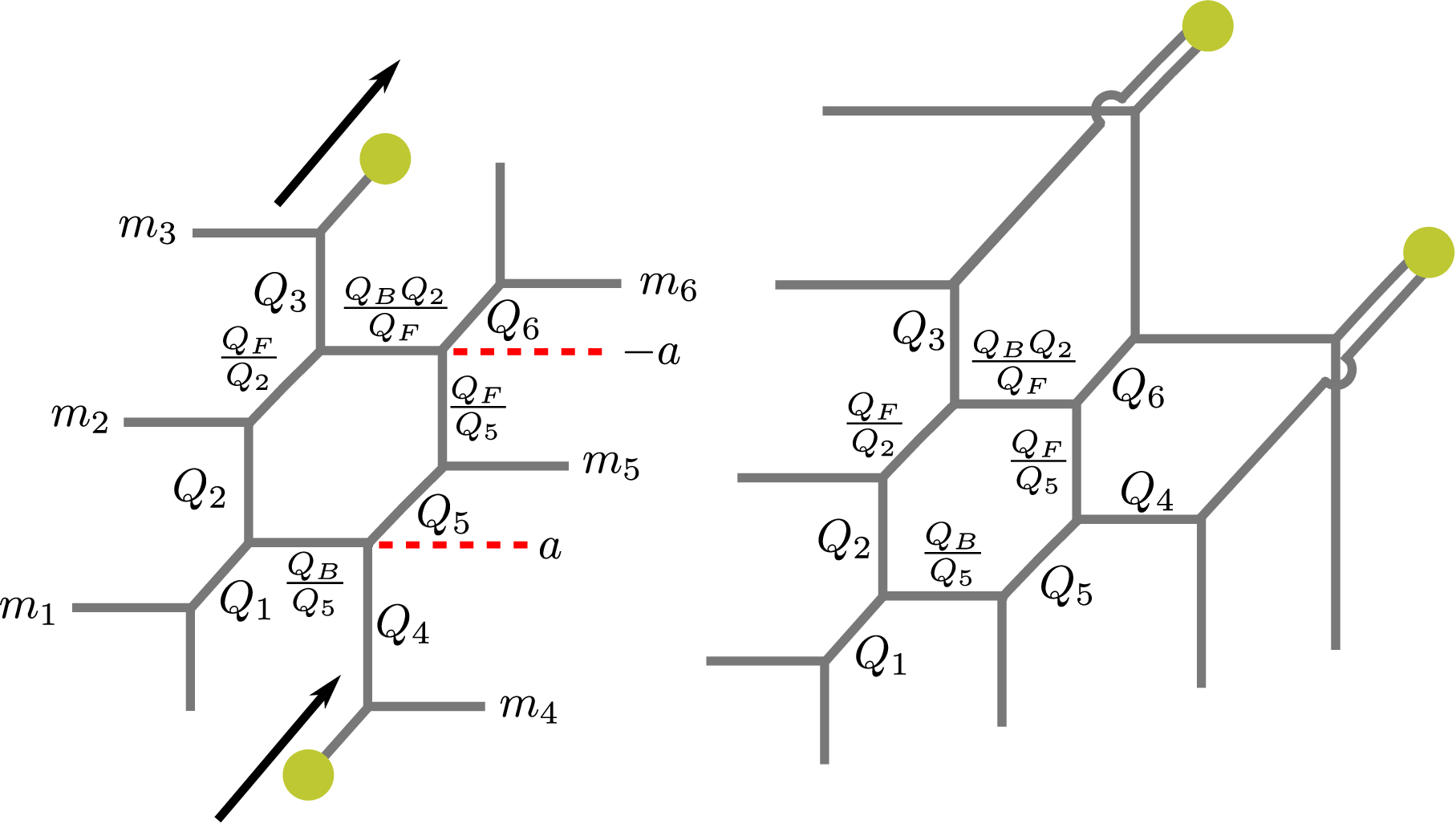}
\caption{The left hand side of this figure illustrates $\SU(2)$ with 6 flavor, where two 5-branes end on 7-branes. We then move the 7-branes as indicated by the arrows and by the Hanany-Witten effect obtain the right hand side of the figure which shows the Higgsed $T_4$. The parametrization of the Higgsed $T_4$ is derived  from the parametrization of the left hand side. }
 \label{fig:E7}
\end{figure}

For the case of $\SU(2)$ with six flavors, we start with the left side of figure~\ref{fig:E7} and by Hanany-Witten effect obtain the right hand side which can also be understood by taking the $T_4$ web diagram and Higgsing one side. We  use the left hand side of figure~\ref{fig:E7} to read off the parametrization and the right hand side to derive the action of the fiber-base duality.
We introduce the fugacities
\be
\label{eq:fugacitiesE7}
y_i=e^{-\beta m_i}, \qquad y_j=e^{\beta m_j}, \qquad y_7=q^{-2}.
\ee
for $i=1, 4 , 5$ and $j=2, 3, 6$. The $\SO(12)$ Weyl transformations act as $y_i\leftrightarrow y_j$ and $y_i\leftrightarrow y_j^{-1}$ for $i,j=1,\ldots, 6$.
Using these fugacities we parametrize the K\"ahler parameters in figure~\ref{fig:E7} as
\begin{align}
&Q_1 =e^{\beta a}y_1,&
& Q_2 = e^{-\beta a}y_2,&
&Q_3 =e^{\beta a}y_3,& &Q_4 = e^{\beta a}y_4, &
\nonumber \\
& Q_5 = e^{-\beta a}y_5^{-1},&
& Q_6 = e^{\beta a}y_6,& &Q_F =e^{-2\beta a},&  &Q_B
= e^{-2\beta a}\prod_{i=1}^7y_i^{-\frac{1}{2}}.&
\end{align}
The duality map is easily taken from the right hand side of figure~\ref{fig:E7} and reads
\begin{eqnarray}
\label{eq:fiberbaseNf6part1}
Q_1 \leftrightarrow Q_1,
\quad
Q_2 \leftrightarrow Q_B Q_5^{-1},
\quad
Q_3 \leftrightarrow Q_4,
\quad
Q_6 \leftrightarrow Q_6,
\quad
Q_F \leftrightarrow Q_B.
\end{eqnarray}
Translated to a transformation involving the fugacities \eqref{eq:fugacitiesE7}, the fiber-base duality becomes
\be
\label{eq:fiberbaseNf6part2}
e^{-\beta a}\rightarrow e^{-\beta a}\prod_{k=1}^7y_k^{-\frac{1}{4}}, \qquad
y_i\rightarrow y_i\prod_{k=1}^7y_k^{-\frac{1}{4}},\qquad y_7\rightarrow y_7\prod_{k=1}^7y_k^{-\frac{1}{2}},
\ee
together with the exchange $y_2\leftrightarrow y_5$ and $y_3\leftrightarrow y_4$. Let us now see how to interpret this transformation as arising from a Weyl reflection of E$_7$. To construct the root system of E$_7$, we start with seven basic weights $h_i$ obeying the scalar products $(h_i,h_j)=\delta_{ij}$ and then build the simple roots of the Lie algebra E$_7$, numbered as shown on the Dynkin diagram of figure~\ref{fig:EDynkin}, as
\be
e_i=h_i-h_{i+1},\qquad e_6=h_5+h_6,\qquad e_7=-\frac{1}{2}\sum_{k=1}^6h_k-\frac{1}{\sqrt{2}}h_7,
\ee
for $i=1,\ldots, 5$. The first six simple roots generate the Dynkin diagram of $\SO(12)$. We can identify the fugacities as $y_i=e^{h_i}$ for $i=1,\ldots, 6 $ together with $y_7=e^{\sqrt{2}h_7}$, so that Weyl reflections with respect to the root $\tilde{\alpha}$ are defined to be acting as $e^{\alpha}\mapsto e^{\alpha-(\alpha,\tilde{\alpha})\tilde{\alpha}}$, where we have used the fact that E$_7$ is simply laced. We can now see that the fiber-base transformation \eqref{eq:fiberbaseNf6part2} can be understood the Weyl reflection corresponding to the additional simple root $e_7$ combined with two Weyl reflections of $\SO(12)$ exchanging $y_2\leftrightarrow y_5$ and $y_3\leftrightarrow y_4$.  From \eqref{eq:fiberbaseNf6part2}, we also find that the the combination $qe^{-\beta a}$
is invariant under both the $\SO(12)$ Weyl transformation (that affect only the $m_i$) and the fiber-base duality and so we define the invariant Coulomb modulus as
\be
\label{eq:invariantCoulombModulusNf6}
\tilde{A}=qe^{-\beta a}.
\ee
The instanton contribution for $N_f=6$ is given by
\begin{eqnarray}
\label{eq:sixflavorinstantons}
Z_{\text{inst}}^{N_f=6}&=&1+\Bigg( - \frac{\fq^{\frac{1}{2}} \ft^{\frac{1}{2}} \chi_{\overline{32}}}{(1-\ft)(1-\fq)} \tilde{a}
+ \frac{ (\fq + \ft) \chi_{{32}}}{(1-\ft)(1-\fq)} \tilde{a}^2 \Bigg)q+\Bigg(- \frac{\fq^{\frac{1}{2}} \ft^{\frac{1}{2}} \chi_{12}}{(1-\ft)(1-\fq)} \tilde{a}
\nonumber \\
&&
+ \Biggl[
\frac{\fq + \ft}{(1 - \fq) (1 - \ft)} (\chi_{66} + 1)
+ \frac{\fq \ft}{(1 - \fq^2) (1 - \ft^2)} (\chi_{495} + 1)
\nonumber \\
&&
 + \frac{\fq \ft (\fq + \ft)}
{(1 - \fq)^2  (1 + \fq) (1 - \ft)^2 (1 + \ft)} \chi_{\overline{32}}{}^2
+ \frac{(1 + \fq \ft) (\fq^2 + \fq \ft + \ft^2)}{\fq \ft (1 - \fq) (1 - \ft) }
\Biggr] \tilde{a}^2 \Bigg)q^2\nonumber\\
&&+\Bigg( \left[
\frac{\fq \ft}{(1-\fq)^2 (1-\ft)^2} \chi_{\overline{32}} \chi_{12}
+ \frac{\fq+\ft}{(1-\fq)(1-\ft)} \chi_{{32}}
\right] \tilde{a}^2 \Bigg)q^3\nonumber\\
&&+\Bigg(
\Biggl[
\frac{\fq\ft(\fq+\ft)}{(1-\fq)^2 (1+\fq) (1-\ft)^2 (1+\ft)} \chi_{12}{}^2
+ \frac{\fq \ft}{(1-\fq^2)(1-\ft^2)} \chi_{66}
\nonumber \\
&&
+ \frac{\fq + \ft}{(1-\fq) (1-\ft)}
\Biggr] \tilde{a}^2\Bigg)q^4+\mathcal{O}(\tilde{a}^3)+\mathcal{O}(q^5),
\end{eqnarray}
where $\tilde{a}=e^{-\beta a}$. Plugging in our expression for the invariant Coulomb modulus  \eqref{eq:invariantCoulombModulusNf6} and multiplying with the perturbative part, we find
\begin{eqnarray}
\label{eq:finalNf6expansion}
Z^{N_f=6}
&=&
 1 - \frac{\fq^{\frac{1}{2}}\ft^{\frac{1}{2}}}{(1-\fq)(1-\ft)} \chi_{56}^{\E_7}\tilde{A}
+ \Bigg[
\frac{\fq+\ft}{(1-\fq)(1-\ft)} \chi_{133}^{\E_7}
+ \frac{\fq \ft}{(1-\fq^2)(1-\ft^2)} (\chi_{1539}^{\E_7} + 1)
\nonumber \\
&&
+ \frac{\fq \ft (\fq+\ft)}{(1-\fq)^2(1+\fq)(1-\ft)^2(1+\ft)} \left( \chi_{56}^{\E_7} \right)^2
+ \frac{(1+\fq\ft)(\fq^2+\fq\ft+\ft^2)}{\fq\ft(1-\fq)(1-\ft)}\Bigg] \tilde{A}^2  \nonumber\\ &&+\mathcal{O}(\tilde{A}^3).
\end{eqnarray}

\subsubsection{$\SU(2)$ with $N_f= 7$}

\begin{figure}
\centering
\includegraphics[width=12cm]{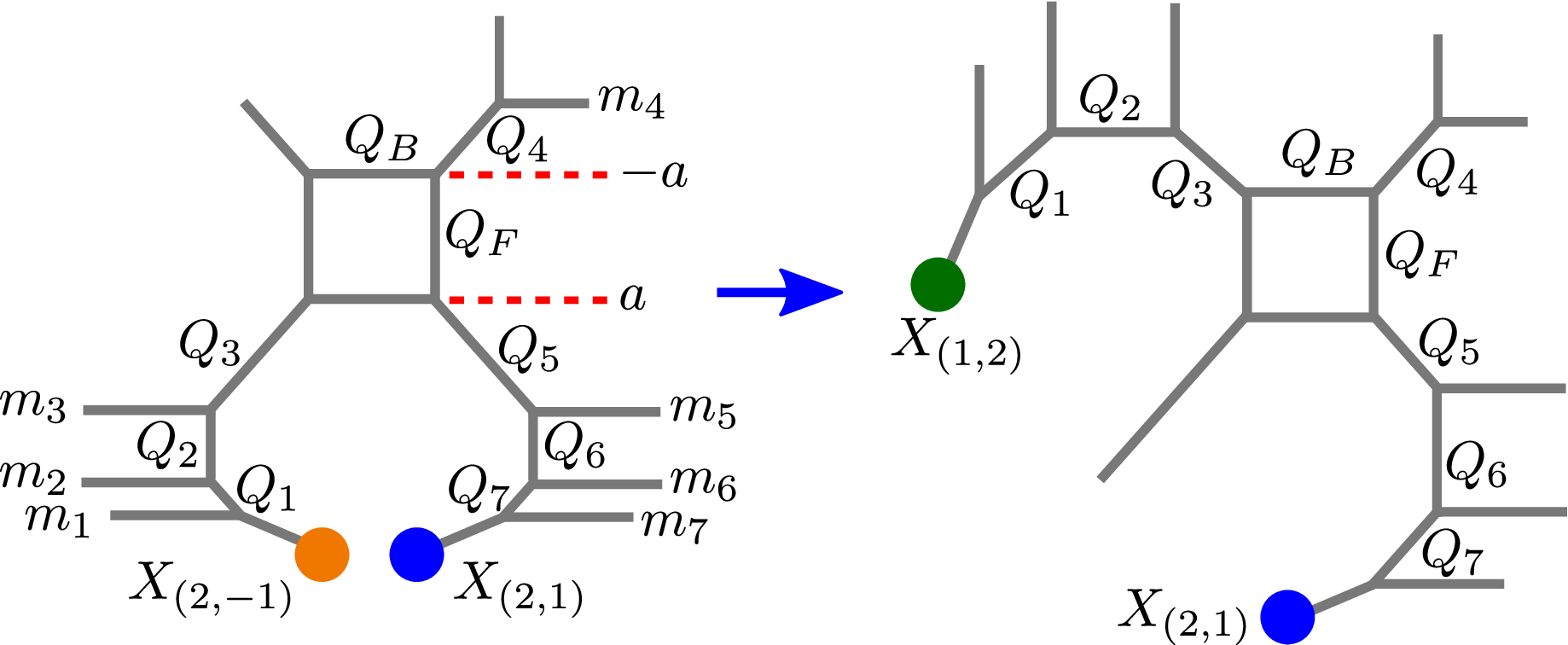}
\caption{This figure shows the brane configuration for $N_f=7$ with two 7-branes. Using the Hanany-Witten effect we obtain a fiber-base invariant configuration.}
\label{fig:E8}
\end{figure}
The final case to consider is the one of $\SU(2)$ with seven flavors. 
The web diagram is shown in figure~\ref{fig:E8}, with the right hand side showing the configuration for which the fiber-base symmetry is apparent.\footnote{F.Y.~thanks Gabi Zafrir for giving us an idea about this diagram.} As in the previous subsections, we introduce the fugacities
\be
y_i=e^{-\beta m_i}\text{ for } i\neq 4,8,\qquad y_4=e^{\beta m_4},\qquad y_8=q^{-2},
\ee
so that the  K\"ahler parameter are given by
\begin{align}
&Q_1=y_1y_2^{-1},& &Q_2=y_2y_3^{-1},& &Q_3=y_3 e^{\beta a},& &Q_4=y_4 e^{\beta a},& && \nonumber\\
&Q_5=y_5 e^{\beta a},& &Q_6=y_5^{-1}y_6,& &Q_7=y_6^{-1} y_7,& &Q_F=e^{-2\beta a}, &Q_B=e^{-2\beta a}\prod_{i=1}^8y_i^{-\frac{1}{2}}.&
\end{align}
From the right hand side of figure~\ref{fig:E8}, we see that the fiber-base duality map acts on the  K\"ahler parameters as
\begin{eqnarray}
Q_1 \leftrightarrow Q_7,
\quad
Q_2 \leftrightarrow Q_6,
\quad
Q_3 \leftrightarrow Q_5,
\quad
Q_4 \leftrightarrow Q_4,
\quad
Q_F \leftrightarrow Q_B,
\end{eqnarray}
which we translate as the following action on the fugacities
\be
\label{eq:FiberBaseE8}
e^{-\beta a}\rightarrow e^{-\beta a}\prod_{k=1}^8y_k^{-\frac{1}{4}},\qquad y_i\rightarrow y_i \prod_{k=1}^8y_k^{-\frac{1}{4}},
\ee
combined with the exchange $y_i$ with $y_{8-i}$ for $i=1,2$ and $3$.
As in the previous subsections, we wish to interpret \eqref{eq:FiberBaseE8} as an E$_8$ Weyl reflection.
The numbering of the simple roots of E$_8$ is shown in figure~\ref{fig:EDynkin} and the roots themselves are given as
\begin{eqnarray}
e_i=h_{i}-h_{i+1}\text{ for } 1\le i \le 6, \qquad e_7=h_6+h_7,\qquad e_8=-\frac{1}{2}\sum_{k=1}^8h_8,
\end{eqnarray}
where as usual $(h_i,h_j)=\delta_{ij}$.  
The fugacities themselves are given simply by $y_i=e^{h_i}$ and the Weyl reflections defined as before. 
We easily see that the fiber-base transformation \eqref{eq:FiberBaseE8} is simply a Weyl reflection by the root $e_8$.
Combined with the $\SO(14)$ Weyl transformations (generated by the simple roots $e_i$ with $i<8$) that exchange $y_i$ with $y_{8-i}$ for $i=1,2$ and $3$,
it generates the E$_8$ Weyl symmetry. Furthermore, we obtain the invariant Coulomb modulus
\begin{eqnarray}
\label{eq:invariantCoulombModulusNf7}
\tilde{A} =q^2 \tilde{a}.
\end{eqnarray}

We now direct our attention to the Nekrasov partition function. The instanton contributions are
\begin{eqnarray}
Z_{\text{inst}}^{N_f=7}&=&1+\Bigg( - \frac{\fq^{\frac{1}{2}} \ft^{\frac{1}{2}}}{(1-q)(1-t)} \chi_{\overline{64}}^{\SO(14)} \tilde{a}  \Bigg)q\nonumber\\
&&+\Bigg( \left[
\frac{(1+\fq\ft)(\fq+\ft)}{\fq^{\frac{1}{2}} \ft^{\frac{1}{2}}(1-\fq)(1-\ft)}
- \frac{\fq^{\frac{1}{2}} \ft^{\frac{1}{2}}}{(1-\fq)(1-\ft)} (\chi_{91}^{\SO(14)} + 1)
\right] \tilde{a}\Bigg)q^2\\
&&- \frac{\fq^{\frac{1}{2}} \ft^{\frac{1}{2}}}{(1-\fq)(1-\ft)} \chi_{64}^{\SO(14)} \tilde{a}q^3 - \frac{\fq^{\frac{1}{2}} \ft^{\frac{1}{2}}}{(1-\fq)(1-\ft)} \chi_{14}^{\SO(14)} \tilde{a} q^4+\mathcal{O}(\tilde{a}^2)+\mathcal{O}(q^5)\,.\nonumber
\end{eqnarray}
Combining the above with the perturbative part \eqref{eq:pertpartgeneralNf} leads to
\be
\label{eq:finalNf7expansion}
Z^{N_f=7} =1 + \Bigg[
 \frac{(1+\fq \ft)(\fq+\ft)}{\fq^{\frac{1}{2}} \ft^{\frac{1}{2}}(1-\fq)(1-\ft)}
- \frac{\fq^{\frac{1}{2}} \ft^{\frac{1}{2}}}{(1-\fq)(1-\ft)}
\chi^{\E_8}_{248}\Bigg] \tilde{A}+\mathcal{O}(\tilde{A}^2),
\ee
thus explicitly showing the E$_8$ invariance of the partition function.

\subsection{Effective coupling}
\label{subsec:effectivecoupling}

In this section we show that the fiber-base invariant Coulomb moduli parameter
$\tilde{A}$ that we introduced in the previous sections should be understood as the effective coupling constant of the theory.

The 5D theories that we have been studying are non-renormalizable and should be viewed as field theories with a cutoff. Even if in the classical theory the cubic term in
the prepotential vanishes,
it can be generated at one loop \cite{Witten:1996qb}.
Since the coefficient of the cubic term is a finite
quantity,  it is independent of the cutoff.  Following Seiberg \cite{Seiberg:1996bd},
the effective coupling constant for the
5D $\SU(2)$ theory with $N_f$ flavor is given by\footnote{This is equation (3.5) in  \cite{Seiberg:1996bd} but with different conventions.}
\begin{eqnarray}
\frac{1}{g_{\rm eff}{}^2}
= \frac{1}{g^2} + 4 \phi
- \frac{1}{4}\sum_i |\phi-m_i|
- \frac{1}{4}\sum_i |\phi+m_i| .
\end{eqnarray}
where $\phi$ is the vacuum expectation value of the real scalar in the 5D vector multiplet. In our article, we consider the parameter region where
$| \tilde{A} |  \ll 1,$
and the situation
$\phi \gg m_i. $
In this case, the effective coupling reduces to
\be
\frac{1}{g_{\rm eff}{}^2}
=\frac{1}{g^2} + 4\phi
- \frac{1}{4}\sum_i (\phi-m_i)
- \frac{1}{4}\sum_i (\phi+m_i)
= \frac{1}{g^2} + \big(4 - \frac{N_f}{2}\big) \phi\ .
\ee
In order to identify this effective coupling
constant as the ``invariant Coulomb moduli parameter'',
we divide by the factor $4-N_f/2$ and obtain
\begin{eqnarray}
\tilde{\phi}
\equiv
\frac{1}{4-\nicefrac{N_f}{2}} \frac{1}{g_{\rm eff}{}^2}
=
\phi + \frac{1}{4-\nicefrac{N_f}{2}} \frac{1}{g^2}\ .
\end{eqnarray}
Since $\phi$ is the real part of $a$, this implies that the invariant Coulomb moduli parameter
should be defined as
\begin{eqnarray}
\label{eq:generaldeftildeA}
\tilde{A} = e^{-\beta \tilde{a}}=e^{- \beta a} q^{\frac{2}{8-N_f}}
\end{eqnarray}
where $q$ is the instanton factor.
The invariant Coulomb moduli parameters introduced in
the previous subsections, see \eqref{eq:shifted0}, \eqref{eq:shifted1}, \eqref{eq:shifted2}, \eqref{eq:shifted3},  \eqref{eq:shifted4}, \eqref{eq:invariantCoulombmodulusNf5}, \eqref{eq:invariantCoulombModulusNf6} and \eqref{eq:invariantCoulombModulusNf7}  all agree with the general solution \eqref{eq:generaldeftildeA}.


\subsection{On the perturbative part of the partition functions }

In this subsection we focus on a subtle
difference between perturbative part in the Nekrasov partition function and the
perturbative part of the topological string partition function.

So far, we have identified the topological string partition function
divided by the contribution from the non-full spin content,
which we denote the normalized topological string partition function,
as the full Nekrasov partition function.
Indeed, it is known that the instanton part of the Nekrasov partition function
is perfectly reproduced from the normalized topological string partition function
\cite{Hayashi:2013qwa, Bao:2013pwa}.
However, there is a subtle difference between
the perturbative part in the Nekrasov partition function and the
perturbative part of the topological string partition function.
The tree level contribution is not included in
the topological string partition function
and moreover the 1-loop piece is also slightly different.

In the following, we clarify these difference at the level of the
effective coupling constant and check that this difference is also
invariant under the enhanced global symmetry.
Since we have already checked that
the normalized topological string partition function
is invariant under the enhanced global symmetry,
we discuss that the full Nekrasov partition function is also
invariant.

For simplicity, we will illustrate our point only for the pure $\SU(2)$ case
partially following discussion in \cite{Eguchi:2003sj}.
The extension to the cases with matter is straightforward.
Following  \cite{Iqbal:2012xm}, we have found in \eqref{eq:partfunction0}
that the perturbative contribution computed from
the topological string partition function is given by
\begin{equation}
Z_{\text{pert}}^{\rm top}=
\mathcal{M} \left( Q_F \right)
\mathcal{M} \left( \frac{\mathfrak{t}}{\mathfrak{q}} Q_F \right)
=\exp \left(
\sum_{n=1}^{\infty} \frac{Q_F{}^n}{n}
\frac{\mathfrak{q}^n + \mathfrak{t}^n}{(1-\mathfrak{q}^n)(1-\mathfrak{t}^n)}
\right)
\end{equation}
where we used \eqref{eq:mainfunctionsdefinitions}.
We can compute its contribution to the prepotential by taking the logarithm and it is
\begin{eqnarray}
\mathcal{F}_{\text{pert}}^{\rm top}
&\colonequals &
\lim_{\epsilon_1 \to 0 \atop \epsilon_2 \to 0}
\epsilon_1 \epsilon_2 \ln Z_{\text{pert}}^{\rm top}
= - \frac{2}{\beta^2} \sum_{n=1}^{\infty} \frac{Q_F{}^n}{n^3}
= - \frac{2}{\beta^2}\text{Li}_3(Q_F)
\end{eqnarray}
where we have used the standard convention \eqref{eq:Omegaqt}.
From this $\mathcal{F}_{\text{pert}}^{\rm top}$ that we obtain from topological string we go ahead and compute the 4D 1-loop effective coupling corrected by the contribution from the Kaluza-Klein tower. We find
\be
\label{eq:tauversion1}
\tau_{\text{pert}}^{\rm top}(a)
= \frac{\partial^2}{\partial a^2} \mathcal{F}_{\text{pert}}^{\rm top}
= - 8 \sum_{n=1}^{\infty} \frac{Q_F{}^n}{n}
= 8 \ln (1-Q_F)
\ee
where we used that $Q_F = e^{- 2 \beta a}$.
As mentioned above, the tree level contribution is not included.
Moreover, there is discrepancy,
already discussed in \cite{Lawrence:1997jr, Eguchi:2003sj}, between \eqref{eq:tauversion1} and
the 1-loop result computed from the field theory side
\cite{Nekrasov:1996cz}.
In order to reproduce the correct result known from perturbation theory, including the tree level part,
we need to add to the calculation \eqref{eq:tauversion1} from topological string
\be
\tau_{\text{missing}}(a)
= - 2 \ln (Q_F Q_B)
= 8 \beta a - 2 \ln q.
\ee
so that we correctly obtain
\be
\label{eq:taupert}
\tau_{\text{pert}}(a)
= \tau_{\text{1-loop}}^{\rm top}(a) + \tau_{\text{missing}}(a)
=  4 \ln ( 2 \sinh \beta a )^2 - 2 \ln q.
\ee
It is remarkable that the missing term $\tau_{\text{missing}}(a)$ is
also invariant under the fiber-base symmetry
$Q_F \leftrightarrow Q_B$.
Thus, the full Nekrasov partition
function is also invariant under the enhanced global symmetry.

Finally, we check that the perturbative contribution
(\ref{eq:taupert}), and not just (\ref{eq:tauversion1}) from the topological string,
reproduces the expected 4D/5D limit.
When we take decompactifying 5D limit,
we should substitute the 5D coupling constant in the instanton factor as
\begin{eqnarray}
\label{eq:5dlim}
q=\exp \left( - \beta \frac{1}{(g_0^{5D})^2} \right)
\end{eqnarray}
 and then take the limit $\beta \to \infty$.
Assuming Re$(a)>0$, we get
\begin{eqnarray}
\lim_{\beta \to \infty} \frac{1}{2 \beta} \tau_{\text{pert}}(a)
= \frac{1}{(g_0^{5D})^2} + 4 a \, ,
\end{eqnarray}
which is exactly the correct answer for the effective coupling constant
for the flat 5D SYM theory that appears already in  \cite{Seiberg:1996bd}.
On the other hand, when we take 4D limit, the instanton factor should be identified as
\begin{eqnarray}
q = (\beta \Lambda)^4 \, .
\end{eqnarray}
Then, we obtain
\begin{eqnarray}
\lim_{\beta \to 0} \tau_{\text{pert}}(a)
&=& 2 \ln \left(
\frac{(2 a)^4}{\Lambda^4}
\right)
\end{eqnarray}
which is the correct 4D effective coupling.


\section{Fiber-base duality for higher rank gauge theory}
\label{sec:symmetryenhancement}

In this section, we consider the case with the higher rank gauge group:
the $\SU(N)^{M-1}$ linear quiver theory with $N+N$ fundamental flavor,
which include $\SU(N)$ with $2N$ fundamental flavor as a special case.
Unlike $\SU(2)$ case, the theory is not self dual under
the fiber-base duality for generic $N$ and $M$.
In this case, the fiber-base duality map is not
part of the enhanced symmetry any more.
However, we will see that the fiber-base duality still plays
an important role to understand the global symmetry enhancement.
Especially, by using the fiber-base duality map,
we derive the invariant Coulomb moduli parameter,
which is invariant under the Weyl symmetry of the enhanced global symmetry.

The relation between the fiber-base duality and the global symmetry
enhancement for slightly different theory has been studied in \cite{Zafrir:2014ywa}
in the context of superconformal index.

\subsection{Notation and the duality map}
\begin{figure}[h]
\begin{center}
\includegraphics[width=10cm]{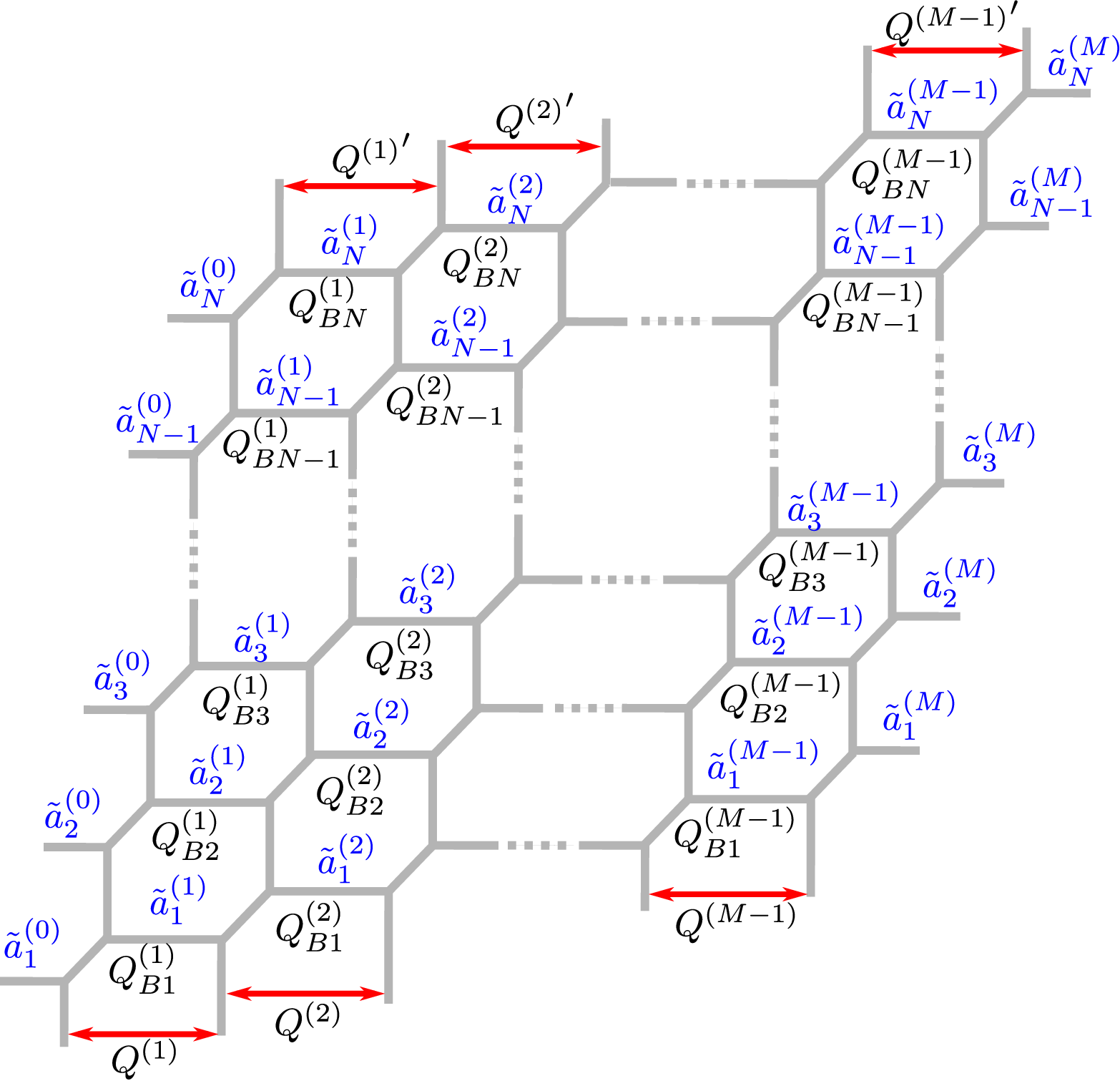}
\end{center}
\caption{This figure shows the parametrization of the web diagram for the linear quiver theory $\SU(N)^{M-1}$.}
\label{fig:parametrizationSUNM}
\end{figure}

We begin by introducing our parametrization of the web diagram of the $\SU(N)^{M-1}$ linear quiver with $N_f = N+N$. Here and in the following we follow \cite{Bao:2011rc} as closely as possible and work with exponentiated distances. We begin with the exponentiated positions $\tilde{a}^{(i)}_{\alpha}$ of the D5 branes, see figure~\ref{fig:parametrizationSUNM}, where $i=0,\ldots, M$ and $\alpha=1,\ldots, N$.
The horizontal exponentiated distances $Q_{B\alpha}^{(i)}$ can be written as follows
\be
\label{eq:defQB}
Q_{B\alpha}^{(i)}=\frac{\left(\prod_{\beta=1}^{\alpha-1}\tilde{a}_{\beta}^{(i)}\right)^2\tilde{a}_\alpha^{(i)}}{\prod_{\beta=1}^{\alpha}\tilde{a}_\beta^{(i-1)}\prod_{\beta=1}^{\alpha-1}\tilde{a}_\beta^{(i+1)}}Q^{(i)}, \quad \alpha=1,\ldots, N,\, i=1,\ldots, M-1.
\ee
The moduli $Q^{(i)}$ and ${Q^{(i)}}'$ are related via the constraint
\be
\frac{Q^{(i)}}{{Q^{(i)}}'}=\prod_{\beta=1}^N\frac{\tilde{a}_\beta^{(i-1)}\tilde{a}_\beta^{(i+1)}}{\big(\tilde{a}_\beta^{(i)}\big)^2}
\ee
and the gauge couplings are given by the geometric average
\be
q^{(i)}=\sqrt{Q^{(i)}{Q^{(i)}}'},\quad i=1,\ldots, M-1.
\ee
The number of independent quantities is easy to count: we have $(M+1)N$ positions $\tilde{a}^{(i)}_{\alpha}$ subject to one condition setting the center of mass, as well as $M-1$ couplings $q^{(i)}$, leaving us with exactly $(M+1)(N+1)-3$ independent parameters.

In the  topological string computations it is convenient to use the parameters $\tilde{a}^{(i)}_{\alpha}$ and $Q_{B\alpha}^{(i)}$. However, in order to relate the theory to its dual, it is convenient to introduce new variables: the ``traceless'' positions
\be
\hat{a}_{\alpha}^{(i)}=\frac{\tilde{a}_{\alpha}^{(i)}}{\left(\prod_{\beta=1}^N\tilde{a}^{(i)}_{\beta}\right)^{\frac{1}{N}}},\quad (i=0,\ldots, M,\quad \alpha=1,\ldots, N,)
\ee
and the bifundamental masses
\be
\label{eq:defbifundamentalmasses}
\tilde{m}^{(i-1,i)}_{\text{bif}}=\left(\prod_{\beta=1}^N\frac{\tilde{a}_{\beta}^{(i)}}{\tilde{a}_{\beta}^{(i-1)}}\right)^{\frac{1}{N}}\quad (i=1,\ldots, M).
\ee
Of course, since there are only $M-2$ bifundamental hypermultiplets, $\tilde{m}^{(0,1)}_{\text{bif}}$ and $\tilde{m}^{(M-1,M)}_{\text{bif}}$ are not actual bifundamentals and were just introduced for convenience. We can of course obtain the $\tilde{a}^{(i)}_{\alpha}$ from the $\hat{a}_{\alpha}^{(i)}$ and $\tilde{m}^{(i-1,i)}_{\text{bif}}$, up to an unimportant global shift. For example, if we were to fix the global position by demanding that $\prod_{\alpha=1}^N\tilde{a}_\alpha^{(0)}=1$, then we get
\be
\tilde{a}_\alpha^{(i)}=\hat{a}_\alpha^{(i)}\prod_{j=1}^i\tilde{m}_{\text{bif}}^{(j-1,j)}.
\ee

For the sake of completion, we remind that for the leftmost brane positions we have the identification $\tilde{a}^{(0)}_\alpha=\tilde{m}_\alpha$, while for the rightmost $\tilde{a}^{(M)}_\alpha=\tilde{m}_{N+\alpha}$. The actual exponentiated fundamental masses are obtained by dividing our by the center of mass of the $D5$ branes connected to the nearest $NS5$ brane, i.e.
\be
\label{eq:deffundamentalmasses}
\tilde{m}_\alpha^{\text{f}}=\frac{\tilde{a}_{\alpha}^{(0)}}{\left(\prod_{\beta=1}^N\tilde{a}_{\beta}^{(1)}\right)^{\frac{1}{N}}},\qquad \tilde{m}_{N+\alpha}^{\text{f}}=\frac{\tilde{a}_{\alpha}^{(M)}}{\left(\prod_{\beta=1}^N\tilde{a}_{\beta}^{(M-1)}\right)^{\frac{1}{N}}},
\ee
for $\alpha=1,\ldots, N$.

In (3.42) of \cite{Bao:2011rc},
we derived the following map\footnote{We remark that due to a difference in the convention in which distances are measured between the current article and \cite{Bao:2011rc}, we need to apply the transformation $\tilde{a}^{(i)}_{\alpha} \to \left( \tilde{a}^{(i)}_{\alpha} \right)^{-1}$ and
$ (\tilde{a}^{(\alpha)}_{i} )_d \to \left( \tilde{a}^{(\alpha)}_{i} \right)_d^{-1}$ to the parameters in order to get equation \eqref{eq:duality-map} from the one in \cite{Bao:2011rc}.} between the parameters of the original $\SU(N)^{M-1}$ theory and those of its dual $\SU(M)^{N-1}$ theory:
\bea
\left( \hat{a}_{i}^{(\alpha)} \right)_d
&=& \left( \tilde{m}_{\text{bif}}^{(i-1,i)} \right)^{\alpha-\frac{N}{2}}
\prod_{\gamma=1}^\alpha
\left(
\frac{\hat{a}_{\gamma}^{(i)} }{\hat{a}_{\gamma}^{(i-1)} } \right)
\left(
\frac{\hat{a}_\gamma^{(0)}}{\hat{a}_\gamma^{(M)}}
\right) ^{\frac{1}{M}}
 \prod_{k=1}^{M}
\left( \tilde{m}_{\text{bif}}^{(k-1,k)} \right)^{\frac{N-2\alpha}{2M}}
\nonumber\\&&\times\prod_{\ell=1}^{i-1} \left( q^{(\ell)} \right)^{-\frac{\ell}{M}}
\prod_{\ell=i}^{M-1} \left( q^{(\ell)} \right) ^{\frac{M-\ell}{M}}  \, , \quad
(i=1,\cdots,M, \quad  \alpha = 0, \cdots, N), \nonumber
\\
\left( \tilde{m}^{(\alpha-1,\alpha)}_{\text{bif}} \right)_d
&=& \left(
\frac{\hat{a}_\alpha^{(M)}}{\hat{a}_\alpha^{(0)}}
\prod_{k=1}^{M} \tilde{m}_{\text{bif}}^{(k-1,k)}
\right)^{\frac{1}{M}} \, , \qquad (\alpha=1,\cdots,N)
\label{eq:duality-map}
\\
\left( q^{(\alpha)} \right)_d
&=& \left( \frac{\hat{a}^{(0)}_{\alpha} \hat{a}^{(M)}_{\alpha}}
{\hat{a}^{(0)}_{\alpha+1} \hat{a}^{(M)}_{\alpha+1}}
 \right)^{1/2} \, , \quad
\qquad (\alpha=1,\cdots,N-1). \nonumber
\eea
The label $d$ at the left hand side of the equation indicates that they are the variables of the dual theory.

\subsection{Invariant Coulomb moduli: a special case }
\label{subsec:specialcase}

We now want to derive the set of parameters which make the enhanced symmetry apparent.
We first consider the special case $M=2$, for which we have the original $\SU(N)$, $N_f=2N$ theory its dual theory with gauge group $\SU(2)^{N-1}$, $N_f=2+2$ and $M-2$ bifundamental hypermultiplets.

On one hand, in the original $\SU(N)$ theory,
the manifest global symmetry is $\U(1) \times \SU(2N) \times \U(1)$.
Corresponding fugacities are
the trace part of the fundamental masses,
\begin{align}
\tilde{M}
=
 \prod_{\alpha=1}^{2N} \tilde{m}^{\text{f}}_{\alpha}
=
\prod_{\beta=1}^N
\frac{\tilde{a}_{\beta}^{(0)}\tilde{a}_{\beta}^{(2)}}{\big(\tilde{a}_{\beta}^{(1)}\big)^2}
=
\left(
\frac{\tilde{m}_{\text{bif}}^{(1,2)}}{\tilde{m}_{\text{bif}}^{(0,1)}}
\right)^{N},
\end{align}
where we used \eqref{eq:defbifundamentalmasses},
the traceless part of the fundamental masses
\be
\label{eq:deftracefund}
\tilde{M}_{\alpha}
= \tilde{m}_{\alpha}^{\text{f}} \tilde{M}^{-\frac{1}{2N}}=\left\{\begin{array}{ll}\frac{\tilde{a}_{\alpha}^{(0)}}{\prod_{\beta=1}^N\big(\tilde{a}_{\alpha}^{(0)}\tilde{a}_{\alpha}^{(2)}\big)^{\frac{1}{2N}}} & \text{ for }\alpha=1,\ldots, N\\\frac{\tilde{a}_{\alpha-N}^{(2)}}{\prod_{\beta=1}^N\big(\tilde{a}_{\alpha}^{(0)}\tilde{a}_{\alpha}^{(2)}\big)^{\frac{1}{2N}}} & \text{ for }\alpha=N+1,\ldots, 2N\end{array}\right.,
\ee
and the instanton fugacity $q^{(1)}$, respectively.
On the other hand, in the dual $\SU(2)^{N-1}$ theory,
the manifest global symmetry is $\SU(2) \times \SU(2) \times \U(1)^{N} \times \U(1)^{N-1}$.
Corresponding fugacities are
the traceless part of the fundamental masses
$( \hat{a}^{(0)}_1 )_d$, $( \hat{a}^{(N)}_1 )_d$,
the bifundamental masses $(\tilde{m}_{\text{bif}}^{(\alpha-1,\alpha)})_d$ $(\alpha=1,\cdots, N)$,
which also include the trace part of fundamental masses
as we mentioned at \eqref{eq:defbifundamentalmasses},
and the instanton fugacity $(q^{(\alpha)})_d$ $(\alpha=1,\cdots, N-1)$.
However, as we discussed in section \ref{sec:7brane},
the expected global symmetry in the case $M=2$, $N\ge 3$ is given by
$\SU(2N) \times \SU(2) \times \SU(2)$.
We will see that the duality map helps us to understand this global symmetry enhancement.

In the following, we use the duality map \eqref{eq:duality-map} with $M=2$ is substituted.
First, using \eqref{eq:deftracefund}, we find that that the first line of \eqref{eq:duality-map} for $\alpha=0$ and  $N$ can be written as
\be
\left( \hat{a}^{(0)}_1 \right)_d
= \tilde{M}^{\frac{1}{4}} \left(q^{(1)}\right) ^{\frac{1}{2}}
= \left( Q^{(1)} \right)^{\frac{1}{2}}
\qquad   \left( \hat{a}^{(N)}_1 \right)_d
= \tilde{M}^{-\frac{1}{4}} \left(q^{(1)}\right)^{\frac{1}{2}}
= \left( Q^{(1)}{}'\right)^{\frac{1}{2}} .
\ee
This indicates that the $\U(1) \times \U(1)$ fugacities in the original theory
maps to the $\SU(2) \times \SU(2)$ fugacities in the dual theory.
Thus, from the view point of the original theory,
$\U(1) \times \U(1)$ global symmetry is enhanced to $\SU(2) \times \SU(2)$.

The second two lines of \eqref{eq:duality-map}
give us the relation between the $\SU(2N)$ fugacities of the original $\SU(N)$
theory and $\U(1)^N \times \U(1)^{N-1}$ fugacities of the dual $\SU(2)^{N-1}$
theory if we rewrite them as follows:
\be
\begin{split}
&\left( \tilde{m}^{(\alpha-1,\alpha)}_{\text{bif}} \right)_d
= \left(
\frac{\tilde{M}_{N+\alpha}}{\tilde{M}_{\alpha}}
\right)^{\frac{1}{2}},
\qquad (\alpha = 1 ,\cdots N)
\\
& \left( q^{(\alpha)} \right)_d
=  \left( \frac{\tilde{M}_{\alpha} \tilde{M}_{N+\alpha}}
{\tilde{M}_{\alpha+1} \tilde{M}_{N+\alpha+1}}
 \right)^{1/2} \, , \qquad (\alpha = 1 ,\cdots N-1)
\end{split}
\ee
Thus, from the view point of the dual theory,
the $\U(1)^N \times \U(1)^{N-1}$ symmetry is enhanced to $\SU(2N)$.

Combining the result above, from the point of view of either theory,
we find that the global symmetry is enhanced to $\SU(2N) \times \SU(2) \times \SU(2)$
as expected. It would be convenient to use
$\tilde{M}_{\alpha}$, $Q^{(1)}$ and $Q^{(1)}{}'$ as the corresponding fugacities.

Finally, we derive the invariant Coulomb moduli parameter
using the maps corresponding to the remaining moduli parameters.
From the the first line of \eqref{eq:duality-map} for $1 \le \alpha \le N-1$,
it is straightforward to derive the following set of equations:
\begin{align}
\frac{( \tilde{M}_2 \tilde{M}_{N+2}) ^{\frac{1}{2}} }
{( \tilde{M}_1 \tilde{M}_{N+1}) ^{\frac{1}{2}} }
\frac{\left( \hat{a}^{(1)}_1 \right)_d{}^2}
{\left( \hat{a}^{(2)}_1 \right)_d}
&= \left( Q^{(1)} \right)^{\frac{1}{2}}
\frac{\hat{a}_1^{(1)}}{\hat{a}_2^{(1)}},
\cr
 \frac{( \tilde{M}_{\alpha+1} \tilde{M}_{N+\alpha+1}) ^{\frac{1}{2}}}
{( \tilde{M}_\alpha \tilde{M}_{N+\alpha}) ^{\frac{1}{2}}}
\frac{\left( \hat{a}^{(\alpha)}_1 \right)_d{}^2}
{\left( \hat{a}^{(\alpha-1)}_1 \right)_d
  \left( \hat{a}^{(\alpha+1)}_1 \right)_d}
&= \frac{\hat{a}_{\alpha}^{(1)} }{\hat{a}_{\alpha+1}^{(1)} } \, ,
\qquad (\alpha = 2,\cdots, N-2)
\cr
\frac{( \tilde{M}_{N-1} \tilde{M}_{2N-1}) ^{\frac{1}{2}}}
{( \tilde{M}_{N} \tilde{M}_{2N}) ^{\frac{1}{2}}}
\frac{\left( \hat{a}^{(N-1)} \right)_d{}^2}
{\left( \hat{a}^{(N-2)} \right)_d}
&= \left( Q^{(1)}{}'\right)^{\frac{1}{2}}
\frac{\hat{a}_{N-1}^{(1)}}{\hat{a}_{N}^{(1)}}
 \, ,
\label{key_eq}
\end{align}
Since $\SU(2N)$ part is identified as part of the manifest global symmetry of the original $\SU(N)$ theory,
it is natural to assume that the Coulomb moduli parameter
$\hat{a}_{\alpha}^{(1)}$ should be invariant under the $\SU(2N)$ part.
Analogously, since $\SU(2) \times \SU(2)$ part is identified as the manifest global symmetry
of the dual $\SU(2)^{N-1}$ theory, the Coulomb moduli parameter $(\tilde{a}_1^{(\alpha)})_d$
of the dual theory should be invariant under the $\SU(2) \times \SU(2)$ part.
Under these assumption, we see that
the right hand side of \eqref{key_eq} is invariant under the $\SU(2N)$
while the left hand side of \eqref{key_eq} is invariant under the $\SU(2) \times \SU(2)$.
Since they are equal, these combinations are invariant under the
full enhanced symmetry $\SU(2N) \times \SU(2) \times \SU(2)$.
Thus, we can identify them as the invariant Coulomb moduli parameters.

In summary, our conclusion is that
we should define the invariant Coulomb moduli parameters for the $\SU(N)$ gauge theory with $N_f=2N$ as
\begin{align}
\label{eq:definvCoulmoduli1}
\tilde{A}_1
\equiv
\frac{\hat{a}_1^{(1)}}{\hat{a}_2^{(1)}} \left( Q^{(1)} \right)^{\frac{1}{2}}\,,\qquad
\tilde{A}_{\alpha}
\equiv \frac{\hat{a}_{\alpha}^{(1)}}{\hat{a}_{\alpha+1}^{(1)}}\, ,\qquad
\tilde{A}_{N-1}
\equiv \frac{\hat{a}_{N-1}^{(1)}}{\hat{a}_{N}^{(1)}} \left( Q^{(1)}{}' \right)^{\frac{1}{2}}\,,
\end{align}
where $ \alpha = 2,\cdots, N-2.$

\subsection{The general case}
%
Since the generalization of what we discussed in
section \ref{subsec:specialcase} to the general linear quiver is straightforward,
we briefly summarize the result here.
The manifest global symmetry for $\SU(N)^{M-1}$ theory is
$\SU(N) \times \SU(N) \times \U(1)^{M} \times \U(1)^{M-1}$
but it enhances to $\SU(N) \times \SU(N) \times \SU(M) \times \SU(M) \times \U(1)$.
The fugacities of $\SU(N) \times \SU(N)$ part is
$\hat{a}_{\alpha}^{(0)}$ and $\hat{a}_{\alpha}^{(M)}$.
The fugacities of $\SU(M) \times \SU(M)$ part is
$(\hat{a}_{i}^{(0)})_d$ and $(\hat{a}_{i}^{(N)})_d$, which satisfies
\begin{eqnarray}
\frac{(\hat{a}_{i}^{(0)})_d}{(\hat{a}_{i+1}^{(0)})_d} = Q^{(i)},
\qquad
\frac{(\hat{a}_{i}^{(N)})_d}{(\hat{a}_{i+1}^{(N)})_d} = Q^{(i)}{}'\, .
\end{eqnarray}
The remaining $\U(1)$ fugacity is given by
$\prod_{k=1}^M \tilde{m}_{\text{bif}}^{(k-1,k)} .$

We can explicitly check
that the following set of Coulomb moduli parameters
are invariant under the enhanced global symmetry
\be
\tilde{A}^{(i)}_{1}=\frac{\hat{a}_1^{(i)}}{\hat{a}_2^{(i)}}
\prod_{j=1}^i \left( \hat{a}_j^{(0)} \right)_d\,,\qquad \tilde{A}^{(i)}_{\alpha}=\frac{\hat{a}_{\alpha}^{(i)}}{\hat{a}_{\alpha+1}^{(i)}}\,, \qquad \tilde{A}^{(i)}_{N-1}=\frac{\hat{a}_{N-1}^{(i)}}{\hat{a}_{N}^{(i)}}
\prod_{j=1}^i \left( \hat{a}_j^{(M)} \right)_d\,,
\ee
where as in \eqref{eq:definvCoulmoduli1} the index $\alpha$ runs over $2,\ldots, N-2$.

\section{Partition functions for higher gauge groups}
\label{sec:flopping}

The purpose of this section is to construct the partition functions for the general linear quiver theories $\SU(N)^{M-1}$. Furthermore, we will demonstrate for the case of $\SU(3)$ with six flavors that the expansion in the invariant Coulomb moduli makes the enhanced symmetry $\SU(6)\times \SU(2)^2$, whose presence we derived in section \ref{sec:7brane} using D7 branes, manifest.

\subsection{The strip decomposition for linear quivers}
\label{subsec:newstrip}

\begin{figure}[h]
\begin{center}
\includegraphics[height=6cm]{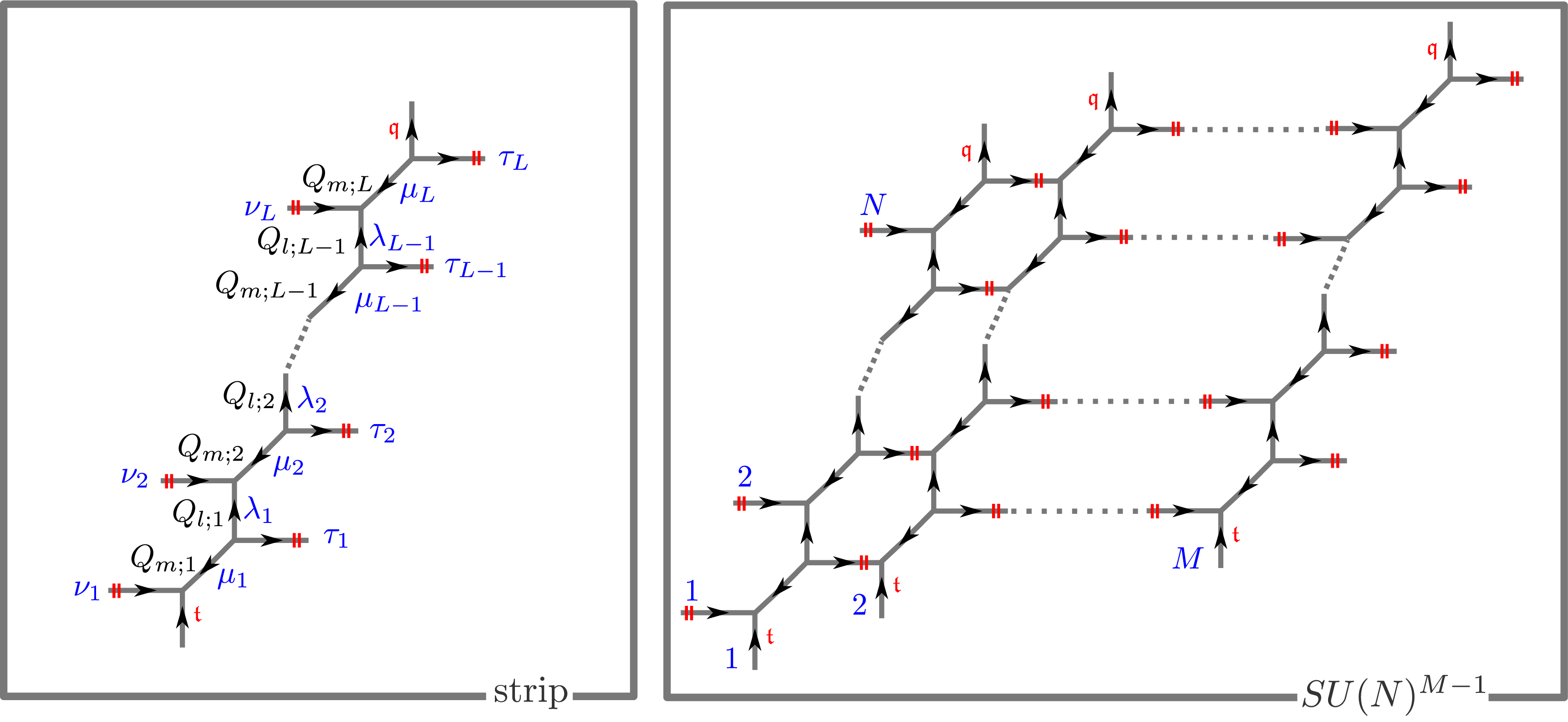}
\end{center}
\caption{This figure illustrates the strip with the same number of left and right flavor branes and shows on the right how an $\SU(N)^{M-1}$ linear quiver theory would be constructed from it. }
\label{fig:newstrip}
\end{figure}

In \cite{Bao:2013pwa}, we computed the topological string amplitude of the $T_N$ junction by cutting the web diagram in $N$ strips. We can perform a similar procedure for the quiver theories, but the strips are different from the ones in \cite{Bao:2013pwa} due to the fact that the number of flavor branes on the left and on the right is the same. The geometry is depicted in figure~\ref{fig:newstrip}.
Setting $\lambda_0=\lambda_L=\emptyset$, we can resum the expression for the topological string partition function as
\bea
\label{eq:newstrip}
Z^{\text{strip}}_{\boldsymbol{\nu}\boldsymbol{\tau}}(\boldsymbol{Q}_m,\boldsymbol{Q}_l,\ft,\fq)&=&\prod_{i=1}^L\big(-Q_{m;i}\big)^{|\mu_i|}\prod_{i=1}^{L-1}\big(-Q_{l;i}\big)^{|\lambda_i|}\prod_{i=1}^LC_{\mu_i^t\lambda_{i-1}^t\nu_i^t}(\fq,\ft)C_{\mu_i\lambda_i\tau_i}(\ft,\fq)\nonumber\\&=&\prod_{i=1}^L\fq^{\frac{||\tau_i||^2}{2}}\ft^{\frac{||\nu_i^t||^2}{2}}\tilde{Z}_{\tau_i}(\ft,\fq)\tilde{Z}_{\nu_i^t}(\fq,\ft)\prod_{i\leq j=1}^L\calR_{\nu_i^t\tau_j}\big(\prod_{k=i}^jQ_{m;k}\prod_{k=i}^{j-1}Q_{l;k}\big)\nonumber\\
&&\times \prod_{i< j=1}^L\calR_{\tau_i^t\nu_j}\big(\prod_{k=i+1}^{j-1}Q_{m;k}\prod_{k=i}^{j-1}Q_{l;k}\big)\nonumber\\&&\times \left(\prod_{i< j=1}^L\calR_{\nu_i^t\nu_j}\big(\sqrt{\frac{\fq}{\ft}}\prod_{k=i}^{j-1}Q_{m;k}Q_{l;k}\big)\calR_{\tau_i^t\tau_j}\big(\sqrt{\frac{\ft}{\fq}}\prod_{k=i}^{j-1}Q_{m;k+1}Q_{l;k}\big)\right)^{-1}.
\eea
By using \eqref{eq:newstrip}, we can write the partition function of the  $\SU(N)^{M-1}$ linear quiver as
\be
Z'=\prod_{r=1}^M(-\boldsymbol{Q}_B^{(r)})^{|\boldsymbol{\nu}^{(r)}|}Z^{\text{strip}}_{\boldsymbol{\nu}^{(r-1)}\boldsymbol{\nu}^{(r)}}(\boldsymbol{Q}_m^{(r)},\boldsymbol{Q}_l^{(r)},\ft,\fq),
\ee
where $\nu_i^{(0)}=\nu_i^{(M)}=\emptyset$, the $\boldsymbol{Q}_B^{(r)}=(Q_{B1}^{(r)},\ldots, Q_{BN}^{(r)})$ are defined in \eqref{eq:defQB} and
\be
\boldsymbol{Q}_m^{(r)}=\left(\frac{\tilde{a}_1^{(r-1)}}{\tilde{a}_1^{(r)}},\ldots, \frac{\tilde{a}_N^{(r-1)}}{\tilde{a}_N^{(r)}} \right),\qquad \boldsymbol{Q}_l^{(r)}=\left(\frac{\tilde{a}_1^{(r)}}{\tilde{a}_2^{(r-1)}},\ldots, \frac{\tilde{a}_{N-1}^{(r)}}{\tilde{a}_N^{(r-1)}} \right),
\ee
with the $\tilde{a}_{\alpha}^{(r)}$ read off from figure~\ref{fig:parametrizationSUNM}. To obtain the proper partition function, we then need to divide by the decoupled non-full spin content.

\subsection{Flopping for  $\SU(3)$}
\label{subsec:floppedsu3}

\begin{figure}[h]
\begin{center}
\includegraphics[height=4cm]{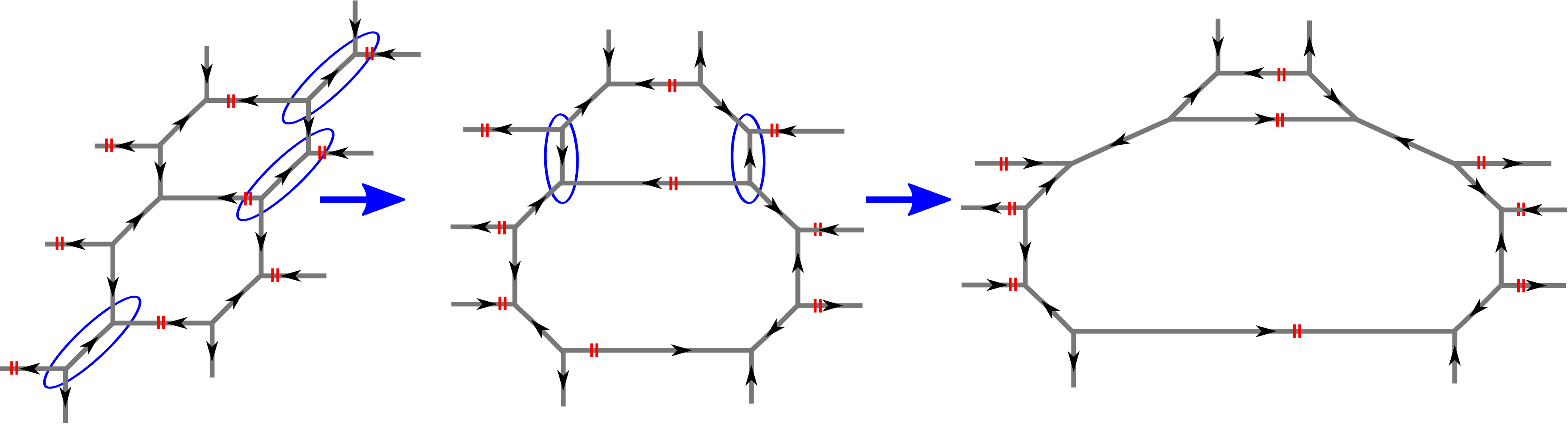}
\end{center}
\caption{This figure illustrates the flopping procedure for  $\SU(3)$ with six flavors. The operations do not change the distances between the external parallel branes.}
\label{fig:su3andflopping}
\end{figure}

\begin{figure}[h]
\begin{center}
\includegraphics[height=5.5cm]{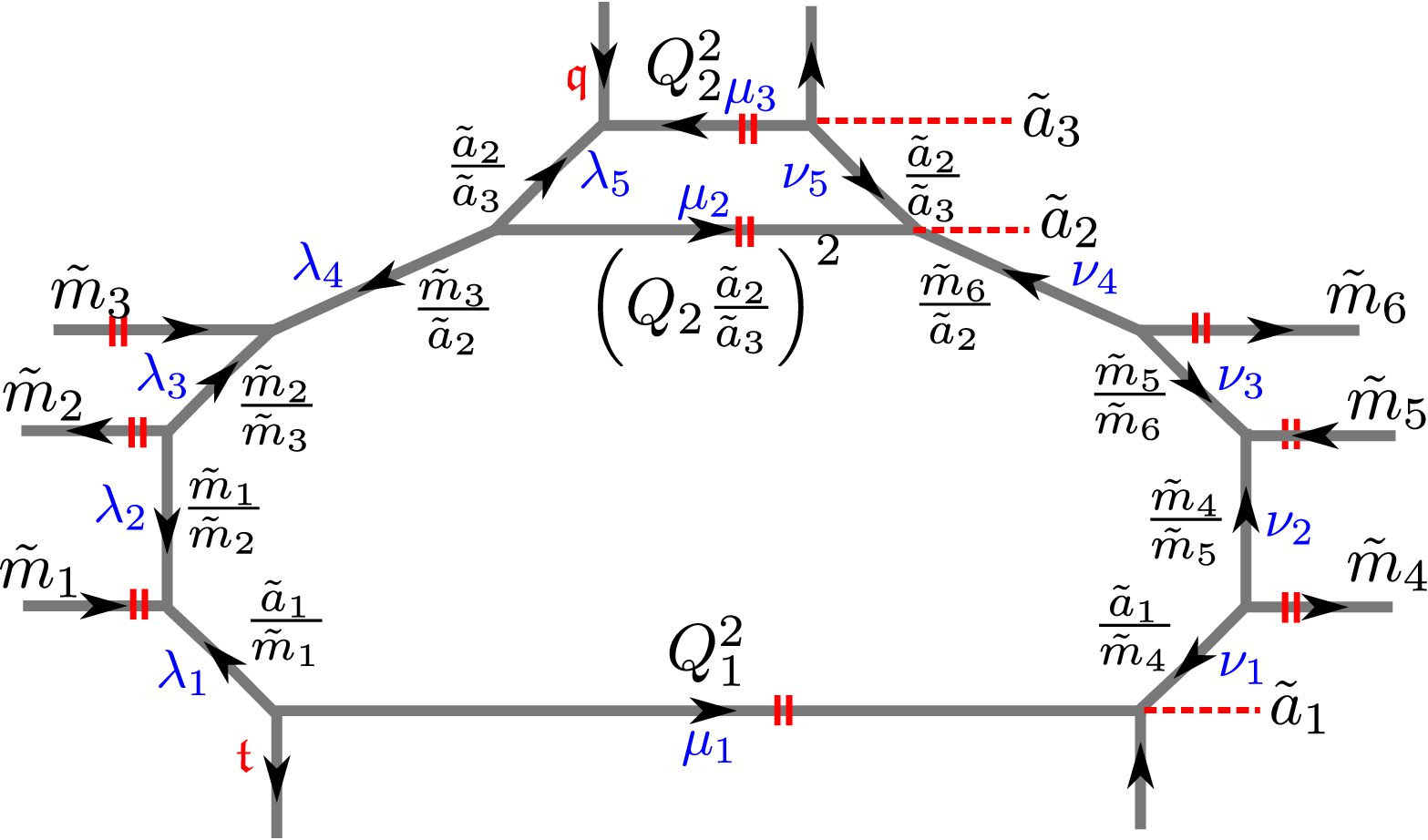}
\end{center}
\caption{This figure illustrates $\SU(3)$ with six flavor, flopped in such a way as to make the global symmetry apparent. The names of the partitions used in the computation of the partition function are highlighted in blue.}
\label{fig:su3flavor}
\end{figure}

In order to be able to see the symmetry on the level of the partition function, we need to expand in positive powers of the invariant Coulomb moduli. In practice, it requires us to flop the diagrams in an appropriate way and we illustrate this in the case of $\SU(3)$ with six flavors. Figure~\ref{fig:su3andflopping} shows how to get to the flopped diagram for  $\SU(3)$ with six flavors, starting from the standard one of figure~\ref{fig:parametrizationSUNM}.
The partition function from the standard diagram is given by a sum over the product of two strips \eqref{eq:newstrip} and equals\footnote{Compared  to section \ref{sec:symmetryenhancement}, here we use the notation $Q_1\equiv Q^{(1)}$ and $Q_2\equiv Q^{(1)'}$.}
\bea
{Z^{N_f=6}_{\SU(3)}}'&=&\sum_{\boldsymbol{\nu}}\left(-Q_1^2\frac{\tilde{a}_1}{\tilde{m}_1}\right)^{|\nu_1|}\left(-Q_2^2\frac{\tilde{m}_3\tilde{m}_5\tilde{m}_6}{\tilde{a}_2\tilde{a}_3^2}\right)^{|\nu_2|}\left(-Q_2^2\frac{\tilde{m}_6}{\tilde{a}_3}\right)^{|\nu_3|}\nonumber\\&&\times Z^{\text{strip}}_{\{\emptyset\emptyset\emptyset\}\{\nu_1\nu_2\nu_3\}}\left(\left\{\frac{\tilde{m}_1}{\tilde{a}_1},\frac{\tilde{m}_2}{\tilde{a}_2},\frac{\tilde{m}_3}{\tilde{a}_3}\right\},\left\{\frac{\tilde{a}_1}{\tilde{m}_2},\frac{\tilde{a}_2}{\tilde{m}_3}\right\},\ft,\fq\right)\nonumber\\&&\times Z^{\text{strip}}_{\{\nu_1\nu_2\nu_3\}\{\emptyset\emptyset\emptyset\}}\left(\left\{\frac{\tilde{a}_1}{\tilde{m}_4},\frac{\tilde{a}_2}{\tilde{m}_5},\frac{\tilde{a}_3}{\tilde{m}_6}\right\},\left\{\frac{\tilde{m}_4}{\tilde{a}_2},\frac{\tilde{m}_5}{\tilde{a}_3}\right\},\ft,\fq\right).
\eea
The flopping is illustrated in figure~\ref{fig:su3andflopping}. We can use (dropping phases) the flopping equation \eqref{eq:Sflopping} to flop the parts
\be
\calS_{\emptyset\nu_1}\big(\frac{\tilde{m}_1}{\tilde{a}_1}\big),\quad \calS_{\nu_2\emptyset}\big(\frac{\tilde{a}_2}{\tilde{m}_5}\big),\quad \calS_{\nu_3\emptyset}\big(\frac{\tilde{a}_3}{\tilde{m}_6}\big),\quad \calS_{\nu_2\emptyset}\big(\frac{\tilde{a}_2}{\tilde{m}_3}\big),\quad \calS_{\nu_2\emptyset}\big(\frac{\tilde{a}_2}{\tilde{m}_6}\big).
\ee
We thus get into the frame, depicted in figure~\ref{fig:su3flavor}, that tell us that the full flavor symmetry should contain $\SU(6)\times \SU(2)\times \SU(2)$.
After removing the non full-spin content the partition function becomes
\bea
Z_{\SU(3)}^{N_f=6}&=&\frac{\prod_{i<j=1}^3\calM\left(\frac{\tilde{a}_i}{\tilde{a}_j}\right)\calM\left(\frac{\ft}{\fq}\frac{\tilde{a}_i}{\tilde{a}_j}\right)}{\calM(Q_2^2)\calM\left(Q_1^2\frac{\ft}{\fq}\right)\prod_{k=1}^6\calM\left(\sqrt{\frac{\ft}{\fq}}\frac{\tilde{a}_1}{\tilde{m}_k}\right)\calM\left(\sqrt{\frac{\ft}{\fq}}\frac{\tilde{m}_k}{\tilde{a}_2}\right)\calM\left(\sqrt{\frac{\ft}{\fq}}\frac{\tilde{m}_k}{\tilde{a}_3}\right)}\nonumber\\&&\times \sum_{\boldsymbol{\nu}}Q_1^{2|\nu_1|}\left(Q_2^2\frac{\tilde{a}_2^2}{\tilde{a}_3^2}\right)^{|\nu_2|}Q_2^{2|\nu_3|}\ft^{||\nu_1^t||^2-||\nu_2^t||^2}\fq^{2||\nu_2||^2+||\nu_3||^2}\prod_{i=1}^3\tilde{Z}_{\nu_i}(\ft,\fq)\tilde{Z}_{\nu_i^t}(\fq,\ft)\nonumber\\&&\times \frac{\prod_{k=1}^6\calN_{\nu_1\emptyset}\left(\sqrt{\frac{\ft}{\fq}}\frac{\tilde{a}_1}{\tilde{m}_k}\right)\calN_{\emptyset\nu_2}\left(\sqrt{\frac{\ft}{\fq}}\frac{\tilde{m}_k}{\tilde{a}_2}\right)\calN_{\emptyset\nu_3}\left(\sqrt{\frac{\ft}{\fq}}\frac{\tilde{m}_k}{\tilde{a}_3}\right)}{\prod_{i<j=1}^3\calN_{\nu_i\nu_j}\left(\frac{\tilde{a}_i}{\tilde{a}_j}\right)\calN_{\nu_i\nu_j}\left(\frac{\ft}{\fq}\frac{\tilde{a}_i}{\tilde{a}_j}\right)}.
\eea
We now want to make the flavor symmetry in the partition function apparent. First, using translation invariance, we get the parameters $\tilde{a}_i$ to obey the condition $\tilde{a}_1\tilde{a}_2\tilde{a}_3=1$. Furthermore, we see that
\be
\frac{Q_1^2}{Q_2^2}=\prod_{i=1}^6\tilde{m_i}.
\ee
We introduce new mass parameters subject to the relation $\tilde{M}_i$ with $\prod_{i=1}^6\tilde{M}_i=1$ via the equation
\be
\tilde{M}_i=\tilde{m}_i\left(\prod_{i=1}^6\tilde{m}_i\right)^{-\frac{1}{6}}=\left(\frac{Q_2}{Q_1}\right)^{\frac{1}{3}}\tilde{m}_i.
\ee
They should make the $\SU(6)$ part of the flavor symmetry manifest. The $\SU(2)\times \SU(2)$ part is contained in the $Q_i$ parameters.  Furthermore, as we shall see in \eqref{eq:definvCoulmoduli1}, we need to define the invariant Coulomb moduli as $\tilde{A}_1=Q_1\frac{\tilde{a}_1}{\tilde{a}_2}$ and $\tilde{A}_2=Q_2\frac{\tilde{a}_2}{\tilde{a}_3}$. We can also write the inverse relation
\be
\tilde{a}_1=\left(\frac{\tilde{A}_1}{Q_1}\right)^{\frac{2}{3}}\left(\frac{\tilde{A}_2}{Q_2}\right)^{\frac{1}{3}},\qquad \tilde{a}_2=\left(\frac{\tilde{A}_1}{Q_1}\right)^{-\frac{1}{3}}\left(\frac{\tilde{A}_2}{Q_2}\right)^{\frac{1}{3}},\qquad \tilde{a}_3=\left(\frac{\tilde{A}_1}{Q_1}\right)^{-\frac{1}{3}}\left(\frac{\tilde{A}_2}{Q_2}\right)^{-\frac{2}{3}}.
\ee
Expanding to first order in $\tilde{A}_i$ and second order in $\ft$ and $\fq$, we get
\bea
\label{eq:expansionSU3sixflavors}
Z_{\SU(3)}^{N_f=6}&=&1+\frac{(\fq+\ft)\chi_{2}^{\SU(2)}(Q_2)}{(1-\fq)(1-\ft)}\tilde{A}_2+\tilde{A}_1\Big[\chi_{2}^{\SU(2)}(Q_1)\Big(\ft+\fq+3\ft\fq+4\ft^2\fq+4\ft\fq^2+6\ft^2\fq^2\nonumber\\&&+\chi^{\SU(6)}_{35}(\ft\fq+2\ft^2\fq+2\ft\fq^2+4\ft^2\fq^2)\Big)-\ft^{\frac{3}{2}}\fq^{\frac{3}{2}}\chi_{2}^{\SU(2)}(Q_2)\big(\chi^{\SU(6)}_{70}+\chi^{\SU(6)}_{20}\big)+\cdots \Big]\nonumber\\&&
+\tilde{A}_1\tilde{A}_2\Big[\chi_{2}^{\SU(2)}(Q_1)\chi_{2}^{\SU(2)}(Q_2)\big(\ft+\fq+5\ft\fq+11\ft^2\fq+11\ft\fq^2+26\ft^2\fq^2\nonumber\\&&+\chi^{\SU(6)}_{35}\big(\ft\fq+3\ft^2\fq+3\ft\fq^2+10\ft^2\fq^2\big))\big)
-\sqrt{\ft\fq}\big(\chi_{20}^{\SU(6)}+(\ft+\fq)\big(\chi_{70}^{\SU(6)}+\chi_{20}^{\SU(6)}\big)\nonumber\\&&+\ft\fq\big(\chi_{56}^{\SU(6)}+(\chi_{70}^{\SU(6)}+\chi_{20}^{\SU(6)})(6+\chi_{3}^{\SU(2)}(Q_2)\big)\big)+\cdots \Big]+\cdots,
\eea
where the characters of $\SU(6)$ are labeled by their dimensions.  While we did not succeed in obtaining the exact $\ft$, $\fq$ dependence of the coefficients in the $\tilde{A}_i$ expansion, \eqref{eq:expansionSU3sixflavors} still demonstrates the appearance of the enhanced symmetry $\SU(6)\times \SU(2)^2$ at the level of the partition function.

\section{Conclusions}
\label{sec:conclusions}

In this paper, we have discussed how the fiber-base duality of some Calabi Yau geometries affects the global symmetry enhancement of the corresponding 5D $\mathcal{N}=1$ gauge theory. 
We have clarified how the masses, the gauge coupling, and the Coulomb moduli parameters
are mapped to each other by the fiber-base duality
by using the idea discussed in \cite{Bao:2013wqa}.
For the case of the $\SU(2)$ gauge theory with $0\leq N_f\leq 7$ flavor,
the fiber-base duality maps the theory onto itself and thus becomes a symmetry. We showed in section \ref{sec:fiberbaseinvariance} how the duality map, combined with
the manifest $\SO(2N_f)$ flavor symmetry, generates the expected $\E_{N_f+1}$ Weyl symmetry. Furthermore, we saw that, while the original Coulomb moduli parameter transforms non-trivially under the fiber-base duality map, we can obtain an invariant one by multiplying with an appropriate power of the gauge coupling constant \eqref{eq:generaldeftildeA}, making the new Coulomb moduli parameter invariant under the whole $\E_{N_f+1}$ enhanced global symmetry.
In addition, in section \ref{subsec:effectivecoupling}, we found that the invariant Coulomb modulus turns out to be proportional to the effective coupling constant of the flat 5D theory.
By expanding the corresponding topological string partition function,
or Nekrasov partition function, in terms of the invariant Coulomb modulus,
we found that it can be written in terms of $\E_{N_f+1}$ characters, thus indicating the invariance of the partition function under the Weyl symmetry of the enhanced $\E_{N_f+1}$ symmetry. This provides a further check of the global symmetry enhancement of the 5D $\calN=1$ supersymmetric gauge theory.

For linear quiver gauge theories of higher rank,
the fiber-base duality implies that the $\SU(N)^{M-1}$ gauge theory
is equivalent to the $\SU(M)^{N-1}$ one. Thus, unlike in the $\SU(2)$ case, the duality cannot be interpreted as a symmetry of one specific theory. However, by carefully observing the manifest flavor symmetries of the two dual theories, we can derive the enhanced flavor symmetry $\SU(N)^2 \times \SU(M)^2\times \U(1)$. For the special case  of the fiber-base duality between $\SU(N)$ with $2N$ flavors and the linear superconformal quiver $\SU(2)^{N-1}$, we show that the enhanced flavor symmetry is $\SU(2N) \times \SU(2)^2$.
In section \ref{sec:symmetryenhancement}, we introduced the set of invariant Coulomb moduli parameters in such a way as to make them invariant under the enhanced global symmetry. Finally,  in section \ref{sec:flopping}, we showed that the topological string partition function, for the case $N=3$ and $M=2$ is explicitly invariant under the Weyl symmetry of the enhanced global symmetry if expanded in the invariant Coulomb moduli.

Although we performed the expansion of the partition functions only for the  $\SU(2)$ theories with $N_f$ flavor as well as for $\SU(3)$ with six flavors, we believe that our ideas have a much wider range of applications.
For a generic toric diagram, or even for a generalized toric diagram
as introduced in \cite{Benini:2009gi}, the expected global symmetry can be systematically analyzed by using 7-branes monodromy.
Then, we propose that we can always define appropriate Coulomb moduli parameters that are invariant under the expected global symmetry so that the corresponding topological string partition function can be expanded in a power series in them with coefficients given in terms of the characters of the global symmetry. Further checks of this proposal are left for future work.

In addition, there is a multitude of other possible directions for future works.
One of them is the study of the relation between our $\E_{N_f+1}$ manifest expansion of the Nekrasov partition function and
the elliptic genus of the E-string or Nekrasov-type partition function that was
recently studied for instance in \cite{Haghighat:2014pva,Sakai:2011xg, Sakai:2012ik, Ishii:2013nba, Sakai:2014hsa,toappear}.
Taking into account that the elliptic genus of the E-string
is also written in  an $\E_8$ manifest way, it is natural to expect that
proper dimensional reduction and/or mass decoupling limit will
directly reproduce our $\E_{N_f+1}$ results.

Since we have shown that the fiber-base duality,
which translates to S-duality in the brane setup picture,
is part of the enhanced $\E_{N_f+1}$ symmetry, it would be also interesting to consider whether there is any relation between this observation and \cite{Iqbal:2001ye} in which the duality between
del-Pezzo surfaces and U-dualities was discussed.
Our result may also be related to the theta function arising 
in the BPS spectrum of M5-branes wrapped on del-Pezzo surfaces \cite{Bonelli:2001jf}.

Recently, an interesting observation was made\footnote{In this paper
the equivalence between the $T_N$ theory and linear quiver theory plays a key role \cite{Hayashi:2014hfa}.} by \cite{Bergman:2014kza}
that the fiber-base duality can be essentially seen as the 5D uplift of the 4D ``$\mathcal{N}=2$ dualities'' of \cite{Gaiotto:2009we}. However,  there are several subtleties that should be studied further. In \cite{Bergman:2014kza},  using the superconformal index the authors showed that $\SU(3)$ with 6 flavors and with Chern-Simons level 1 is dual to $\SU(2)\times \SU(2)$ with 3 flavors coupled to the first color group and 1 flavor to the second, denoted as $3+\SU(2)\times \SU(2)+1$. In
 \cite{Bao:2013wqa} and in this paper, we showed that the fiber-base dual of  $\SU(3)$ with 6 flavors and with Chern-Simons level 0 is  $2+\SU(2)\times \SU(2)+2$. Ideas used in this paper will definitely be useful to further investigations in this direction.

Still another interesting future work involves  translating the result of this paper in the language of the AGT-W correspondence \cite{Alday:2009aq,Wyllard:2009hg}. Thanks to the 5D version of the AGT-W relation \cite{Awata:2009ur,Awata:2010yy} (see also \cite{Schiappa:2009cc,Mironov:2011dk,Itoyama:2013mca,Bao:2011rc,Nieri:2013yra,Bao:2013pwa,Nieri:2013vba,Aganagic:2013tta,Aganagic:2014oia,Taki:2014fva}), the partition functions of the 5D gauge theories are related to the correlation functions of the $q$-deformed Liouville and Toda SCFTs in 2D. The fiber-base duality implies, as we already suggested in \cite{Bao:2013wqa}, that gauge theories with different gauge groups are equivalent, meaning that the correlation functions of different Toda theories that are connected via the duality map have to agree. This implies that the appropriately normalized Toda correlators are invariant under the enhanced symmetry  as discussed for instance in \cite{Mitev:2014isa}.
Finally, since we have shown that not only the superconformal index
but also the Nekrasov partition function is invariant under the enhanced symmetry, one would expect the enhanced symmetry to be present at the level of chiral half of the correlation function as well.


\section*{Acknowledgments}

We thank Seok Kim,  Sung-Soo Kim,  Kimyeong Lee, Sara Pasquetti, Kazuhiro Sakai,  Yuji Tachikawa, Yutaka Yoshida and Gabi Zafrir for insightful comments and discussions. 
We are grateful to Chiung Hwang for giving us the explicit  result
of the Nekrasov partition function for the $\SP(1)$ theories with $N_f$ flavor.

FY is thankful to the following workshops and conferences, which were useful to complete this work:
``Liouville, Integrability and Branes (10)'' at APCTP,
``IV Workshop on Geometric Correspondences of Gauge Theories'' at SISSA,
``Autumn Symposium on String/M Theory'' at KIAS.
FY and MT are grateful to the workshops:
``String and Fields'' at YITP, Kyoto University,
``2014 Summer Simons Workshop in Mathematics and Physics''.
MT is supported by RIKEN iTHES Project.
EP is partially supported by the Marie Curie grant FP7-PEOPLE-2010-RG. VM acknowledges support from the Simons Center for Geometry and Physics, Stony Brook University, as well as of the C.N.\ Yang Institute for Theoretical Physics,  where some of the research for this paper was performed. V.M.\ acknowledges the support of the Marie Curie International Research Staff Exchange Network UNIFY of the European Union's Seventh Framework Programme [FP7-People-2010-IRSES] under grant agreement n°269217, which allowed him to visit Stony Brook University.

\appendix

\section{Special functions and identities}
\label{app:special}

\subsection{Definitions and basics}

For the reader's convenience, we collect here the definitions of the special functions used in the main text. First we mention those that are given by finite products
\begin{equation}
\label{eq:mainfunctionsdefinitions}
\begin{split}
\tilde{Z}_{\nu}(\ft,\fq)=& \prod_{i=1}^{\ell(\nu)}\prod_{j=1}^{\nu_i}\left(1-\ft^{\nu^t_j-i+1}\fq^{\nu_i-j}\right)^{-1},\\
\calN_{\lambda\mu}(Q;\ft,\fq)= &\prod_{i,j=1}^{\infty}\frac{1-Q\ft^{i-1-\lambda_j^t}\fq^{j-\mu_i}}{1-Q\ft^{i-1}\fq^j}\\=&\prod_{(i,j)\in\lambda}(1-Q\ft^{\mu_j^t-i}\fq^{\lambda_i-j+1})\prod_{(i,j)\in\mu}(1-Q\ft^{-\lambda_j^t+i-1}\fq^{-\mu_i+j}).
\end{split}
\end{equation}
In the above, the notation $(i,j)\in \mu$ means that $i$ runs over the rows of $\mu$, i.e. $i=1,\ldots, \ell(\mu)$, while $j$ runs over the columns, i.e. $j=1,\ldots, \mu_i$.
Then we also need $\calM$ that is given by an infinite product or by a Plethystic exponential as
\be
\calM(Q;\ft,\fq)=\prod_{i,j=1}^{\infty}(1-Q \ft^{i-1}\fq^{j})^{-1}=\exp\left[\sum_{n=1}^{\infty}\frac{Q^n}{n}\frac{\fq^n}{(1-\ft^n)(1-\fq^n)}\right]=\text{PE}\left[\frac{Q\fq}{(1-\ft)(1-\fq)}\right],
\ee
where the product converges for all $Q$ if $|\ft|<1$ and $|\fq|<1$ and the Plethystic exponential converges for all $\ft$ and all $\fq$ provided that $|Q|<\fq^{-1+\theta(|\fq|-1)}\ft^{\theta(|\ft|-1)}$ with $\theta(x)=1$ if $x>0$ and zero otherwise. The function $\calM$ can be defined for all $Q$, $\ft$ and $\fq$  if we require that
\be
\label{eq:inversionidentities}
\calM(Q;\ft^{-1},\fq)=\frac{1}{\calM(Q\ft;\ft,\fq)}, \qquad \calM(Q;\ft,\fq^{-1})=\frac{1}{\calM(Q\fq^{-1};\ft,\fq)}.
\ee
We shall often use the combined function
\be
\calR_{\lambda\mu}(Q;\ft,\fq)= \prod_{i,j=1}^{\infty}\left(1-Q \ft^{i-\frac{1}{2}-\lambda_j}\fq^{j-\frac{1}{2}-\mu_i}\right)=\calM(Q\sqrt{\frac{\ft}{\fq}};\ft,\fq)^{-1}\calN_{\lambda^t\mu}(Q\sqrt{\frac{\ft}{\fq}};\ft,\fq).
\ee

The following exchange properties can be useful
\be
\label{eq:exchangerelationMN}
\calM(Q;\fq,\ft)=\calM(Q\frac{\ft}{\fq};\ft,\fq), \quad \calN_{\lambda\mu}(Q;\fq,\ft)=\calN_{\mu^t\lambda^t}(Q\frac{\ft}{\fq};\ft,\fq),\quad \calR_{\lambda\mu}(Q;\fq,\ft)=\calR_{\mu\lambda}(Q;\ft,\fq).
\ee

\subsection{Topological vertex}

We use the $\Omega$ deformation parameters as
\begin{equation}
\label{eq:Omegaqt}
\mathfrak{q}=e^{-\beta \epsilon_1},
\qquad
\mathfrak{t}=e^{+\beta \epsilon_2},\qquad x=\sqrt{\frac{\fq}{\ft}}=e^{-\beta \frac{\epsilon_+}{2}}, \qquad y=\sqrt{\fq\ft}=e^{-\beta \frac{\epsilon_-}{2}},
\end{equation}
where $\epsilon_{\pm}=\epsilon_1\pm\epsilon_2$.
The refined topological vertex is given by
\begin{eqnarray}
\label{eq:topvertex}
C_{\lambda \mu \nu} (\ft,\fq)
&=& \fq^{\frac{||\mu||^2+||\nu||^2}{2}}
\ft^{-\frac{||\mu^t||^2}{2}} \tilde{Z}_{\nu}(\ft,\fq)
\nonumber \\
&& \qquad \times
\sum_{\eta} \left( \frac{\fq}{\ft} \right) ^{\frac{|\eta|+|\lambda|-|\mu|}{2}}
s_{\lambda^t/\eta} (\ft^{-\rho} \fq^{-\nu})
s_{\mu/\eta} (\fq^{-\rho} \ft^{-\nu^t}).
\end{eqnarray}
We remind that for a partition $\nu$, the vector  $\ft^{-\rho}\fq^{-\nu}$ is given by
\be
\ft^{-\rho}\fq^{\nu}=(\ft^{\frac{1}{2}}\fq^{-\nu_1},\ft^{\frac{3}{2}}\fq^{-\nu_2},\ft^{\frac{5}{2}}\fq^{-\nu_3}, \ldots).
\ee
The framing factors are defined as
\be
\label{framingFactors}
f_{\nu}(\ft,\fq)
= (-1)^{|\nu|} \ft^{\frac{||\nu^t||^2}{2}} \fq^{-\frac{||\nu||^2}{2}},
\qquad
\tilde{f}_{\nu} (\ft,\fq)
= \left( \frac{\ft}{\fq} \right) ^{\frac{|\nu|}{2}}
f_{\nu}(\ft,\fq).
\ee

\subsection{Flopping review}
\label{subsec:floppingreview}

The procedure of flopping involves sending the K\"ahler parameter $Q$ of one of the branes of the web diagram to $Q^{-1}$. This of course modifies the geometry of the web diagram and in particular the  K\"ahler parameters $Q_i$ of all the branes adjacent to the one being flopped are sent to $Q_iQ$, see figure~\ref{fig:floppingillustrated}.

In our previous article \cite{Bao:2013pwa}, we use the functions defined in \eqref{eq:mainfunctionsdefinitions} to write the topological string partition functions. In order to make the invariance under flopping $Q\rightarrow Q^{-1}$ as nice as possible, we will now introduce a new one.

First, let us consider the function $\calN_{\lambda\mu}$. Using the identities
$\sum_{(i,j)\in \lambda}i=\frac{1}{2}\left(||\lambda^t||^2+|\lambda|\right)$ and $\sum_{(i,j)\in \lambda}\mu_i=\sum_{i=1}^{\min\{\ell(\lambda),\ell(\mu)\}}\lambda_i\mu_i$
we obtain after a straightforward computation the expression
\be
\label{eq:QinvN}
\calN_{\lambda\mu}(Q^{-1};\ft,\fq)=\big(-Q^{-1}\sqrt{\frac{\fq}{\ft}}\big)^{|\lambda|+|\mu|}\ft^{\frac{-||\lambda^t||^2+||\mu^t||^2}{2}}\fq^{\frac{||\lambda||^2-||\mu||^2}{2}}\calN_{\mu\lambda}(Q\frac{\ft}{\fq};\ft,\fq).
\ee
Up to the zeta function regularization, we have
\be
\label{eq:QinvM}
\calM(Q^{-1};\ft,\fq)=\Big(-Q^{-1}\sqrt{\frac{\fq}{\ft}}\Big)^{\frac{1}{12}}\calM(Q\frac{\ft}{\fq},\ft,\fq).
\ee
Setting $\ft=\fq$ in \eqref{eq:QinvM}, we reproduce formula (42) in \cite{Iqbal:2004ne} for the unrefined case. Combining both \eqref{eq:QinvN} and \eqref{eq:QinvM}, we arrive at the following compact expression for the functions $\calR_{\lambda\mu}$ of \eqref{eq:mainfunctionsdefinitions}:
\be
\label{eq:QinvR}
\calR_{\lambda\mu}(Q^{-1};\ft,\fq)=\big(-Q\big)^{\frac{1}{12}-|\lambda|-|\mu|}\ft^{\frac{||\mu^t||^2-||\lambda||^2}{2}}\fq^{\frac{||\lambda^t||^2-||\mu||^2}{2}}\calR_{\mu^t\lambda^t}(Q;\ft,\fq).
\ee
We see that if we define a new function as
\be
\label{eq:deffunctionS}
\calS_{\lambda\mu}(Q;\ft,\fq)=(-1)^{|\lambda|}Q^{-\frac{|\lambda|+|\mu|}{2}}\ft^{\frac{||\lambda^t||^2}{2}}\fq^{\frac{||\mu||^2}{2}}\calR_{\lambda^t\mu}(Q;\ft,\fq),
\ee
then we have the following nice behavior under flopping:
\be
\label{eq:Sflopping}
\calS_{\lambda\mu}(Q^{-1};\ft,\fq)=(-Q)^{\frac{1}{12}}\calS_{\mu\lambda}(Q;\ft,\fq).
\ee

\section{On the Coulomb moduli expansion}
\label{app:notecoulombmoduliexpansion}

In order to show the symmetry enhancement, we expanded in section \ref{sec:fiberbaseinvariance} the Nekrasov partition function in terms of the invariant Coulomb moduli parameter $\tilde{A}$.
For simplicity, we first assumed that the coefficients of $\tilde{A}^{m}$ are completely determined by the $k$ instanton contributions with 
$k \le m/2$ for $N_f=0$,
$k \le m$ for $1 \le N_f \le 4$,
$k \le 2m$ for $N_f = 5,6$
and $k \le 4m$ for $N_f=7$.
We checked this assumption experimentally by expanding to a few more orders with Mathematica.
In this appendix, we give an analytical justification of this assumption, which is valid at least up to $N_f \le 4$.

\subsection{Pure SYM}

We warm up with the pure SYM case, for which it is straightforward to prove our assumption.
We  want to study the $\tilde{A}$ expansion of the topological string amplitude  \eqref{pureSYM} that is given by an expression of the form
\begin{eqnarray}
\sum_{\mu_1,\mu_2,\nu_1,\nu_2} Q_B^{|\mu_1|+|\mu_2|} Q_F^{|\nu_1|+|\nu_2|}
\, G(\fq,\ft;\mu_1,\mu_2,\nu_1,\nu_2) \, .
\label{pure-fn}
\end{eqnarray}
Using  the parametrization  \eqref{eq:defQFQB} of $Q_B$ and $Q_F$ in terms of $\tilde{A}^2$, we see that the $\tilde{A}^{2n}$ coefficient is obtained from the terms with
\begin{eqnarray}
|\mu_1|+|\mu_2|+|\nu_1|+|\nu_2| = n.
\label{n-Young}
\end{eqnarray}
Performing the summation over the Young diagram
$\nu_1$ and $\nu_2$ in (\ref{pure-fn}), we reproduce the standard form of the Nekrasov partition function, which is an expansion in terms of the instanton factor. Since the instanton factor is included only in $Q_B$,
the $k$-instanton contribution to the Nekrasov partition function
comes from the terms in (\ref{pure-fn}) satisfying
\begin{eqnarray}
|\mu_1|+|\mu_2| = k.
\label{k-Young}
\end{eqnarray}
The terms satisfying
(\ref{n-Young}) include only the terms satisfying
(\ref{k-Young}) with $k \le n$.
Therefore, in order to obtain $\tilde{A}^{2n}$ terms
from the Nekrasov partition function,
it is enough to consider the $k$-instanton contribution with $k \le n$.

\subsection{The four flavor case}
\label{4flavorInst}

The generalization of the statement above for the cases with up to two flavors is straightforward.
The situation is slightly different for $N_f \ge 3$.
Here we will discuss the case with $N_f=4$ as an example.
The main difference stems from the fact that the toric diagram has external lines that are parallel.

\begin{figure}
\centering
\includegraphics[width=10cm]{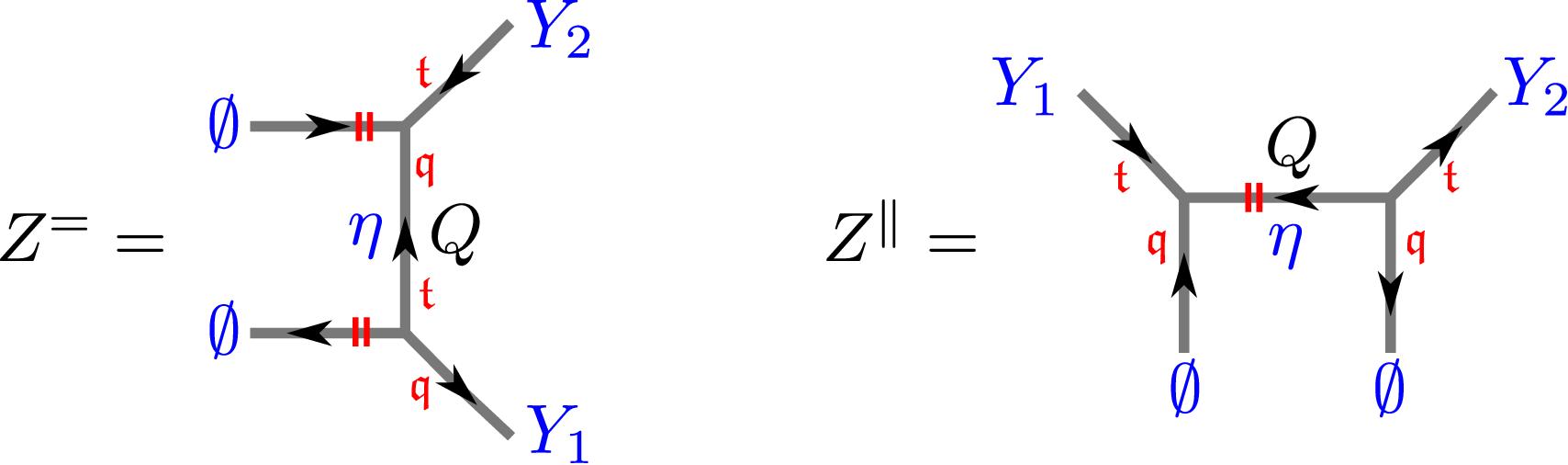}
\caption{Subamplitude}
\label{subamp}
\end{figure}

In order to compute the Nekrasov partition function in the context of the $\tilde{A}$ expansion,
it is convenient to first define the following
``normalized'' sub-amplitudes
\begin{eqnarray}
Z^{=}_{Y_1 Y_2}(Q;\ft,\fq) \colonequals
\frac{Z'{}^{=}_{Y_1 Y_2}(Q;\ft,\fq)}{Z'{}^{=}_{\emptyset \emptyset}(Q;\ft,\fq)},
\qquad
Z^{||}_{Y_1 Y_2}(Q;\ft,\fq) \colonequals
\frac{Z'{}^{||}_{Y_1 Y_2}(Q;\ft,\fq)}{Z'{}^{||}_{\emptyset \emptyset}(Q;\ft,\fq)}\, ,
\label{n-Hopf-sub}
\end{eqnarray}
where
\begin{eqnarray}
Z'{}^{=}_{Y_1 Y_2}(Q;\ft,\fq)
&\equiv&\sum_{\eta} (-Q)^{| \eta |}
C_{Y_1{}^T \eta^T \emptyset}(\ft,\fq)
C_{\eta Y_2 \emptyset}(\ft,\fq)
\tilde{f}_{\eta}(\fq,\ft){}^{-1}\ ,
\nonumber \\
Z'{}^{||}_{Y_1 Y_2}(Q;\ft,\fq)
&\equiv&\sum_{\eta} (-Q)^{| \eta |}
C_{\emptyset Y_1{}^T \eta^T}(\ft,\fq)
C_{Y_2 \emptyset \eta}(\ft,\fq)
f_{\eta}(\fq,\ft){}^{-1}\ .
\label{Hopf-sub}
\end{eqnarray}
and the refined topological vertex is given in \eqref{eq:topvertex}.
These functions correspond to the amplitudes of the two toric diagrams
depicted in figure~\ref{subamp}.

Such topological amplitudes, but without the framing factors $f_{\eta}(\fq,\ft)$ and $\tilde{f}_{\eta}(\fq,\ft)$ (defined in \eqref{framingFactors}),
were computed in \cite{Gukov:2007tf} in the context of the refined Hopf link and its relation to the S-matrix of the refined Chern-Simons theory \cite{Aganagic:2011sg}.
Using a slightly modified version of the computation in \cite{Gukov:2007tf},
we can show that (\ref{n-Hopf-sub}) are also
degree $|Y_1|+|Y_2|$ polynomials of $Q$
even though the summation for $\eta$ in (\ref{Hopf-sub})
is taken over all possible partitions.
Higher order terms cancel against each other
and the Taylor expansion terminates at a finite order.
Knowing this property, we see that we need
only the terms in the sums of (\ref{Hopf-sub}) with
\begin{eqnarray}
|\eta| \le |Y_1| + |Y_2|
\label{poly}
\end{eqnarray}
in order to compute (\ref{n-Hopf-sub}) explicitly.

It is straightforward to see graphically that
the toric diagram for $\SU(2)$ with four flavor, depicted in the left of figure~\ref{fig:4fullyflopped}, can be constructed by combining the sub-diagrams in figure~\ref{subamp}, and that it reads
\begin{align}
\label{whole}
Z^{N_f=4} &= \sum_{Y_1,  Y_2, Y_3, Y_4}
(-Q_1{}^{-1})^{|Y_1|} (-Q_2{}^{-1})^{|Y_2|}
(-Q_3{}^{-1})^{|Y_3|} (-Q_4{}^{-1})^{|Y_4|}\nonumber\\
&\times Z^{=}_{Y_1 Y_2}(Q_1 Q_2 Q_F;\ft,\fq)
Z^{||}_{Y_2 Y_3}(Q_2 Q_3 Q_B;\ft,\fq) \\
&\times Z^{=}_{Y_3 Y_4}(Q_3 Q_4 Q_F;\ft,\fq) Z^{||}_{Y_4 Y_1}(Q_1 Q_4 Q_B;\ft,\fq).\nonumber
\end{align}
We use the ``normalized'' amplitude (\ref{n-Hopf-sub})
rather than (\ref{Hopf-sub}) itself
because, as we have discussed in \cite{Bao:2013wqa}, we should remove the decoupled degrees of freedom, referred to as the  ``non-full spin content'', coming from the parallel external legs.

Since the $\tilde{A}$ dependence of the parameters are $Q_i^{-1} \propto \tilde{A}$, and $Q_F, Q_B \propto \tilde{A}^2$ as given in \eqref{eq:param4flavor}, the $\tilde{A}^m$ contribution in (\ref{whole})
comes from the terms with
\begin{eqnarray}
|Y_1|+|Y_2|+|Y_3|+|Y_4| = m.
\label{4f-pA}
\end{eqnarray}
Thus, the summand in (\ref{whole}) is a polynomial of
degree $(|Y_2|+|Y_3|)+(|Y_1|+|Y_4|)$ in  $Q_B$ due to (\ref{poly}).
Since the instanton factor is included only \eqref{eq:paramfourflavor} in
the K\"ahler parameter $Q_B$, it means that  only the $k \le m$ instanton contributions are included. Therefore, we find that in order to obtain the $\tilde{A}$ expansion up to the order of $m$ from the Nekrasov partition function, only the $k$-instanton contributions with $k \le m$ are necessary.

\subsection{Computation of the normalized partition function $Z^{=}$}

In the previous subsection \ref{4flavorInst}, we used the property
that the sub-amplitude $Z^{=}_{Y_1,Y_2}(Q;\fq,\ft)$ defined in (\ref{n-Hopf-sub}) is a finite polynomial in $Q$ of order $|Y_1|+|Y_2|$.
In this subsection, we show this property  following the discussion in  \cite{Iqbal:2011kq}.
Writing down $Z'{}^{=}_{Y_1,Y_2}(Q;\fq,\ft)$ explicitly,
by using the definition of the refined topological vertex, we obtain
\begin{eqnarray}
\label{Z=}
Z'{}^{=}_{Y_1 Y_2}(Q;\ft,\fq)
&=&
\fq^{\frac{||Y_2||^2}{2}}
\ft^{-\frac{||Y_2^t||^2}{2}}
\left( \frac{\fq}{\ft} \right) ^{\frac{|Y_1|-|Y_2|}{2}}
\sum_{\eta,\eta',\eta''}
\left( \sqrt{\frac{\ft}{\fq}} Q \right)^{|\eta|}
\left( \frac{\fq}{\ft} \right) ^{\frac{|\eta'|+|\eta''|}{2}}
\nonumber \\
&& \qquad
\times s_{Y_1/\eta'} (\ft^{-\rho} )
s_{\eta^t/\eta'} (\fq^{-\rho} )
s_{\eta^t/\eta''} (\ft^{-\rho} )
s_{Y_2/\eta''} (\fq^{-\rho} )
\, .
\end{eqnarray}
As in  \cite{Iqbal:2011kq}, we moreover define the open string partition function
\begin{eqnarray}
\tilde{Z}'{}^{=}(Q;\ft,\fq;\fx,\fy)
&\equiv& \sum_{Y_1,Y_2}
\left(
\fq^{-\frac{||Y_2||^2}{2}}
\ft^{\frac{||Y_2^t||^2}{2}}
\left( \frac{\ft}{\fq} \right) ^{\frac{|Y_1|-|Y_2|}{2}}
Z'{}^{=}_{Y_1 Y_2}(Q;\ft,\fq) \right)
s_{Y_1}(\fx) s_{Y_2}(\fy)
\nonumber \\
&=& \sum_{Y_1,Y_2} \sum_{\eta,\eta',\eta''}
\left( \sqrt{\frac{\ft}{\fq}} Q \right)^{|\eta|}
\left( \frac{\fq}{\ft} \right) ^{\frac{|\eta'|+|\eta''|}{2}}
\nonumber \\
&& \times
s_{Y_1/\eta'} (\ft^{-\rho} )
s_{\eta^t/\eta'} (\fq^{-\rho} )
s_{\eta^t/\eta''} (\ft^{-\rho} )
s_{Y_2/\eta''} (\fq^{-\rho} )
s_{Y_1}(\fx) s_{Y_2}(\fy),
\label{Z=tilde}
\end{eqnarray}
where we divided by the prefactor $\fq^{\frac{||Y_2||^2}{2}}
\ft^{-\frac{||Y_2^t||^2}{2}}
\left( \fq/\ft \right) ^{\frac{|Y_1|-|Y_2|}{2}}$ in front of the summation in (\ref{Z=}) for simplicity. Performing the sums we obtain
\begin{eqnarray}
\tilde{Z}'{}^{=}(Q;\ft,\fq;\fx,\fy)
&=&
\prod_{i,j=1}^{\infty}(1 - \fx_i \ft^{j-\frac{1}{2}})^{-1}
(1 - \fy_i \fq^{j-\frac{1}{2}})^{-1}
(1 - Q \fq^{i-1} \ft^i)^{-1}
\nonumber \\
&&
\times \prod_{i,j=1}^{\infty}(1 - Q \fx_i \ft^{j-\frac{1}{2}})^{-1}
(1 - Q \fy_i \fq^{j-\frac{1}{2}})^{-1}
\left( 1 - \sqrt{ \frac{\fq}{\ft} } Q \fx_i \fy_j \right)^{-1}
\end{eqnarray}
which implies that the normalized version obeys
\begin{eqnarray}
\tilde{Z}{}^{=}(Q;\ft,\fq;\fx,\fy)
\equiv \frac{\tilde{Z}'{}^{=}(Q;\ft,\fq;\fx,\fy)}{\tilde{Z}'{}^{=}(Q;\ft,\fq;0,0)}
&=&
\prod_{i,j=1}^{\infty}(1 - \fx_i \ft^{j-\frac{1}{2}})^{-1}
(1 - \fy_i \fq^{j-\frac{1}{2}})^{-1}
(1 - Q \fx_i \ft^{j-\frac{1}{2}})^{-1}\nonumber \\
&&
\times\prod_{i,j=1}^{\infty}(1 - Q \fy_i \fq^{j-\frac{1}{2}})^{-1}\left( 1 - \sqrt{\frac{\fq}{\ft}} Q \fx_i \fy_j \right)^{-1}
\, .
\end{eqnarray}
We can immediately expand this product  in powers of $\fx_i$ and $\fy_j$ and find  that their coefficients are polynomials of $Q$ of degree less or equal to the sum of the power of $\fx_i$ and $\fy_j$.
This means that when we expand $\tilde{Z}^{=}$ in terms of Schur functions exactly as in (\ref{Z=tilde}),
the coefficient $Z^{=}_{Y_1,Y_2}(Q;\fq,\ft)$ that multiplies $s_{Y_1}(\fx)s_{Y_2}(\fy)$ has to be a polynomial in $Q$ of degree $|Y_1|+|Y_2|$.

\subsection{Computation of  the normalized partition function  $Z^{||}$}

Using the definition of the refined topological vertex \eqref{eq:topvertex},
the amplitude defined in (\ref{Hopf-sub}) is explicitly given by
\begin{eqnarray}
Z'{}^{||}_{Y_1 Y_2}(Q;\ft,\fq)
&=&
(-1)^{|Y_1|} f_{Y_1} (\fq,\ft)^{-1}
\left( \frac{\fq}{\ft} \right) ^{\frac{1}{2}(|Y_1|+|Y_2|)}
\sum_{\eta} (-Q)^{| \eta |}
f_{\eta}(\ft,\fq){}^{-1}
\nonumber \\
&& \,\, \times
P_{\eta}(\ft^{-\rho}; \fq,\ft) P_{\eta^T}(\fq^{-\rho}; \ft,\fq)
s_{Y_1{}^T} (\ft^{-\rho} \fq^{-\eta}) s_{Y_2{}^T} (\ft^{-\rho} \fq^{-\eta})\ .
\end{eqnarray}
Factoring out the overall factor, we concentrate on
\begin{eqnarray}
Z_{Y_1 Y_2}^{''||}
&=&
\sum_{\eta} (-Q)^{| \eta |}
P_{\eta}(\ft^{+\rho}; \fq,\ft) P_{\eta^T}(\fq^{-\rho}; \ft,\fq)
s_{Y_1{}^T} (\ft^{-\rho} \fq^{-\eta}) s_{Y_2{}^T} (\ft^{-\rho} \fq^{-\eta})\, ,
\label{Z}
\end{eqnarray}
where we absorbed the framing factor $f_{\eta}(\ft,\fq){}^{-1}$
in the Macdonald polynomial $P_{\eta}(\ft^{-\rho}; \fq,\ft)$.
The case without this framing factor is already computed\footnote{See also \cite{Iqbal:2011kq}.} by \cite{Awata:2009sz} and the calculation in our case is done analogously. We thus define the operator,
\begin{eqnarray}
H^{r}(\fx,\ft^{-1}) \equiv e_r (\ft^{- \rho}) \sum_{\lambda, \ell(\lambda) \le r }
P_{\lambda} (\fx(p);\ft)
\frac{\ft^{-|\lambda|} \langle P_{\lambda} , P_{\lambda} \rangle _{r,\ft}''}%
{\big(\langle P_{\lambda} , P_{\lambda} \rangle _{\ft}\big)^2}
P_{\lambda} (\fx(p^*); \ft).
\end{eqnarray}
Here, $\fx(p)$ indicates that the Macdonald function,
which is symmetric in the variable $\fx$,
should be rewritten in terms of the power sum function $p_n(\fx) = \sum_i \fx_i^n$.
The derivative operator $p^*$ is defined as
\begin{eqnarray}
p_n^* = \frac{1-\fq^n}{1-\ft^n} n \frac{\partial}{\partial p_n}\,.
\end{eqnarray}
The inner product $\langle * , * \rangle _t $ is given as
\begin{eqnarray}
\langle P_{\lambda} (\fx,\ft) , P_{\mu}(\fx,\ft) \rangle _{\ft}
= \delta_{\lambda \mu} \prod_{j \ge 1} \prod_{i=1}^{m_j} \frac{1}{1-\ft^i}\, ,
\end{eqnarray}
while the other inner product $\langle * , * \rangle _{r,\ft}''$
is given in (2.5) of  \cite{Awata:2009sz}  as
\begin{eqnarray}
\langle P_{\lambda} (\fx,\ft) , P_{\mu}(\fx,\ft) \rangle _{r,\ft}''
= \delta_{\lambda\mu} v_{\lambda}(\ft)^{-1 } [r]!_{\ft}\, ,
\end{eqnarray}
where $[N]!_{\ft}=[1]_{\ft}[2]_{\ft} \cdots [r]_{\ft}$ with $[N]_{\ft} = (1-\ft^N)/(1-\ft)$
and $v_{\lambda} (\ft) = \prod_{j \ge 0} [m_j]!_\ft$
with $m_j = \# \{ \lambda_i | \lambda_i = j \} $.
It was shown in  \cite{Awata:2009sz} that the operators $H^r$ commute with each other and that the Macdonald polynomials are their common  eigenfunctions with the eigenvalue given by
\begin{eqnarray}
H^r(\fx,\ft^{-1}) P_{\lambda}(\fx; \fq, \ft)
 =  e_r(\ft^{-\rho}q^{-\lambda}) P_{\lambda}(\fx; \fq, \ft).
\end{eqnarray}
The Schur function is expanded in terms of the elementary symmetric function $e_\mu$ as
\begin{eqnarray}
s_{\lambda} (\fx) = \sum_{\mu \le \lambda} V_{\lambda, \mu} e_{\mu^T} (\fx)
\end{eqnarray}
where $e_{\lambda} \equiv e_{\lambda_1} e_{\lambda_2} \cdots $.
Then it is straightforward to show that
\bea
Z^{''||}_{Y_1 Y_2}
&=&
\sum_{Y_1 \le \nu} \sum_{Y_2 \le \sigma}
V_{Y_1{}^T, \nu}  V_{Y_2{}^T, \sigma}
H^{\nu^T}(\fx, \ft^{-1})
H^{\sigma^T}(\fx, \ft^{-1})
 \nonumber\\&&\times \left.\exp \left( - \sum_{n=1}^{\infty}
\frac{1}{n!} Q^n
p_n(\fq^{-\rho}) p_n(\fx)
\right)\right| _{\fx=\ft^{+\rho}}
\eea
where we defined $H^{\lambda} = H^{\lambda_1} H^{\lambda^2} \cdots$ and
$p_n(\ft^{-\rho}) = \ft^{\frac{n}{2}}/(1-\ft^n)$. Taking into account that $H^{\lambda}$ is a linear combination
of derivatives $\frac{\partial}{\partial p_n}$,
we find that the normalized amplitude $Z{}^{||}_{Y_1,Y_2}(Q;\fq,\ft)$
is a degree $|Y_1|+|Y_2|$ polynomial in $Q$.

\section{On character formulas}
\label{app:characters}

In this appendix, for the convenience of the reader we summarize the formulas for the characters as well as our conventions that we have used in section \ref{Nfg5}.

We begin with the $N_f=5$ case. We define the $\SO(10)$ characters as
\begin{align}
&\chi_{10}^{\SO(10)} = \sum_{I=1}^{10} \tilde{M}_I,&
&\chi_{16}^{\SO(10)}  = \sum_{\{s_i = \pm \} \atop \sum s_i = 1 \,\, { \rm mod} \,\, 4} \prod_{i=1}^5 \tilde{m}_i{}^{\frac{s_i}{2} },&
\nonumber\\
&\chi_{\overline{16}}^{\SO(10)} 
=  \sum_{\{s_i = \pm \} \atop \sum s_i = -1 \,\,  {\rm mod} \,\, 4} \prod_{i=1}^5 \tilde{m}_i{}^{\frac{s_i}{2} }, &
 &
\chi_{120}^{\SO(10)} 
= \sum_{ I<J<K =1}^{10} \tilde{M}_I \tilde{M}_J \tilde{M}_K,&\\
&\chi_{45}^{\SO(10)} = \sum_{I<J=1}^{10} \tilde{M}_I\tilde{M}_J,& &\chi_{144}^{\SO(10)}=\chi_{10}^{\SO(10)}\chi_{\overline{16}}^{\SO(10)}-\chi_{16}^{\SO(10)},&\nonumber
\end{align}
where we used the convention that 
$\tilde{M}_I = 
\tilde{m}_I $ for $I=1,\ldots 5$ and 
$\tilde{M}_I=\tilde{m}_{I-6}$ for $I=6,\ldots, 10$. Furthermore, we have used the exponentiated masses $\tilde{m}_i=e^{-\beta m_i}$. 
The perturbative contributions for five flavors case are given in \eqref{eq:pertpartgeneralNf}  $\chi^{\SO(2N_f)}_{\rm fund}= \chi^{\SO(2N_f)}_{10}$ and with $\chi_{\rm antisym}^{\SO(2N_f)}=\chi_{45}^{\SO(2N_f)}$.
For our expansion \eqref{eq:finalNf5expansion} up to $\tilde{A}^2$, the only $\E_6$ character decomposition formulas that we need are
\begin{align}
\label{eq:E6characters}
\chi_{27}^{\E_6}&=q^{-\frac{4}{3}}+\chi^{\SO(10)}_{10}q^{\frac{2}{3}}+\chi^{\SO(10)}_{16}q^{-\frac{1}{3}},\quad \chi_{\overline{27}}^{\E_6}=q^{\frac{4}{3}}+\chi^{\SO(10)}_{10}q^{-\frac{2}{3}}+\chi^{\SO(10)}_{\overline{16}}q^{\frac{1}{3}},\\ \chi_{351}^{\E_6}&=\chi^{\SO(10)}_{10}q^{\frac{2}{3}}+\chi^{\SO(10)}_{16}q^{-\frac{1}{3}}+\chi^{\SO(10)}_{\overline{16}}q^{\frac{5}{3}}+\chi^{\SO(10)}_{45}q^{-\frac{4}{3}}+\chi^{\SO(10)}_{120}q^{\frac{2}{3}}+\chi^{\SO(10)}_{144}q^{\frac{1}{3}}.\nonumber
\end{align}
Thus, multiplying the perturbative part \eqref{eq:pertpartgeneralNf} with the instanton contributions \eqref{eq:fiveflavorinstantons}, replacing $\tilde{a}=e^{-\beta a}$ with $\tilde{A}$ by using \eqref{eq:invariantCoulombmodulusNf5} and using the character identities \eqref{eq:E6characters} leads to the five flavor partition function expressed in $\E_6$ characters \eqref{eq:finalNf5expansion}. 

Next, for the $N_f=6$ case, we  define the $\SO(12)$ characters as
\begin{align}
&\chi_{12}^{\SO(12)}
= \sum_{I=1}^{12} \tilde{M}_I,&
&\chi_{32}^{\SO(12)}
= \sum_{\{s_i = \pm \} \atop \sum s_i = 2 \,\, { \rm mod} \,\, 4} \prod_{i=1}^6 \tilde{m}_i{}^{\frac{s_i}{2} },&
& \chi_{\overline{32}}^{\SO(12)}
=  \sum_{\{s_i = \pm \} \atop \sum s_i = 0 \,\,  {\rm mod} \,\, 4} \prod_{i=1}^6 \tilde{m}_i{}^{\frac{s_i}{2} },&
\nonumber \\
&\chi_{66}^{\SO(12)} = \sum_{ I<J =1}^{12} \tilde{M}_I \tilde{M}_J,&
 &\chi_{77}^{\SO(12)}=\sum_{ I\leq J =1}^{12} \tilde{M}_I \tilde{M}_J-1,& & \chi_{495}^{\SO(12)}
= \sum_{I_1<\cdots<I_4=1}^{12} \prod_{k=1}^4\tilde{M}_{I_k},&
\end{align}
as well as $\chi_{{352}}^{\SO(12)}=\chi_{12}^{\SO(12)}\chi_{\overline{32}}^{\SO(12)}-\chi_{{32}}^{\SO(12)}$. Here, we have used the convention that
$\tilde{M}_I = 
\tilde{m}_I $ for $I=1,\ldots 6$ and 
$\tilde{M}_I=\tilde{m}_{I-7}$ for $ I=7,\ldots, 12$. 
In order to write down the partition function \eqref{eq:finalNf6expansion}, we need the following $\E_7$ character decomposition
\be
\begin{split}
\label{eq:E7characters}
\chi_{56}^{\E_7}&=\chi_2^{\SU(2)}(q)\chi^{\SO(12)}_{12}+\chi^{\SO(12)}_{\overline{32}}\, ,\\
\chi_{133}^{\E_7}&=\chi^{\SU(2)}_3(q)+\chi^{\SU(2)}_2(q)\chi^{\SO(12)}_{{32}}+\chi^{\SO(12)}_{66}\, ,\\
\chi_{1539}^{\E_7}&=\chi^{\SU(2)}_3(q) \chi^{\SO(12)}_{66}+\chi^{\SU(2)}_2(q)\left( \chi^{\SO(12)}_{32}+ \chi^{\SO(12)}_{{352}}\right)+\chi^{\SO(12)}_{495}+\chi^{\SO(12)}_{77}+1\, ,
\end{split}
\ee
where $\chi_2^{\SU(2)}(q)=q+q^{-1}$ and $\chi_3^{\SU(2)}(q)=q^2+1+q^{-2}$.
Thus, multiplying the perturbative part \eqref{eq:pertpartgeneralNf} with the instanton contributions \eqref{eq:sixflavorinstantons}, using the invariant Coulomb modulus \eqref{eq:invariantCoulombModulusNf6} and the the character identities \eqref{eq:E7characters} leads to the six  flavor partition function expressed in $\E_7$ characters \eqref{eq:finalNf6expansion}. 

Finally, for the $N_f=7$ case we define the $\SO(14)$ characters as
\begin{align}
&\chi_{14}^{\SO(14)}
= \sum_{I=1}^{14} \tilde{M}_I,& &\chi_{64}^{\SO(14)}
= \sum_{\{s_i = \pm \} \atop \sum s_i = -1 \,\, { \rm mod} \,\, 4} \prod_{i=1}^7 \tilde{m}_i{}^{\frac{1}{2} s_i},&
\nonumber\\
&\chi_{\overline{64}}^{\SO(14)}
=  \sum_{\{s_i = \pm \} \atop \sum s_i = 1 \,\,  {\rm mod} \,\, 4} \prod_{i=1}^7 \tilde{m}_i{}^{\frac{s_i}{2} },&
&\chi_{91}^{\SO(14)}
= \sum_{1\le I<J \le 14} \tilde{M}_I \tilde{M}_J,&
\end{align}
where we defined
$\tilde{M}_I = \tilde{m}_I $ for $I=1,\ldots 7$ and 
$\tilde{M}_I=\tilde{m}_{I-8}$ for $ I=8,\ldots, 14$.
The only $\E_8$ character decomposition that we need for \eqref{eq:finalNf7expansion} is 
\be
\chi^{\E_8}_{248}=1+q\chi^{\SO(14)}_{64}+q^{-1}\chi^{\SO(14)}_{\overline{64}}+\chi^{\SO(14)}_{91}+(q^2+q^{-2})\chi^{\SO(14)}_{14}.
\ee

\bibliographystyle{JHEP}
\providecommand{\href}[2]{#2}\begingroup\raggedright\endgroup

\end{document}